\documentclass[3p]{elsarticle}

\usepackage[utf8]{inputenc}
\usepackage{amsmath}
\usepackage{amsfonts}
\usepackage{amssymb}
\usepackage{siunitx}
\usepackage{graphicx}
\usepackage{subcaption}
\usepackage{float}
\usepackage{appendix}
\usepackage{bm} % bold typeset for math
\usepackage{algorithm}
\usepackage{cases}
\usepackage[noend]{algpseudocode}
\usepackage{xcolor}
\usepackage{indentfirst}
\usepackage{url}
\usepackage{caption}
\usepackage{wrapfig}
\usepackage{dsfont}
\usepackage{natbib}

\begin{document}

\definecolor{myGreen}{rgb}{0.3,0.7,0}
\newcommand{\commentOut}[1]{}
\newcommand{\frederic}[1]{\textcolor{red}{#1}}
\newcommand{\elyce}[1]{\textcolor{magenta}{#1}}
\newcommand{\raphael}[1]{\textcolor{myGreen}{#1}}
\newcommand{\bluetext}[1]{\textcolor{blue}{#1}}

\newcommand{\added}[1]{\textcolor{blue}{#1}}

\newcommand{\revised}[1]{\textcolor{black}{#1}}
\definecolor{revGreen}{rgb}{0.3,0.7,0.15}
\newcommand{\Irevised}[1]{\textcolor{black}{#1}}

\newcommand{\todo}[1]{\textcolor{magenta}{#1}}

% Custom variables: ------------------------
% Domains 
\newcommand{\domain}{\Omega}
\newcommand{\domainAtN}{\Omega^{n}}
\newcommand{\domainAtNP}{\Omega^{n+1}}

\newcommand{\interface}{\Gamma}
\newcommand{\interfaceAtN}{\Gamma^{n}}
\newcommand{\interfaceAtNP}{\Gamma^{n+1}}

\newcommand{\liqDomain}{\Omega_{l}}
\newcommand{\liqDomainAtN}{\Omega_{l}^{n}}
\newcommand{\liqDomainAtNP}{\Omega_{l}^{n+1}}
\newcommand{\rVec}{\bm{r}}

\newcommand{\solDomain}{\Omega_{s}}
\newcommand{\solDomainAtN}{\Omega_{s}^{n}}
\newcommand{\solDomainAtNP}{\Omega_{s}^{n+1}}

\newcommand{\temperaturel}{T_{l}}
\newcommand{\temperatures}{T_{s}}

\newcommand{\Tl}{T_{l}}
\newcommand{\rhol}{\rho_{l}}
\newcommand{\kl}{k_{l}}
\newcommand{\cpl}{C_{p,l}}
\newcommand{\alphal}{\alpha_{l}}

\newcommand{\Ts}{T_{s}}
\newcommand{\rhos}{\rho_{s}}
\newcommand{\ks}{k_{s}}
\newcommand{\cps}{C_{p,s}}
\newcommand{\alphas}{\alpha_{s}}

\newcommand{\vint}{\textbf{V}_{\Gamma}}
\newcommand{\normal}{\textbf{n}}

\newcommand{\vns}{\bm{v}}
\newcommand{\mul}{\mu_{l}}
\newcommand{\press}{P}

\newcommand{\divergence}[1]{\nabla \cdot #1}
\newcommand{\laplace}{\nabla^{2}}
\newcommand{\advection}[2]{#1 \cdot \nabla #2}
\newcommand{\ddt}[1]{\frac{\partial #1}{\partial t}}
% Non-dimensionalized:

\newcommand{\vNSNondim}{\bar{\vNS}}
\newcommand{\vInterfaceNondim}{\bar{\vInterface}}
\newcommand{\PNondim}{\bar{P}}
\newcommand{\rNondim}{\bar{\rVec}}
\newcommand{\kappaNondim}{\bar{\kappa}}

\newcommand{\liqTempNondim}{\theta_{l}}
\newcommand{\solTempNondim}{\theta_{s}}

\newcommand{\tNondim}{\bar{t}}
\newcommand{\nablaNondim}{\bar{\nabla}}
\newcommand{\divergenceNondim}{\nablaNondim \cdot}
\newcommand{\laplaceNondim}{\nablaNondim^{2}}

% Derivatives:
\newcommand{\dFd}[1]{\partial F_{#1}}
\newcommand{\dTwoFd}[1]{\partial^2 F_{#1 #1}}

% Timesteps:
\newcommand{\dtn}{\Irevised{\Delta t_{n+1}}}
\newcommand{\dtnm}{\Irevised{\Delta t_{n}}}

\newcommand{\re}{\textrm{Re}}
\newcommand{\st}{\textrm{St}}

\newcommand{\citea}[1]{\citeauthor{#1}}
\newcommand{\etal}{\emph{et al.}}
% -------------------------------------------

\title{A Sharp Numerical Method for the Simulation of Stefan Problems with Convective Effects}

\author[ME]{Elyce Bayat\corref{cor1}}
\author[ME]{Raphael Egan}
\author[ME]{Daniil Bochkov}
\author[ME]{Alban Sauret}
\author[ME,CS]{Frederic Gibou}

\address[ME]{Department of Mechanical Engineering, University of California, Santa Barbara, CA 93106, USA}
\address[CS]{Department of Computer Science, University of California, Santa Barbara, CA 93106, USA}

\cortext[cor1]{Corresponding author: ebayat@ucsb.edu}

\begin{abstract}
We present a numerical method for the solution of interfacial growth governed by the Stefan model coupled with incompressible fluid flow. An algorithm is presented which takes special care to enforce sharp interfacial conditions on the temperature, the flow velocity and pressure, and the interfacial velocity. The approach utilizes level-set methods for sharp and implicit interface tracking, hybrid finite-difference/finite-volume discretizations on adaptive quadtree grids, and a pressure-free projection method for the solution of the incompressible Navier-Stokes equations. The method is first verified with numerical convergence tests using a synthetic solution.  Then, computational studies of ice formation on a cylinder in crossflow are performed and provide good quantitative agreement with existing experimental results, reproducing qualitative phenomena that have been observed in past experiments. Finally, we investigate the role of varying Reynolds and Stefan numbers on the emerging interface morphologies and provide new insights around the time evolution of local and average heat transfer at the interface. 
\end{abstract}

\begin{keyword}
Level-set method \sep Quadtree \sep Stefan \sep Navier-Stokes
\end{keyword}

\maketitle

\section{Introduction}
Interfacial evolution driven by heat and mass transfer is governed by the classical Stefan problem, which describes such evolution driven by diffusive processes. An area of interest is the extension of the Stefan problem in the presence of convective processes, and the effect that such processes may have on the evolution of the interface. The consideration of convective effects leads to a complex coupling between the interface shape and fluid flow, giving rise to a fascinating coevolution of interface morphology and corresponding flow dynamics \cite{ristroph2012sculpting, ristroph2018sculpting, mac2015shape, huang2021a, wang2021how}. The advective transport of either energy or species can have a unique effect on the interface shape and vice versa; the changes in interface can lead to changes in flow and heat transfer behavior.  This type of phenomena is present in a wide variety of applications -- from water, energy \cite{epstein1983complex, hirata2000crystal}, 
 and metallurgical systems \cite{melissari2004identification, kumar2010heat} to the formation of natural landscapes \cite{ristroph2012sculpting, ristroph2018sculpting, wang2021how}. Similar problems relating to this coupling between the time evolution of the geometry and the flow dynamics \cite{gallaire2017fluid, moore2017shapes} may be found in erosion/deposition \cite{ristroph2012sculpting}, flow through porous media \cite{chiu2020viscous}, and even in biofilm growth \cite{telgmann2004influence}.

Regimes where diffusion and convection both play comparable roles in the \revised{interfacial} growth and heat or mass transfer behavior are not well-suited for analytical approaches and demand computational solutions. Numerical solution of the classical Stefan problem (purely diffusion-driven) has been tackled using various approaches, including 
phase-field \cite{Langer:80:Instability-and-Patt, Karma;Rappel:97:Quantitative-Phase-F, Nestler;Danilov;Galenko:05:Crystal-growth-of-pu, Karma;Rappel:96:Phase-Field-Modeling, Elder;Grant;Provatas;etal:01:Sharp-interface-limi, Boettinger;Warren;Beckermann;etal:02:Phase-Field-Simulati}, 
Volume of Fluids \cite{Benson:92:Computational-method,Benson:02:Volume-of-Fluid-Inte, DeBar:74:Fundamentals-of-the-,Noh;Woodward:76:SLIC-simple-line-int,Youngs:84:An-Interface-Trackin}, 
Front Tracking \cite{Al-Rawahi:02:Numerical-Simulation}, 
integral  \cite{myers2011application}, 
moving grids \cite{beckett2001a,javierre2006a}, 
finite element \cite{Fedoseyev;Alexander:97:An-Inverse-Finite-El,Bars;Worster:06:Solidification-of-a-}, 
and level-set  \cite{chen1997simple, yang2005sharp, javierre2006a, Gibou;Fedkiw:05:A-fourth-order-accur, Gibou:2003aa, Chen;Min;Gibou:09:A-numerical-scheme-f} methods. 
For solution of the coupled Stefan problem with fluid dynamics, some recent approaches include finite element \cite{Zabaras;Ganapathysubramanian;Tan:06:Modelling-dendritic-}, immersed boundary \cite{huang2021a},  \revised{level-set \cite{udaykumar2003sharp}}, and  front-tracking methods \cite{vu2016numerical,vu2017numerical,vu2018fully}.   

In this study, \revised{we utilize sharp interface numerical methods, \emph{i.e.} methods that numerically preserve the discontinuity in discontinuous quantities.} We develop an accurate and efficient method which combines three main aspects  -- adaptive grids for more efficient computation, sharp interface representation with accurate gradients at the interface for calculation of the interfacial velocity, and accurate application of boundary conditions at the interface to best capture the coupled nature of interface motion and flow.  Firstly, the problem is multiscale in nature -- the interfacial dynamics must be resolved at a relatively fine length scale, and it is beneficial to resolve other physics at varying length scales  (ie. fluid recirculation zones, vortices, etc). \revised{Indeed, } Huang \etal \cite{huang2021a} notes the need for adaptive grids in order to study regimes of interest, often with thin boundary layers, without sacrificing computational cost. Thus, we use adaptive grids in order to achieve modest computational cost whilst still resolving the length scales of interest. Secondly, as the evolution of the interface is governed by gradients at the interface, we seek a method with accurate calculation of such gradients. In addition, we apply boundary conditions at the interface to avoid $O(1)$ error in gradients typical when numerically smearing the solution profile near the boundary. In particular, we make use of and build upon recent elliptic and parabolic solvers developed on Quad-/Oc-tree grids, which provide second-order accurate gradients in the maximum norm, their extension to Stefan-type problems and Navier-Stokes solvers on such grids as well as the level-set method on adaptive grids in parallel \cite{Gibou;Min;Fedkiw:13:High-resolution-shar, Chen;Min;Gibou:09:A-numerical-scheme-f,Helgadottir;Gibou:11:A-Poisson--Boltzmann,Min;Gibou:07:A-second-order-accur,Mirzadeh;Theillard;Gibou:10:A-Second-Order-Discr, Mirzadeh;Gibou:14:A-conservative-discr,Gibou;Fedkiw;Cheng;etal:02:A-Second-Order-Accur,Theillard;Gibou;Pollock:14:A-Sharp-Computationa,Chen;Min;Gibou:07:A-Supra-Convergent-F,Mirzadeh;Theillard;Helgadottir;etal:12:An-Adaptive-Finite-D,Papac;Gibou;Ratsch:10:Efficient-symmetric-, Brun;Guittet;Gibou:12:A-local-level-set-me, guittet2015stable, egan2021direct}. Lastly, the ability to apply accurate interfacial boundary conditions enables us to enforce the conservation of mass across the interface, which relates interfacial velocity and fluid velocity, rather than approximating this boundary condition as no-slip as is done in previous work \cite{vu2016numerical, huang2021a}. While the no-slip approximation is often a fair one, this allows for a coupling between interfacial velocity and fluid velocity that is truer to the physics and allows for the possibility of tackling regimes in the future where the no-slip approximation may not hold. 

This paper is organized as follows. First, we describe the physical model used for the Stefan-type problem with convective effects in Sec.~\ref{sec:problem_description}. In Sec.~\ref{sec:numerical_approach}, we outline the algorithm used in our numerical approach for the solution of this problem, and provide an overview of the numerical methods utilized. Sec.~\ref{sec:numerical_verification} details the results from a numerical verification test used to capture the convergence behavior of the numerical approach. Sec.~\ref{sec:okada_compare}  describes several numerical experiments conducted to validate the performance of the solver against past experimental results. In Sec.~\ref{sec:investigating}, we investigate the role of the Reynolds and Stefan numbers on the relationship between the interface morphology, flow dynamics, and heat transfer near the interface, and offer a new scaling law for average Nusselt number $\overline{Nu_d}$ at the interface as it relates with $Re$ and $St$. \revised{In Sec.~\ref{sec:porousmedia}, we illustrate an application for the presented method in the context of an evolving porous media.} Lastly, we briefly summarize the current state of our numerical method and outline possibilities for future work in Sec.~\ref{sec:conclusion}.

\section{Problem Description} \label{sec:problem_description}
\subsection{Governing equations}

We consider a physical domain $\domain$, which may be separated into liquid and solid subdomains denoted $\liqDomain$ and $\solDomain$, respectively. Throughout the paper, we use the subscript $l$ to denote fluid quantities and the subscript $s$ to denote solid quantities. In the present derivation, we focus on the temperature-driven solidification process; however it is noted that an analogous derivation may be done in the case of a concentration-driven dissolution/precipitation process \cite{huang2021a}.

\begin{wrapfigure}{r}{4cm}
\vspace{-1cm}
\centerline{
\includegraphics[width=0.2\textwidth]{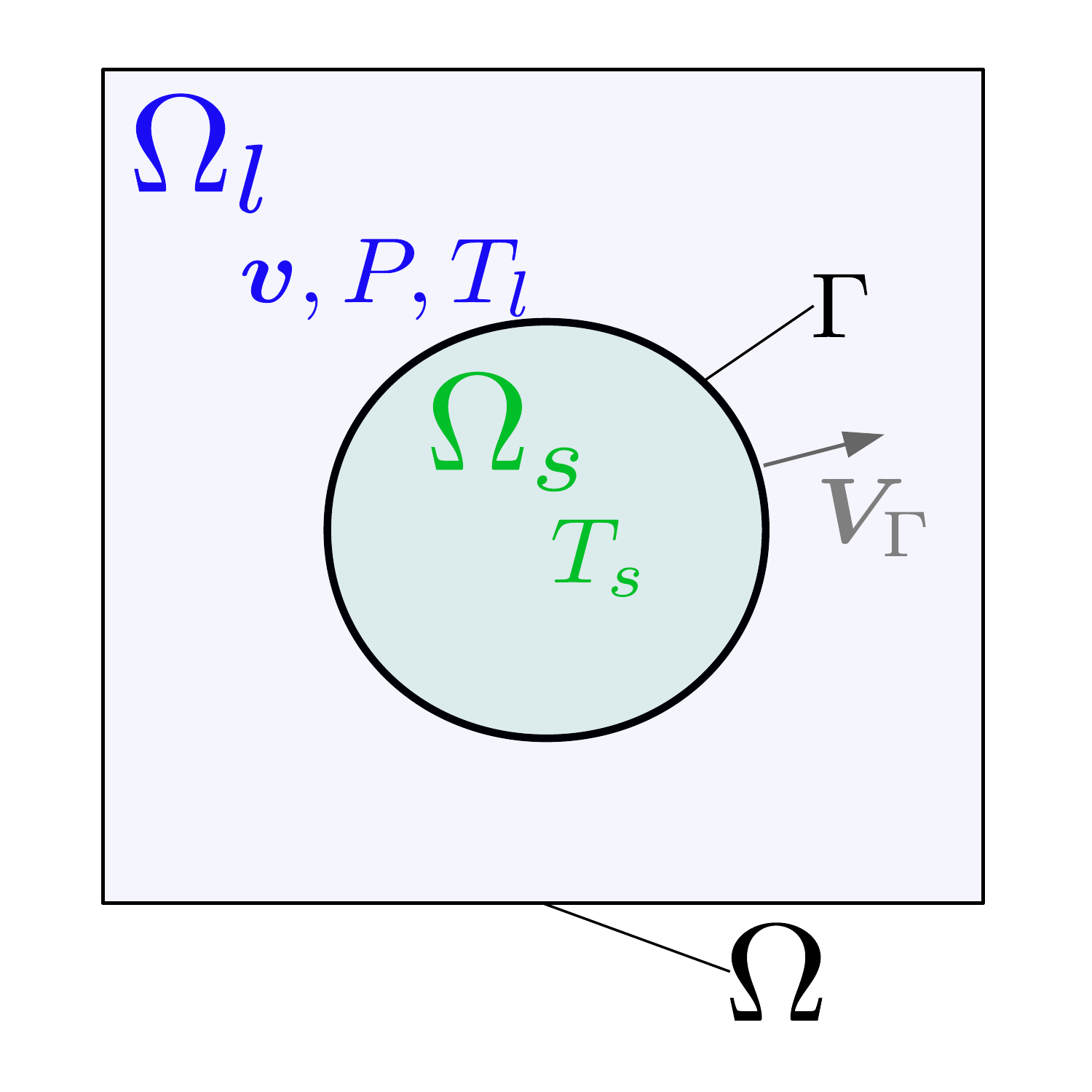}}
\caption{\revised{Example of notations used. The physical domain $\domain$, is subdivided into the liquid and solid subdomains $\liqDomain$ and $\solDomain$, respectively. The subdomains are separated by the evolving interface $\Gamma$. } }
\label{fig:physprobdomain}
\vspace{-1.3cm}
\end{wrapfigure}

We model the solidification process at the macroscopic level and thus consider the transition between the liquid and solid regions to be sharp, where the sharp interface between the two regions is denoted as $\interface$. Assuming a Newtonian, incompressible flow in the liquid domain $\liqDomain$, conservation of mass and momentum yield the incompressible Navier-Stokes equations given by
\begin{subequations}
\begin{equation*}
\nabla \cdot \vns = 0
\end{equation*}
\begin{equation*}
    \rhol \left( \ddt{\vns} + \advection{\vns}{\vns} \right) = \mul \laplace{\vns} - \nabla \press + \rhol \bm{f},
\end{equation*}
\end{subequations}
which govern the fluid velocity and pressure, where $\rho_l$ is the fluid density, $\mu_l$ is the fluid dynamic viscosity, and $\bm{f}$ may describe the presence of external forces such as gravity. Writing the energy conservation for the liquid domain and assuming that the internal energy is a linear function of the temperature in the range of interest yields an advection-diffusion equation in temperature in the fluid, given by
\begin{equation*}
    \ddt{\Tl} + \advection{\vns}{\Tl} = \alphal \laplace{\Tl} + g_{l},
\end{equation*}
where $\alpha_l$ is the fluid thermal diffusivity, and $g_{l}$ may describe external heat sources per unit volume divided by the fluid heat capacity per unit volume $\rhol c_{\mathrm{v}, l}$ in the fluid. We note that for heat transport in the fluid we neglect viscous dissipation, which is typically negligible except in the case of certain high speed flows \cite{deen1998analysis}. In the solid domain, the heat transfer is given by the standard heat equation
\begin{equation*}
    \ddt{\Ts} = \alphas \laplace{\revised{\Ts}} + g_{s},
\end{equation*}
where $g_{s}$ represents an analogous heat source in the solid and $\alphas$ is the solid thermal diffusivity. On the interface $\interface$, the energy balance reduces to the Stefan condition, given by 
\begin{equation*}
    \normal \cdot \vint = -\frac{(\kl \nabla \Tl - \ks \nabla \Ts)\cdot \normal}{\rhos L},\\
\end{equation*}
where $\kl$ and $\ks$ are the thermal conductivities of the fluid and solid respectively, $L$ is the latent heat of fusion, and $\normal$ denotes the unit normal which points outward from the liquid subdomain \cite{deen1998analysis}. The sharp interface $\interface = \interface (t)$ which separates the liquid and the solid subdomains will then evolve with the normal velocity $\normal \cdot \vint $ governed by this conservation of energy across interface. The interfacial condition for the temperature is derived from the Gibbs-Thomson relation \cite{dantzig2016solidification}, which captures the effects of the interface curvature on the melt temperature by relating $T_{\interface}$ with the melt temperature of the given species, $T_{m}$, the curvature of the interface, $\kappa$, the latent heat of fusion $L$, and the interfacial energy of the solid-liquid interface, $\gamma_{sl}$. This expression is given by
\begin{equation*}
T_{\interface} = T_{m}(1 - \sigma \kappa),
\end{equation*}  
where $\sigma = \gamma_{sl} / (\rhos L)$ \cite{dantzig2016solidification}. To determine the interfacial boundary condition for the fluid velocity, we again consider a mass balance across the interface, which reduces to the condition
\begin{equation*} 
\normal \cdot \vns  = \vint \cdot \normal \left(1 - \frac{\rhos}{\rhol} \right).
\label{eq:ns_bc}
\end{equation*}

\subsection{Non-dimensionalized form}
It is convenient to express the physical model in a non-dimensionalized form, noting the characteristic length, velocity, and time scales as $d$, $u_{\infty}$, and $d / u_{\infty}$, respectively. The characteristic temperature difference is defined as $\Delta T = T_{\infty} - T_{0}$, where $T_{\infty}$ and $T_{0}$ are characteristic temperatures of the system being modeled. We may therefore rescale temperature as $\hat{T} = (T - T_{0})/(T_{\infty} - T_{0})$, velocity as $\hat{\bm{v}} = \vns/u_{\infty}$, time as $\hat{t} = t u_{\infty}/d$, length as $\hat{\bm{r}} = \bm{r}/d$, and curvature as $\hat{\kappa} = d \kappa$. This leads to the following non-dimensional groups: the Reynolds number: $Re = \rho_l u_\infty d /\mu_l$, the Prantdl number: $Pr = \mu_l /(\rho_l \alpha_l)$, the Peclet number: $Pe = u_{\infty} d /\alpha_{l}$, the Stefan number: $St = c_{p,s}/ (L \Delta T)$.  Dropping the $\hat{}$ notation for convenience and taking source terms as zero, we arrive at the following dimensionless system expressed in Eq. \eqref{eq:finalIncompressible} - \eqref{eq:finalNSInterfaceBC}.

\begin{align}
\nabla \cdot \vns &= 0, & \textrm{in $\liqDomain$} \label{eq:finalIncompressible} \\
\ddt{\vns} + \advection{\vns}{\vns} &= \frac{1}{Re} \nabla^{2}\vns - \nabla P, & \textrm{in $\liqDomain$} \label{eq:finalNS} \\
\ddt{\temperaturel} + \advection{\vns}{\temperaturel} &= \frac{1}{Pe} \laplace{\temperaturel}, & \textrm{in $\liqDomain$} \label{eq:finalHeatLiq}  \\
 \ddt{\temperatures} &= \frac{1}{Pe} \frac{\alpha_s}{\alpha_l} \laplace{\temperatures}, & \textrm{in $\solDomain$}\label{eq:finalHeatSol} \\
\normal \cdot \vint &= \frac{St}{Pe} \frac{\alphas}{\alphal}  \left(\nabla \temperatures -  \frac{k_l}{k_s} \nabla \temperaturel \right) \cdot \normal, & \textrm{on $\interface$} \label{eq:finalStefan} \\
\revised{\Tl|_\Gamma = \Ts|_\Gamma =  } T_{\interface} &= T_{m} \left(1 - \kappa \frac{\sigma}{d} \right) - \kappa \frac{\sigma}{d}   \frac{T_{0}}{\Delta T}, & \textrm{on $\interface$}   \label{eq:finalGibbsThomson} \\
\normal \cdot \vns|_{\interface} &= \normal \cdot \vint \left(1 - \frac{\rhos}{\rhol} \right), & \textrm{on $\interface$}. \label{eq:finalNSInterfaceBC} 
\end{align}

For the remainder of the manuscript, we will refer to the problem in its non-dimensionalized form. It is important to note that this model can be applied to an analogous dissolution problem, where the transported scalar field is concentration, and a similar interfacial velocity can be derived based on species transport across the interface rather than energy.

\section{Numerical Approach} \label{sec:numerical_approach}

To tackle the solution of this multiphysics coupled problem, a host of challenges arise. To adequately capture the evolution of the free boundary, which may develop to be quite irregular, and in some cases may present changes in topology, we use the level-set method introduced in \cite{Osher;Sethian:88:Fronts-propagating-w} (see \cite{Gibou;Fedkiw;Osher:18:A-review-of-level-se} for a recent review on Quad-/Oc-tree). Considering the multiscale nature of the problem, it is advantageous to utilize an adaptive grid approach, which provides a more manageable computational cost while still allowing the capturing of relevant physics in regions of interest like sharp variation of temperature near the interface, evolution of flow structures like vortices, etc. In this work, we use the parallel level-set framework introduced in \cite{mirzadeh2016parallel}, which makes use of the scalable \texttt{p4est} library \cite{burstedde2011p4est} enabling the dynamic management of a collection of adaptive quad-/oc-trees. Additionally, the boundary conditions for the temperature and the fluid's variables are imposed in a sharp manner, \emph{i.e.} at the interface.

The algorithm we introduce is provided in section \ref{sec:algorithm}; the following sections describe the details of the numerical methodologies.

\subsection{General algorithm to solve the model} \label{sec:algorithm}
Given the initial conditions for temperatures, flow velocity,  grids at $t_{-1}$, $t_{0}$,  and an initial interface location, we proceed for each $n$-th time level (starting with $n=0$) as follows:
\begin{enumerate}

\item The energy equations (or equivalent) given by equations \eqref{eq:finalHeatLiq} and \eqref{eq:finalHeatSol} are solved for $\temperaturel^{n+1}$ and $\temperatures^{n+1}$. Advective terms are discretized using the semi-Lagrangian formulation detailed in Sec.~\ref{sec:SL} and backtrace values are found using the quadratic non-oscillatory interpolation method described in Sec.~\ref{sec:interp}. The diffusive terms are treated implicitly and discretized spatially using the finite-difference approach described in Sec.~\ref{sec:poissontype}. 

\item The interfacial velocity $\vint^{n+1}$ is computed as per equation \eqref{eq:finalStefan} from the jump in the fields $\temperaturel^{n+1}$, $\temperatures^{n+1}$. Values of $\temperaturel^{n+1}$ and $\temperatures^{n+1}$ are extended in a narrow band across the interface using a PDE extrapolation method in order to compute the jump. 

\item A pressure-free projection method is used to solve the Navier-Stokes equations \eqref{eq:finalIncompressible},\eqref{eq:finalNS} for $\vns^{n+1}$, $\press^{n+1}$.  Advective terms are again discretized with a semi-Lagrangian approach. Diffusive terms are treated implicitly and discretized using finite-volume methods. \revised{The interfacial boundary condition, $(\vns^{n+1} \cdot \normal)|_{\interface} = (\vint^{n+1} \cdot \normal)(1 - \rhos/\rhol )$, is approximated in a component-wise fashion, which is described in further detail in Sec.~\ref{subsec:NSBCexplain}.}

\item The interface $\interface^{n+1}$ is advanced to find $\interface^{n+2}$ and thus $\liqDomain^{n+2}$ and $\solDomain^{n+2}$ by advecting the level-set function $\phi^{n+1}$ under $\vint^{n+1}$ with $\Delta t_{n+1}$. The adaptive timestep $\Delta t_{n+1}$ is computed via the expression $\Delta t_{n+1} = \text{CFL}\times \text{min}(\Delta x)/\text{max}(\vns^{n+1}, \vint^{n+1})$ , where the $\text{CFL}$ (Courant–Friedrichs–Lewy)  coefficient is selected for accuracy. Further details of the level set method can be found in Sec.~\ref{sec:levelSetMethods}. 

\item The current grid, $\texttt{grid}^{n+1}$, is refined and coarsened according to desired criteria, such as a uniform band around the interface $\Gamma^{n+2}$, or resolution of vortex structures defined by certain ranges of vorticity. Further details on the adaptive quadtree grids used are described in Sec. \ref{sec:quadtrees}. 

\item All fields are transferred to the new grid via interpolation, using the methods detailed in Sec.~\ref{sec:interp}.

\item Proceed to the next time level
\end{enumerate}

\subsection{Level-set method for interface representation}\label{sec:levelSetMethods}

To track the interface $\Gamma$ in a sharp manner, we make use of the level-set method \cite{Osher;Sethian:88:Fronts-propagating-w}, which represents the interface as the zero-contour of a higher dimensional function, $\phi (t,\rVec)$, with signed distance property. Thus, the interface and the liquid and solid subdomains can be represented as
\begin{gather*}
\interface = [\rVec \in \domain | \phi = 0 ], \quad
\liqDomain = [\rVec \in \domain | \phi<0 ], \quad
\solDomain = [\rVec \in \domain | \phi>0 ].
\end{gather*}

The level-set function provides a straightforward access to the computation of the normal $\normal$ to the interface and the interface curvature $\kappa$ as
\begin{align*}
\normal = \frac{\nabla \phi}{|\nabla \phi|}, \quad
\kappa = \nabla \cdot \normal.
\end{align*} 

To advance the interface, the level-set function is evolved under the external velocity field, $\vint$, by solving the equation
\begin{equation*} 
\ddt{\phi} + \advection{\vint}{\phi} = 0,
\end{equation*} 
which is solved using the semi-Lagrangian method detailed in Sec.~\ref{sec:SL}. The level-set function is reinitialized at each time step to restore the signed distance property from any degradation that may occur under advection. This is done by solving the reinitialization equation \cite{Sussman;Smereka;Osher:94:A-level-set-approach} in fictitious time $\tau$:
\begin{equation*}
    \frac{\partial \phi}{\partial \tau} + \texttt{sgn}(\phi^{n}_{\tau = 0})(|\nabla \phi^{n}| - 1) = 0,
\end{equation*} 
for a few time steps. Here, $\texttt{sgn}$ is the signum function. In the present work, the reinitialization step is solved using a TVD-RK2 time-stepping scheme with a Godunov Hamiltonian discretization of $|\nabla \phi|$, and the sub-cell fix of \cite{russo2000remark}, extended to the case of adaptive grids in \cite{min2006supra}.

\subsection{Adaptive quadtree grids}\label{sec:quadtrees}

In order to accurately capture the various multiscale physical phenomena in the presented problem, \emph{ie.} interfacial dynamics, flow structures, with an accessible level of computational cost, we utilize adaptive Cartesian quadtree grids for discretization of the computational domain. 

\begin{figure}[H]
\centering
	\begin{subfigure}[t]{0.32\textwidth}
		\centering
		\includegraphics[width=\textwidth]{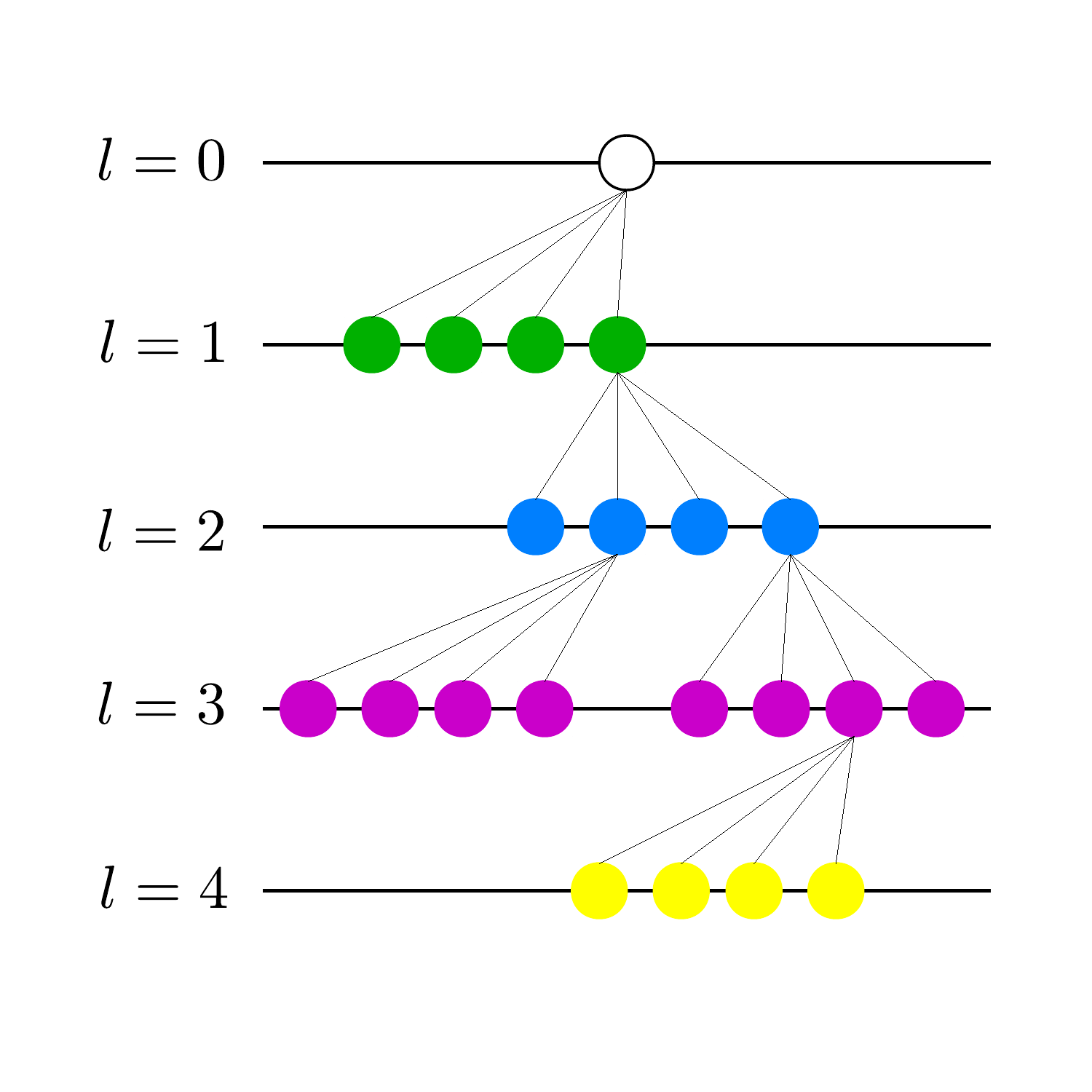} 
		\caption{}
	\end{subfigure}
	\begin{subfigure}[t]{0.32\textwidth}
		\centering
		\includegraphics[width=\textwidth]{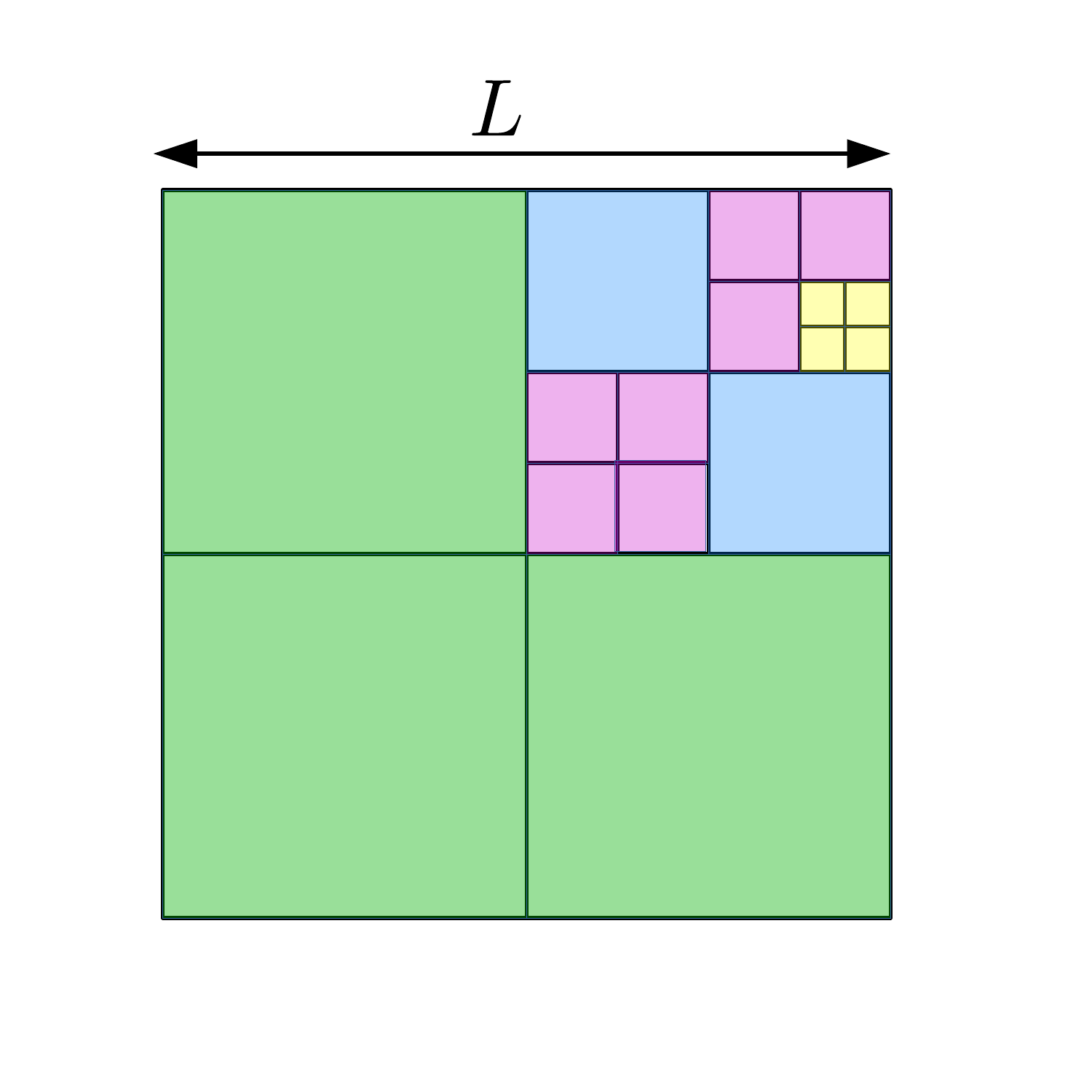} 
		\caption{}
	\end{subfigure}
	\begin{subfigure}[t]{0.32\textwidth}
		\centering
		\includegraphics[width=\textwidth]{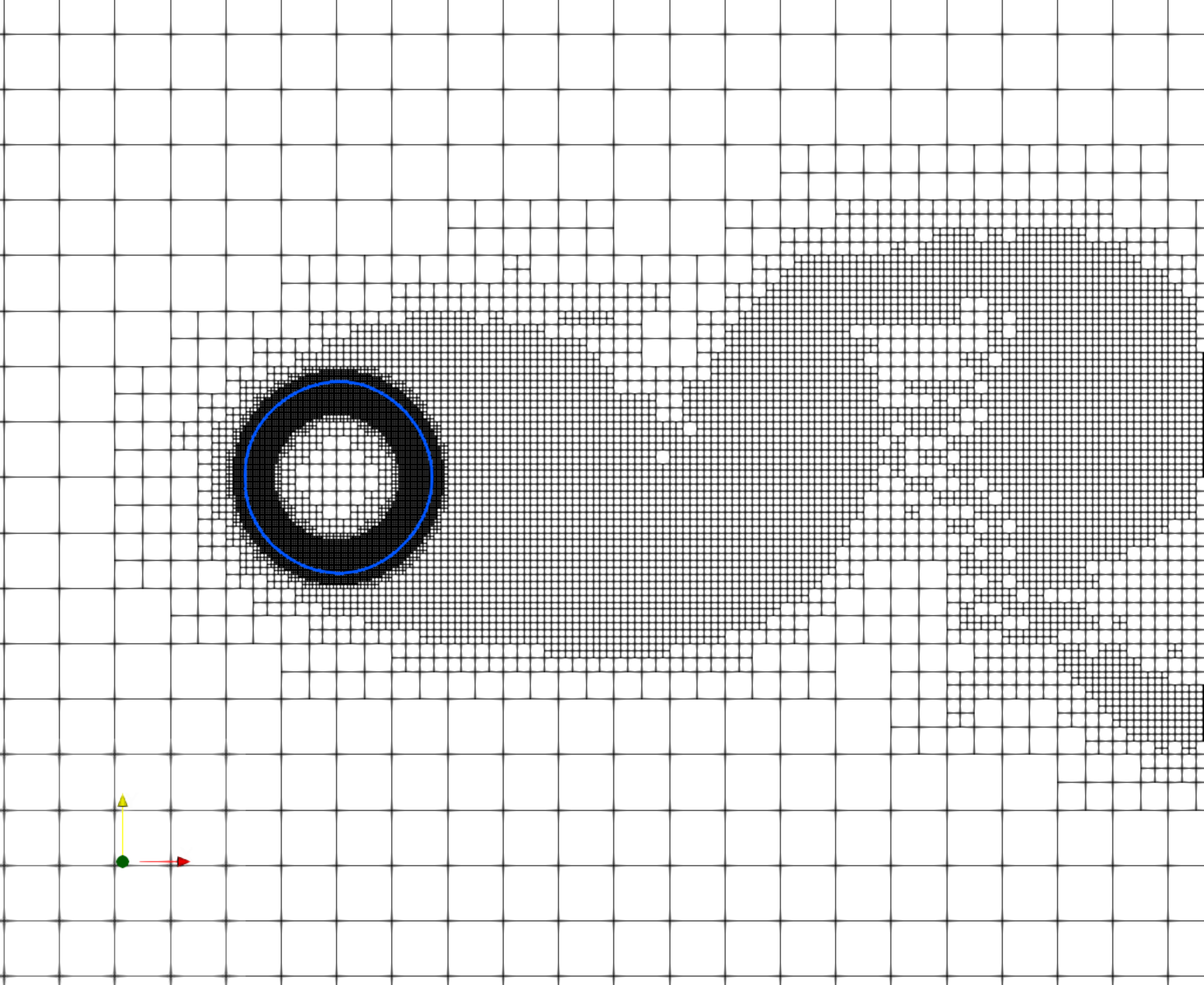} 
		\caption{}
	\end{subfigure}
\caption{Illustration of a quad-tree grid hierarchical structure (a) associated to a computational grid (b). (c) Example of adaptive Cartesian quadtree structure, with custom refinement around the interface,  as well as according to areas where properties such as fluid vorticity or gradients of temperature are above user-chosen thresholds.}
\label{fig:quadtree_structure}
\end{figure}

Initially, the ``root" (also considered level 0) of the quadtree data structure is one cell which describes the entire computational domain. The root cell may then be split into four equal-sized ``child" cells, yielding the next level of refinement (level 1). Cells may be split in this manner (a) recursively to varying levels of refinement, denoted by $l$, and (b) selectively according to a set of given criteria, \emph{i.e.} distance to the interface, or fluid vorticity. The size of a cell is thus given by $L/2^{l}$, where $L$ is the size of the root cell, and $l$ is the level of refinement. Limits are placed on the lowest and highest allowable refinement levels of the grid, usually denoted by $l_{\text{min}}$ and $l_{\text{max}}$. This process is illustrated and exemplified in Fig.~\ref{fig:quadtree_structure}.

\begin{figure}[H]\centering
\includegraphics[height=0.4\textwidth]{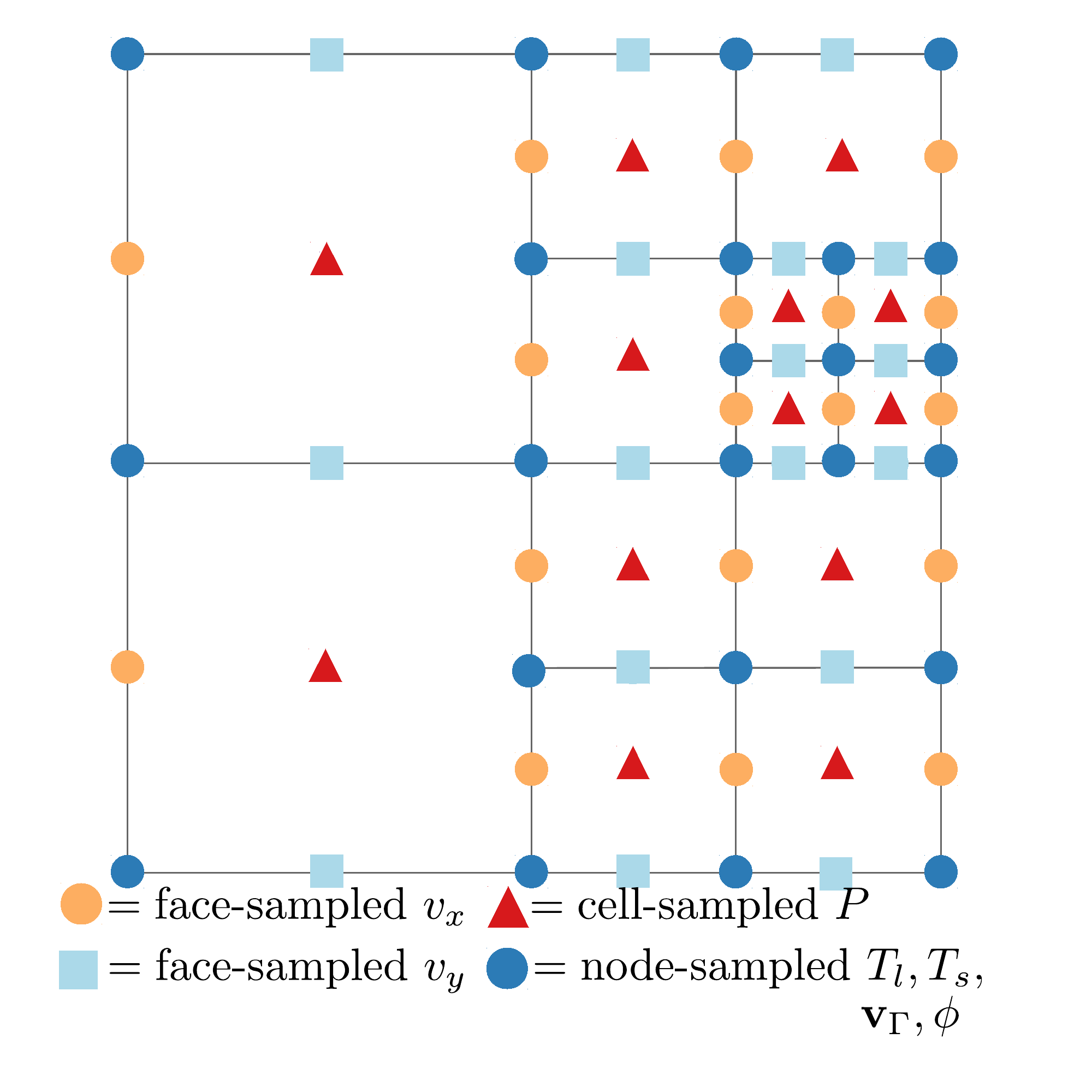} 
\caption{\revised{Illustration of data layout used on adaptive quadtree grids.}}
\label{fig:dataLayout}
\end{figure}

\revised{The solution of the Navier-Stokes equations uses the standard MAC grid arrangement (further described in section \ref{sec:NSProjectionMethod}) to ensure numerical stability of the projection method. This method stores the velocity components $\vns$ at the face-centers of the grid, and the pressure $P$ (and pressure-related Hodge variable $\Phi$) at the cell-centers. However, the temperature fields $\temperaturel$ and $\temperatures$ and the level-set function $\phi$ are stored at the nodes, allowing for the use of higher accuracy node-sampled methods which can provide second order accurate results for the solution and its gradients (described in \ref{sec:NodeBasedSpatialDiscretization}) The interface velocity $\vint$, computed from the gradients of temperature, is also node-sampled. }  An illustration of this data layout is given in Fig.~\ref{fig:dataLayout}.

\subsection{Node-based spatial discretization on quadtrees} \label{sec:NodeBasedSpatialDiscretization}
\begin{figure}[H]
\centering
\includegraphics[height=0.3\textwidth]{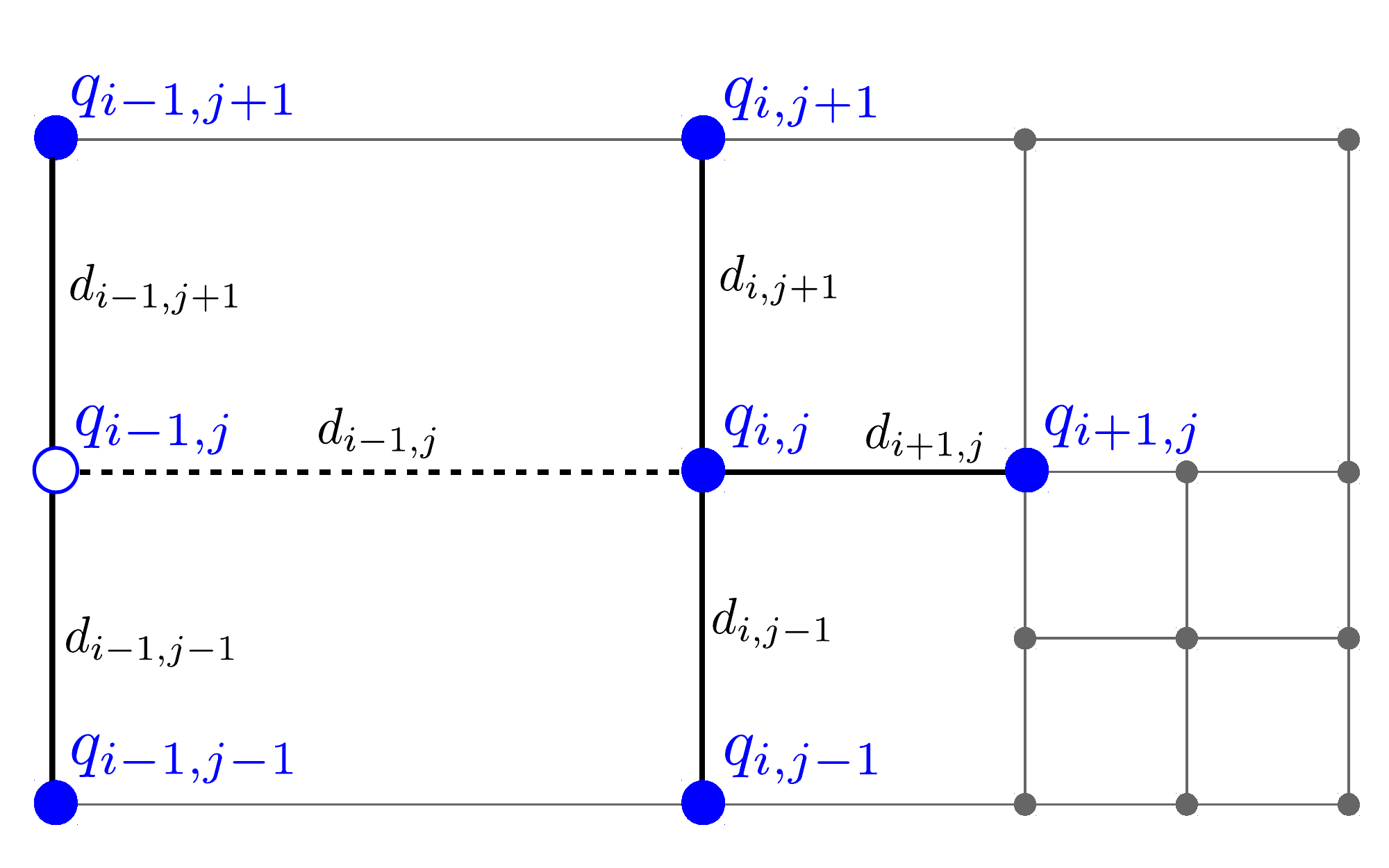} 
\caption{Example discretization on quadtree grid in T-junction case. The missing data of a field $q$ at the grid node $(x_{i-1}, y_j)$ is denoted by $q_{i-1, j}$, illustrated by the white colored grid node with blue edge color. Its distance to the node $(x_i, y_j)$ is denoted by $d_{i-1,j}$. }
\label{fig:quadtree_disc}
\end{figure}

\newcommand{\dFTwoDTwoD}[5]{\left(\frac{#2 - #1}{#4} - \frac{#1 - #3}{#5}\right) \left(\frac{2}{#4 + #5}\right)}
When solving for the advection equation on quadtree grids, \cite{strain2000fastSL, J.Strain:99:Tree-methods-for-mov} showed that a node-based data structure is advantageous when using semi-Lagrangian methods. Such an approach was extended to second-order accuracy for the level-set method and the Navier-Stokes equations in Refs. \cite{min2007second,min2006NS,xiu2001semiSL}. In addition,  \cite{chen2007supra,min2006supra,Min;Gibou:06:A-supra-convergent-f} introduced solvers for the Poisson and heat equations on irregular domains on node-based quad-/oc-trees and showed that both the solution and its gradients are second-order accurate in the $L^\infty$ norm. This is an advantage in the case of Stefan problems since it is the gradient of the solution that drives the overall accuracy. 
When considering the spatial discretization on a node-based quadtree grid structure, the inevitable case will arise in which the node about which we are discretizing is missing a direct neighbor in one of the Cartesian directions (the case referred to as a T-junction). An example of such a scenario is illustrated in Fig.~\ref{fig:quadtree_disc}, which provides the definition of the notations used in the following.

In cases such as these standard formulas may be used in both Cartesian directions, with the exception that the missing value ($q_{i-1,j}$  in this case) is defined using quadratic interpolation, as introduced in \cite{min2007second}:
\begin{align*}
q_{i-1,j}		&= \frac{d_{i-1,j-1}\,q_{i-1,j+1} + d_{i-1,j+1}\,q_{i-1,j-1}}{d_{i-1,j-1} + d_{i-1,j+1}} \\
           	& - (\frac{d_{i-1,j-1}\,d_{i-1,j+1}}{d_{i-1,j+1}+d_{i-1,j-1}}) \left(\frac{q_{i,j+1}-q_{i,j}}{d_{i,j+1}} + \frac{q_{i,j-1}-q_{i,j}}{d_{i,j-1}} \right). 
\end{align*}

This definition of a ghost value is always possible in a node-based setting and is third-order accurate. It then enables one to approximate the first-order and the second-order derivatives with second-order and first-order accuracy, respectively, using standard finite difference formulas as if the grid did not have T-junction node. For example, considering the layout in Fig.~\ref{fig:quadtree_disc}, the discretizations of the first- and second-order derivatives of $q$ (denoted as $q_x$ and $q_{xx}$, respectively) at the node $(i, j)$ are:
\begin{align*}
q_x|_{i,j} &\approx \frac{q_{i+1,j} - q_{i,j}}{d_{i+1,j}} \frac{d_{i-1,j}}{d_{i+1,j} + d_{i-1,j}} + \frac{q_{i,j} - q_{i-1,j}}{d_{i-1,j}} \frac{d_{i+1,j}}{d_{i+1,j} + d_{i-1,j}}, \\
q_{xx}|_{i,j} &\approx \dFTwoDTwoD{q_{i,j}}{q_{i+1,j}}{q_{i-1,j}}{d_{i+1,j}}{d_{i-1,j}}.
\end{align*}
Discretizing the partial derivatives in a dimension-by-dimension fashion allows one to write the discretizations needed to approximate the Stefan model, including the implicit treatment of the parabolic parts of the equations and the level-set equations.

\subsection{Semi-Lagrangian discretizations of advective terms}\label{sec:SL}

\newcommand{\rd}{\rVec_{d}}
\newcommand{\rbackn}{\rd^{n}}
\newcommand{\rbacknm}{\rd^{n-1}}
\newcommand{\thetaBackNM}{q^{n}_{d}}
\newcommand{\thetaBackNMM}{q^{n-1}_{d}}
\newcommand{\velfield}{\bm{u}}
\newcommand{\gridpoint}{\rVec^{n}}
\newcommand{\intpointn}{\rVec^{n}_{*}}
\newcommand{\intpointnm}{\rVec^{n-1}_{*}}
\newcommand{\timen}{t_{n+1}}
\newcommand{\timenm}{t_{n}}
\newcommand{\timenmm}{t_{n - 1}}
\newcommand{\vIntermediate}{\velfield(t_{n+\frac{1}{2}},\intpointn)}
\newcommand{\atInterface}{\textcolor{red}{_{|_{\bm{\Gamma}}}}} % \bm{n}
\newcommand{\atInterfaceRaph}{\textcolor{myGreen}{_{|_{\bm{\Gamma}}}}} % \bm{n}
\newcommand{\lvln}{n + 1}
\newcommand{\lvlnm}{n}
\newcommand{\lvlnmm}{n - 1}

\newcommand{\SLFormulationFirstOrder}[1]{\frac{#1^{\lvln} - #1_{d}^{\lvlnm}}{\Delta \timen}}

\newcommand{\SLFormulation}[1]{\alpha \frac{#1^{\lvln} - #1_{d}^{\lvlnm}}{\Delta \timen} + \beta \frac{#1_{d}^{\lvlnm} - #1_{d}^{\lvlnmm}}{\Delta \timenm} }

\newcommand{\SLFormulationNSIntermediate}[1]{\alpha \frac{#1^{*} - #1_{d}^{\lvlnm}}{\Delta \timen} + \beta \frac{#1_{d}^{\lvlnm} - #1_{d}^{\lvlnmm}}{\Delta \timenm} }

The need for discretization of advective terms arises multiple times in the given system of equations: once for the advection of the level-set function used to evolve the interface, once to discretize the advective terms in the Navier-Stokes equations, and once to capture the advection of temperature in the liquid subdomain. For each of these cases, we make use of a semi-Lagrangian discretization with a Backward Difference Formula. To illustrate this, let us consider the case of some scalar field $q$, advected under a given velocity field $\velfield$:
\begin{equation}
q_t +   (\velfield \cdot \nabla) q = \psi,
\label{eq:SLexampleadvection}
\end{equation}
with source term $\psi$. We may write a discretization for equation \eqref{eq:SLexampleadvection}  by tracing the values of $q$ through the characteristic curve of $q$ advected under $\velfield$. These backtraced values of $q$, denoted by $\thetaBackNM$ and $\thetaBackNMM$ , are found by evaluating $q$ at departure points $\rVec_d $ along the characteristic curve that passes through the point $\rVec(t = t_{n+1})$. Thus, 
\begin{gather*}
\thetaBackNM = q (\timen,\rbackn), \quad \thetaBackNMM = q (\timenm,\rbacknm).
\end{gather*} 
The first-order discretization using only one departure point is given by
\begin{equation*}
\SLFormulationFirstOrder{q}  = \psi^{\lvln}.
\end{equation*}

The second-order method, using two departure points, becomes
\begin{equation*}
\SLFormulation{q} =\psi^{\lvln},
\end{equation*}
where $\alpha $ and $\beta$ are Backward Difference Formula coefficients. 

\begin{figure}[H] 
	\centering
	\begin{subfigure}[t]{0.3\textwidth}
		\centering
		\includegraphics[height=\textwidth]{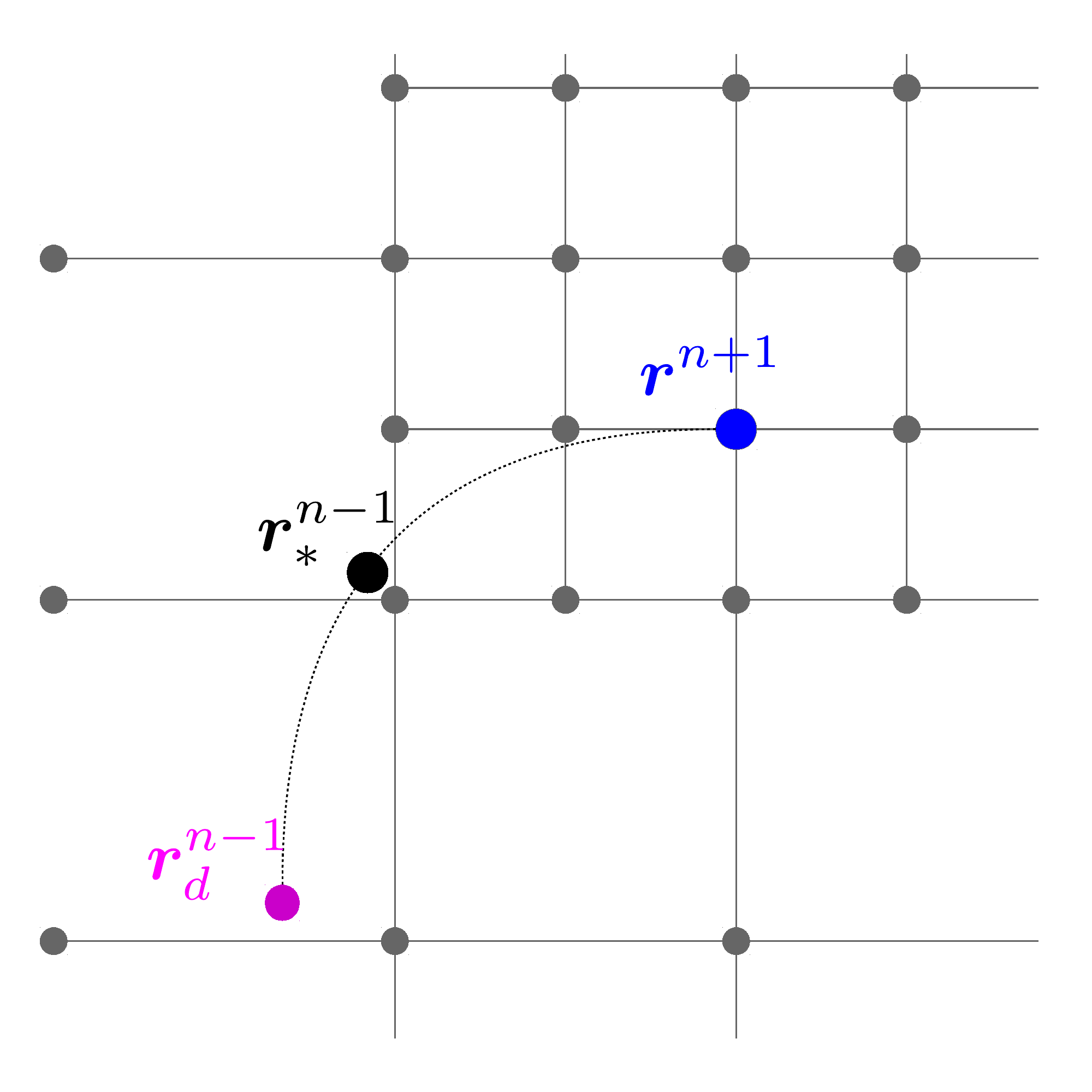} 
		\caption{}
	\end{subfigure}
	~
	\begin{subfigure}[t]{0.3\textwidth}
		\centering
		\includegraphics[height=\textwidth]{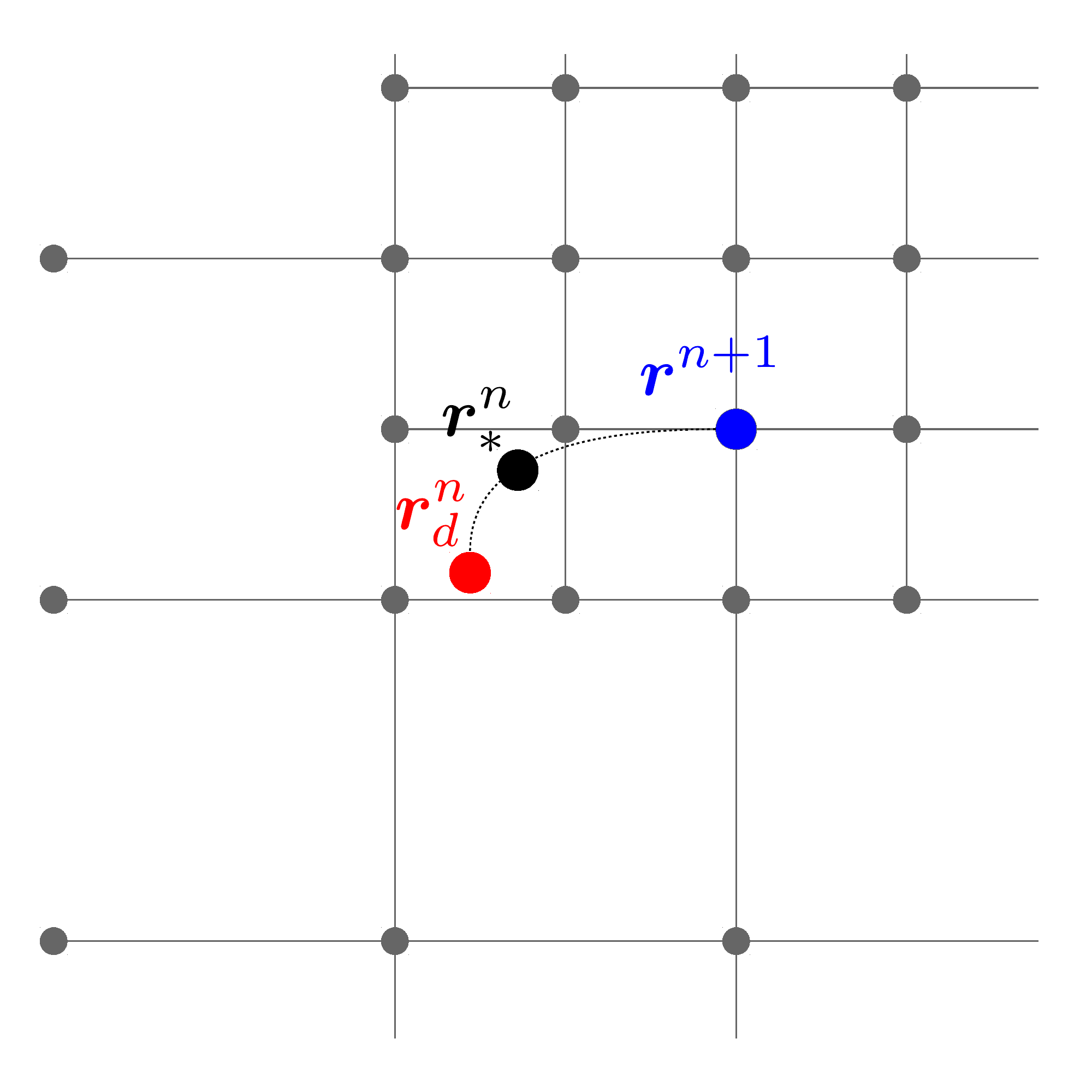} 
		\caption{}

	\end{subfigure}
	~
	\begin{subfigure}[t]{0.3\textwidth}
		\centering
		\includegraphics[height=\textwidth]{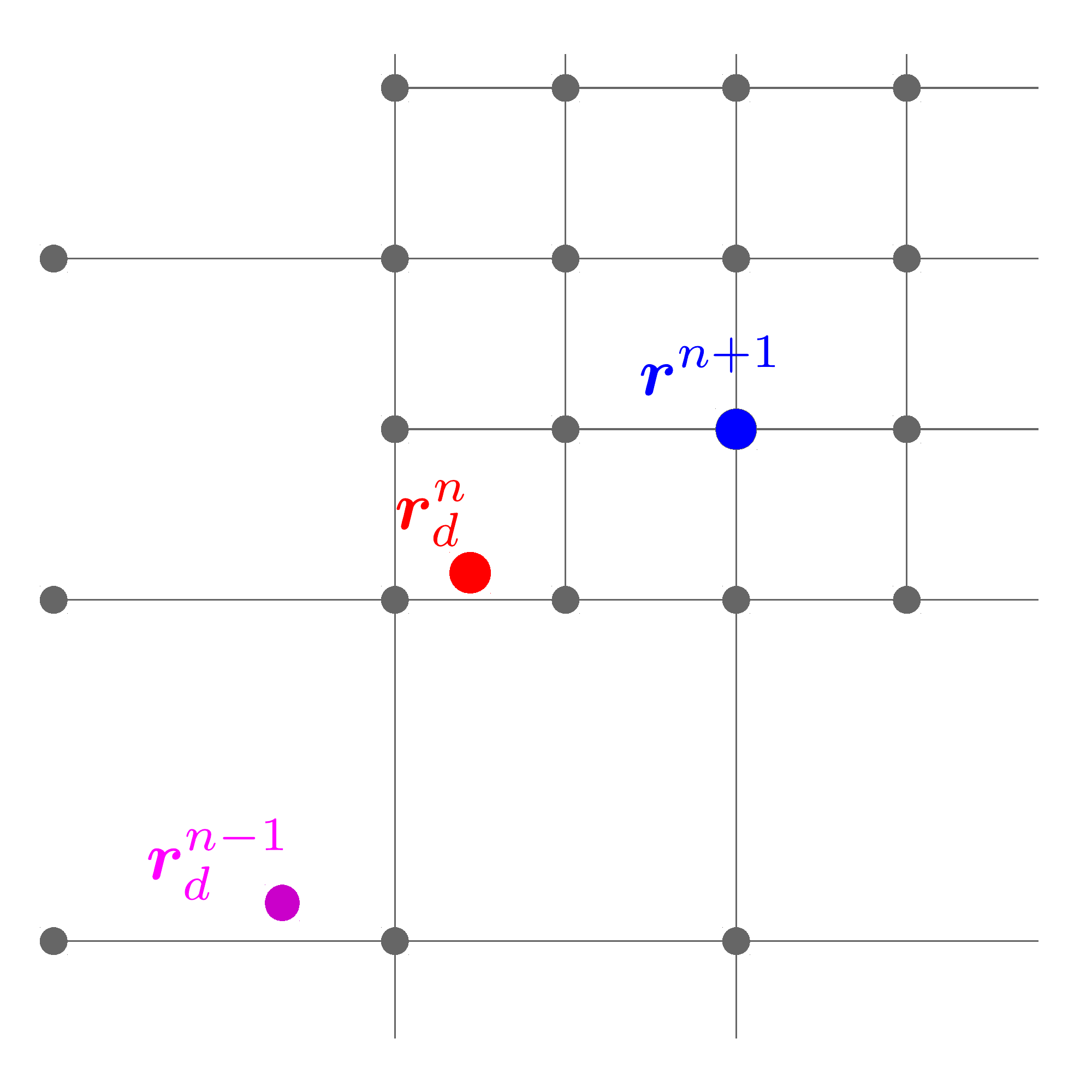} 
		\caption{}
	\end{subfigure}
\caption{Illustration of procedure to acquire departure points for Semi-Lagrangian discretization. (a) Computing departure point $\rbacknm$, (b) computing departure point $\rbackn$, and (c) departure points $\rbacknm$ and $\rbackn$.}
\label{fig:SL_backtrace_pts}
\end{figure}
To obtain the backtraced values $q^{\lvln}_{d}$ and $q^{\lvlnm}_{d}$, we first must obtain the departure points (see Fig.~\ref{fig:SL_backtrace_pts}), computed from solving the characteristic equation d$\rVec/$d$s = \velfield$, where $s$ is the direction along the curve. 
The departure points therefore can be found by tracing this characteristic curve backward in time using the midpoint rule. Given a point $\bm{r}^{n+1}$ for which we wish to compute departure points, we define intermediate points $\intpointn$ and $\intpointnm$ and compute
\begin{gather*}
\begin{split}
\intpointn = \gridpoint - \left( \frac{\Delta \timen}{2} \right) \velfield(\timen,\gridpoint), \quad
\rbackn = \gridpoint - \left( \Delta \timen \right) \vIntermediate,
\end{split}
\end{gather*} 
and 
\begin{gather*}
\begin{split}
\intpointnm =  \gridpoint - \left( \Delta \timen \right) \velfield (\timenm,\gridpoint), \quad
\rbacknm = \gridpoint - \left( \Delta \timen + \Delta \timenm \right) \velfield (\timenm,\intpointnm).
\end{split}
\end{gather*} 

The intermediate velocity $\vIntermediate$ is evaluated at the location $\intpointn$ via multilinear interpolations from the velocity fields at times $\timenm$ and $\timenmm$, and is given by
\begin{equation*}
\vIntermediate = \left(1 + \frac{\Delta \timen}{2 \Delta \timenm} \right) \velfield ( \timen,\intpointn ) + \left( \frac{\Delta \timen}{2 \Delta \timenm} \right) \velfield (\timenm,\intpointn) + O(\Delta t^2).
\end{equation*}

As the computed departure points rarely ever fall at the precise location of grid nodes, the backtraced values of $q$ are evaluated at the departure points using the quadratic non-oscillatory interpolation method of Sec.~\ref{sec:interp}. In the first order case, we simply select $\alpha = 1$. For the second order formula, we select $\alpha$ and $\beta$ as
\begin{align}
\begin{split}
\alpha = \frac{2 \Delta \timen + \Delta \timenm}{\Delta \timen + \Delta \timenm}, \quad
\beta = -\frac{\Delta \timen}{\Delta \timen + \Delta \timenm},
\end{split}
\label{eq:BDF_coeff}
\end{align}
to ensure second-order accuracy, as demonstrated in \cite{guittet2015stable}.

For the three varying cases for which this discretization method is used, we may simply select $\alpha$, $\beta$, $\Phi$, and $\psi$ appropriately. For example:
\begin{enumerate}
\item For the level-set function advection under the interfacial velocity field, we select $q = \phi$, $\velfield = \vint$, $\psi = 0$, $\alpha = 1$, which yields the update $ \phi^{n} = \phi^{n-1}_{d}$
\item For the Navier-Stokes advection, we select $q$ to be each of the components of $\vns$, $\velfield = \vns$, $\psi = $ viscous terms, and $\alpha$, $\beta$ as given by equation \eqref{eq:BDF_coeff}
\item For advection of the scalar field $\temperaturel$ in the fluid, we choose $q = \temperaturel$, $\velfield = \vns$, $\psi = $ diffusion terms, and $\alpha$, $\beta$ as given by equation \eqref{eq:BDF_coeff},
\end{enumerate}
where $\vint$, $\vns$, and $\temperaturel$ are the interface velocity, fluid velocity, and fluid temperature, respectively. To evaluate the backtraced values for the fluid velocity components, the face-sampled velocity fields are interpolated from the faces to the nodes at the end of each timestep to allow for node-based calculation of semi-Lagrangian discretization terms. This is done because (a) the fluid temperature equation requires the computation of node-based advection terms and (b) for the Navier-Stokes advection, the node-based interpolation method is much less expensive than the face and cell based method.

We also note that, although a backtraced point should follow the velocity field and therefore does not land inside the solid region, numerical approximations may be responsible for this case to occur. However, we are extrapolating the fields inside the solid and those are used in such pathological cases.

\subsection{Transport equations on node-sampled fields}\label{sec:poissontype}
To write the full discretization for the advection-diffusion equations for the temperature in the liquid \eqref{eq:finalHeatLiq} and solid \eqref{eq:finalHeatSol} phases, we discretize the diffusive terms implicitly in time and with the approximations of Sec.~\ref{sec:quadtrees} in space; the advective terms are discretized using the the semi-Lagrangian discretization described in Sec.~\ref{sec:SL}. Thus, the heat transport equations can be discretized as
\newcommand{\liqTempN}{\temperaturel^{n+1}}
\newcommand{\liqTempBackNM}{{\temperaturel}_{,d}^{n}}
\newcommand{\liqTempBackNMM}{{\temperaturel}_{,d}^{n-1}}
\newcommand{\solTempN}{\temperatures^{n+1}}
\newcommand{\solTempNM}{\temperatures^{n}}
\newcommand{\bothTempN}{T_{l,s}^{n+1}}
\newcommand{\bothTempBackNM}{T_{l, s, \text{ } d}^{n}}
\newcommand{\bothTempBackNMM}{{T}_{l,s ,\text{ } d}^{n-1}}

\begin{gather*}
\alpha \frac{{\bothTempN}_{i,j} - \bothTempBackNM}{\dtn} + \beta \frac{\bothTempBackNM - \bothTempBackNMM}{\dtnm} = g_{l, s}^{\lvln} + \\
\gamma \dFTwoDTwoD{{\bothTempN}_{i,j}}{{\bothTempN}_{i+1,j}}{{\bothTempN}_{i-1,j}}{d_{i+1,j}}{d_{i-1,j}} +\\ \gamma \dFTwoDTwoD{{\bothTempN}_{i,j}}{{\bothTempN}_{i,j+1}}{{\bothTempN}_{i,j-1}}{d_{i,j+1}}{d_{i,j-1}},
\label{eq:tempLiqDiscretization}
\end{gather*}
where $\gamma$ is $\frac{1}{Pe}$ in the fluid case and $\frac{1}{Pe} \frac{\alphas}{\alphal}$ in the solid case. In the fluid region, the points $\bothTempBackNM$ and $\bothTempBackNMM$ denote the Semi-Lagrangian backtraced values for which the departure points are computed using the fluid velocities $\vns^{\lvlnm}$ and $\vns^{\lvlnmm}$. For the solid region, these points are simply the value at the node, as there is no advective effect in the solid. The coefficients $\alpha$ and $\beta$ are chosen for the fluid region as discussed in Sec.~\ref{sec:SL}, and are taken to be $1$ and $0$ respectively in the solid. 

\begin{wrapfigure}{r}{5cm}
\centering
\vspace{-0.5cm}
\includegraphics[width=0.4\textwidth]{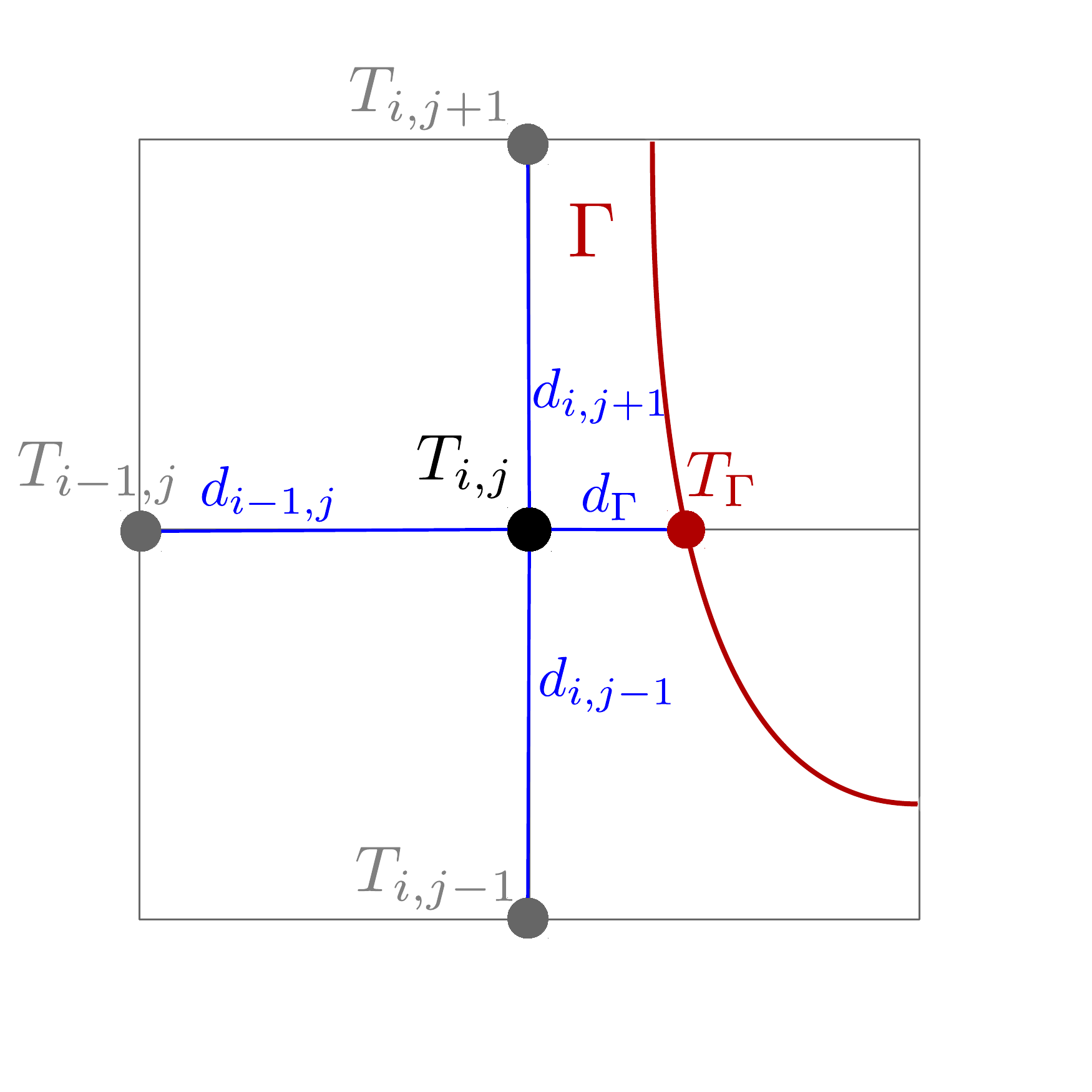} 
\vspace{-1cm}
\caption{Illustration of discretization stencil used to apply Dirichlet boundary conditions at the interface.} 
\label{fig:shortley_weller}
\vspace{-1.5cm}
\end{wrapfigure}Special care must be taken to impose the Gibbs-Thomson boundary condition \eqref{eq:finalGibbsThomson} for the temperature at the solid-liquid interface. We use a sharp treatment based on the Shortley-Weller method that is described in \cite{shortley1938numerical,chen2007supra,Gibou;Fedkiw:05:A-fourth-order-accur}, which essentially includes boundary points into the discretization stencil and shortens the corresponding discretization arm length appropriately such that it ends on the interface $\Gamma$.  For example, in the case below shown in Fig.~\ref{fig:shortley_weller},  the spatial discretization on $T$ incorporates the interfacial condition $T|_\Gamma$, becoming
\begin{eqnarray*}
\revised{\laplace{T^{n+1}_{i,j}}=}  
\dFTwoDTwoD{T^{n+1}_{i,j}}{T^{n+1}_\Gamma}{T^{n+1}_{i-1,j}}{d_{\Gamma}}{d_{i-1,j}} +\\  \dFTwoDTwoD{T^{n+1}_{i,j}}{T^{n+1}_{i,j+1}}{T^{n+1}_{i,j-1}}{d_{i,j+1}}{d_{i,j-1}}.
\end{eqnarray*}
The location of the interface is found by finding the zero of a quadratic interpolation of the level-set function in each spatial direction. This method has been proven to result in second-order accurate solutions and their gradients \cite{yoon2016convergence}.

\subsection{Solving the Navier-Stokes equations} \label{sec:NSProjectionMethod}
For solving the incompressible Navier-Stokes equations given by equation \eqref{eq:finalIncompressible}-\eqref{eq:finalNS}, we take a projection method approach on a MAC sampling. The classical form of the projection method was first introduced by \cite{chorin1968numerical}, and different variations are well-outlined for the interested reader in \cite{brown2001projectionmethods}. In our case, a pressure-free method based on \cite{kim1984application} and developed in a parallel adaptive grid environment by \cite{guittet2015stable,egan2021direct} is used. 

We write the time discrete form of the momentum equation \eqref{eq:finalNS}, making use of the semi-Lagrangian method described in Sec.~\ref{sec:SL}, giving
\begin{equation}
\SLFormulation{\vns} = \frac{1}{Re} \laplace \vns^{n+1}  \Irevised{- \nabla P}.
\label{eq:momentum_discretized}
\end{equation}
Making use of the Hodge decomposition, $\vns = \vns^{*} - \nabla \Phi$ into equation \eqref{eq:momentum_discretized} yields  
\begin{equation*}
\frac{\alpha}{\dtn}(\vns^{*} - \nabla \Phi) + \vns_{d}^{n} (\frac{\beta}{\dtnm} - \frac{\alpha}{\dtn}) + \vns_{d}^{n-1}(\frac{\beta}{\dtnm}) = \frac{1}{Re} \laplace (\vns^{*} - \nabla \Phi) \Irevised{- \nabla P}.
\end{equation*}
The momentum equation can now be decoupled into two separate equations -- a pressure-free momentum equation for the intermediate velocity field $\vns^{*}$ given by equation \eqref{eq:sub_momentum}, and a relation between the fluid pressure $P$ and the Hodge variable $\Phi$ given by equation \eqref{eq:sub_pressure_hodge}. 
\begin{subequations}
\begin{equation}
\SLFormulationNSIntermediate{\vns} = \frac{1}{Re} \laplace \vns^{*},
\label{eq:sub_momentum}
\end{equation}
\begin{equation}
\press^{n+1} = \frac{\alpha}{\dtn} \Phi^{n+1} - \frac{1}{Re} \laplace \Phi^{n+1}.
\label{eq:sub_pressure_hodge}
\end{equation}
\end{subequations}
Substituting the Hodge decomposition into the incompressibility condition then gives the equation that must be satisfied by $\Phi$: 
\begin{equation}\label{eq:sub_incompressibility}
\laplace \Phi^{n+1} = \divergence \vns^{*}.
\end{equation}

\newcommand{\bdry}{\partial \Omega}

Lastly, conditions for $\vns^{*}$ and $\Phi$ on a given boundary $\bdry$ must be carefully selected to ensure compatibility with the physical conditions imposed on $\vns$. First considering $\vns^{*}$, we use the Hodge decomposition to write the Dirichlet case as $\vns^{*}|_{\bdry} = \vns^{n+1}|_{\text{BC}} + (\nabla \Phi^{n+1})|_{\bdry}$, and the Neumann case as $(\nabla \vns^{*} \cdot \normal)|_{\bdry} = (\nabla \vns^{n+1} \cdot \normal)|_{\text{BC}} + (\nabla \nabla \Phi^{n} \cdot \normal)|_{\bdry}$. Now, conditions on the Hodge variable $\Phi$ can be selected to best enforce the conditions on $\vns$. For boundaries where a Dirichlet condition on $\vns$ is applied, we write $(\nabla \Phi \cdot \normal)|_{\partial \Omega} = 0$.   In the Neumann case, we note that because the calculation of the Hodge variable is only second-order accurate, the term $\nabla \nabla \Phi$ cannot be approximated accurately. Instead, we discard it and simply use $(\nabla \vns^{*} \cdot \normal)|_{\bdry} = (\nabla \vns^{n+1} \cdot \normal)|_{\text{BC}}$ as an approximation. This selection enforces equality between $\vns$ and $\vns^{*}$ in the normal direction; however, it does not enforce equality in the tangential direction and thus introduction of spurious slip is possible. It is also of note that the conditions imposed on $\vns^{*}$ require the application of a boundary condition on the Hodge variable at time level $n+1$, which is unknown. Instead, we use $\Phi^{n}$ as an approximation, and iterate on the solution of $\vns^{*}$ and $\Phi$ until the gradient of the computed Hodge variable is no longer changing within a specified tolerance. In practice, this typically serves as a relatively good approximation and requires very few iterations, as was demonstrated in \cite{guittet2015stable}. The general procedure for solving for the fluid is as follows:
\begin{enumerate}
\item Solve equation \eqref{eq:sub_momentum} for the intermediate velocity field $\vns^{*}$, \label{projectionStep1}
\item Solve equation \eqref{eq:sub_incompressibility} for the Hodge variable $\Phi^{n+1}$ that will be used to enforce incompressibility, \label{projectionStep2}
\item Project the intermediate velocity field onto the divergence free space: $\vns^{n+1} = \vns^{*} - \nabla \Phi^{n+1}$, \label{projectionStep3} 
\item Recover the pressure $\press^{n+1}$ via equation \eqref{eq:sub_pressure_hodge}, \label{projectionStep4}  
\end{enumerate}
where steps \ref{projectionStep1}-\ref{projectionStep2} are iterated until convergence of $\nabla \Phi$ within a given tolerance. The procedure given above relies on the spatial discretizations of $\frac{1}{Re} \laplace \vns^{*}$ for the face-sampled intermediate velocity field $\vns^{*}$ in Step 1, and of $\laplace \Phi$ for the cell-sampled Hodge variable $\Phi$ and $\nabla \cdot \vns^{*}$ for the face-sampled intermediate velocity in Step 2, which are described in detail in the following sections.

\subsubsection{Discretization of the viscous term}
The viscous term of  \eqref{eq:sub_momentum}, given by $\frac{1}{Re}\laplace \vns^{*}$, is discretized using an implicit finite volume approach which makes use of control volumes constructed from Voronoi cells. We take advantage of the fact that for a fluid with uniform viscosity, the components of the velocity field are decoupled and therefore can be solved for separately.

\begin{figure}[H] 
	\centering
	\begin{subfigure}[t]{0.45\textwidth}
		\centering
		\includegraphics[height=0.8\textwidth]{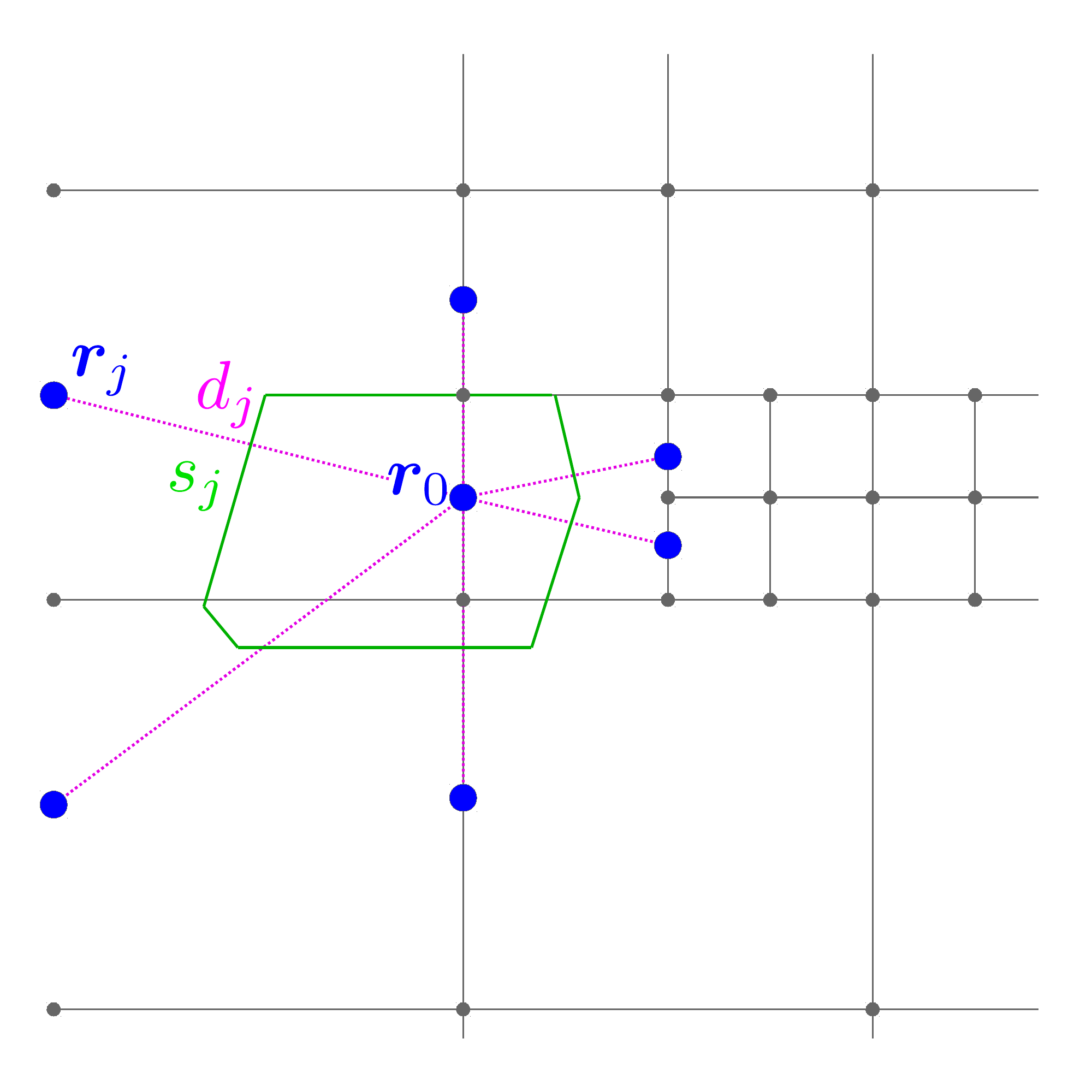} 
		\caption{}
	\end{subfigure}
	~
	\begin{subfigure}[t]{0.45\textwidth}
		\centering
		\includegraphics[height=0.8\textwidth]{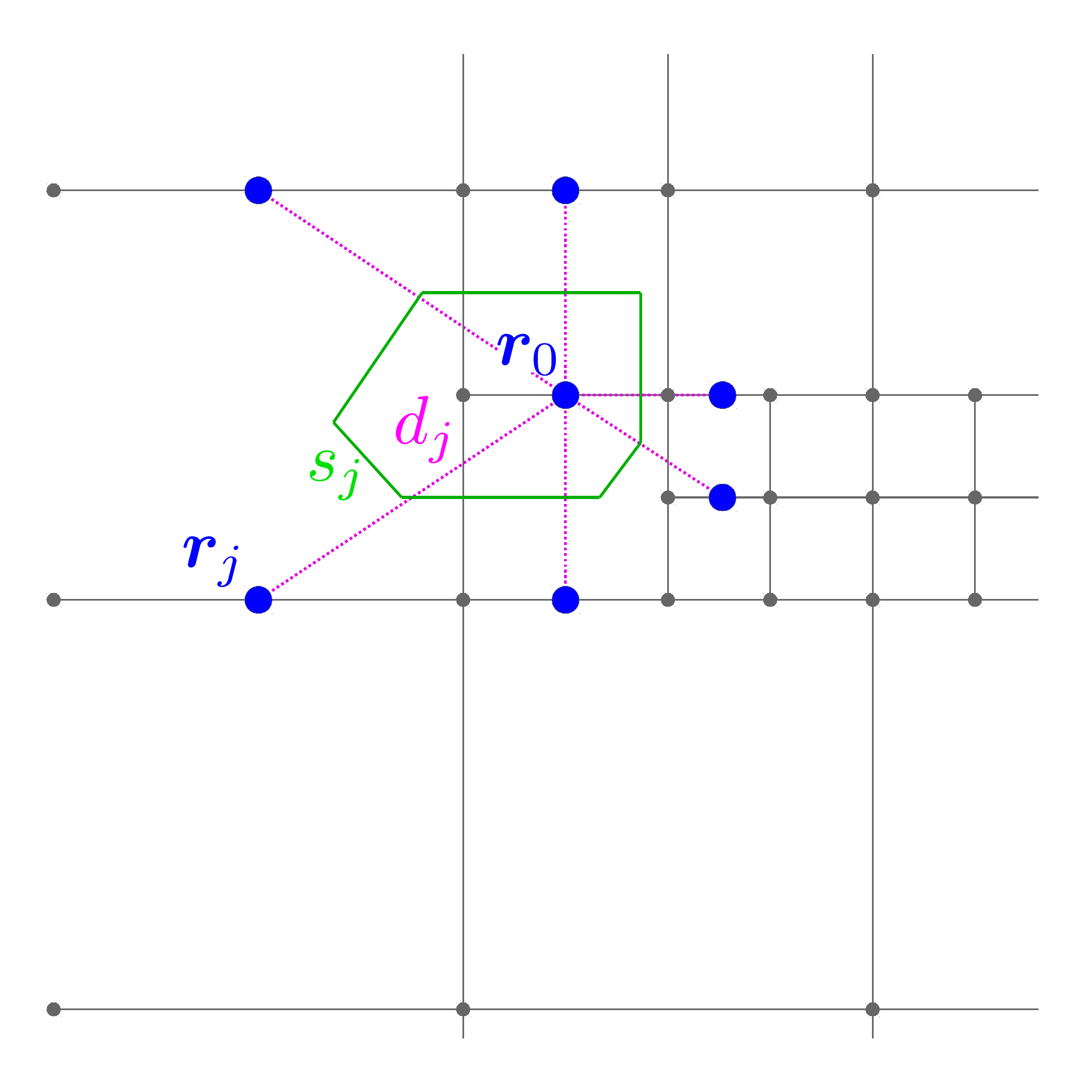} 
		\caption{}
	\end{subfigure}
	~
\caption{Example illustrations of Voronoi cell constructions for face-sampled velocity discretization. (a) A Voronoi cell to discretize x-component of velocity abound point $\rVec_{0}$. (b) A Voronoi cell to discretize y-component of velocity abound point $\rVec_{0}$.  }
\label{fig:voronoi_construction}
\end{figure}

The edges of the Voronoi cells which join two data points are constructed from the bisector line of the segment which connects the two data points. This construction is convenient in part because it yields a flux that is orthogonal the edge connecting the two points. Additionally, the cells are created such that the edges of the control volume next to the interface lie on the interface. An example of such a cell is shown in Fig.~\ref{fig:voronoi_construction}. Let us consider the point $\rVec_{0}$, the location of a face center about which we wish to build the discretization for the corresponding velocity component of $\vns^{*}$.  We construct the Voronoi cell $C$ about the point $\rVec_{0}$ by selecting $j$ neighboring points $\rVec_{j}$. The distance $d_{j}$ denotes the distance between points $\rVec_{0}$ and $\rVec_{j}$, and $s_{j}$ is the length of the edge created by bisecting $d_{j}$. An illustration of this construction is given in Fig.~\ref{fig:voronoi_construction}.

Suppose that we are writing the discretization for the $x$-component of $\vns^{*}(\rVec_{0}) = (u^{*}, v^{*})$, then the momentum equation using the semi-Lagrangian approximation for the advection part gives:
\begin{equation*}
\int_{C} \SLFormulationNSIntermediate{u} \text{ }\rm{d}V = \int_{C}\frac{1}{Re} \laplace u^{*}  \text{ } \rm{d}V= \frac{1}{Re}\int_{\partial C}\nabla u^* \cdot \bm{n} \text{ }\rm{d}S
\end{equation*}
The right-hand side can then be discretized using the Voronoi cell construction, yielding the final form
\begin{equation*}
\text{Vol}(C) \left( \SLFormulationNSIntermediate{u} \right) = \frac{1}{Re} \sum_{j \in \text{Voro}(\rVec_{0})} \frac{s_{j} (u^*(\rVec_{j}) - u^*(\rVec_{0}))}{d_{j}},
\end{equation*}
where $\text{Vol}(C)$ refers to the volume of the current Voronoi cell $C$ and $\text{Voro}(\rVec_{0})$ is the set of points that make up the given Voronoi cell.

Neumann boundary conditions on $\vns$ are applied by incorporating $\nabla \vns \cdot \normal |_{\Gamma}$ naturally into the flux discretization on the edges of cells which are in contact with the boundary. To apply the Dirichlet interface condition on fluid velocity, which relates the fluid velocity to the velocity of the evolving interface given by equation \eqref{eq:finalNSInterfaceBC}, a finite-difference approach utilizing discretizations of the form discussed in Sec.~\ref{sec:quadtrees} is used.  A uniform grid is enforced in a band around the interface, and in the presence of the interface, the discretization collapses to a uniform finite difference discretization which takes into account the boundary condition value via the  Shortley-Weller method \cite{shortley1938numerical,chen2007supra}.

\subsubsection{Discretizations for projection step}
\label{sec:hodgesubsec}

\newcommand{\cell}[1]{C_{#1}}
\newcommand{\ngbdf}[1]{\text{ngbd}_{f}(#1)}

\newcommand{\sbar}[1]{s_{#1}}
\newcommand{\ngbdfCn}[2]{\text{ngbd}_{f_{#1\cell{#2}}}}
\newcommand{\ngbdC}[1]{\text{ngbd}_{\cell{}}(#1)}

The discretization of equation \eqref{eq:sub_incompressibility} satisfied by the Hodge variable $\Phi$ follows the approach introduced in \cite{Losasso;Fedkiw;Osher:06:Spatially-Adaptive-T}, which is a second-order accurate extension of the discretization of \cite{Losasso;Gibou;Fedkiw:04:Simulating-Water-and}.

To solve the projection step \eqref{eq:sub_incompressibility}, we use a cell-based finite-volume approach with a second order accurate discretization of Cartesian derivatives at the cell's faces. The discretization considers the integration form of  equation \eqref{eq:sub_incompressibility} over a finite volume cell $\cell{}$: 
\begin{equation}
\int_{\partial \cell{}} \nabla \Phi \cdot \bm{n}_{f} \text{ }\rm{d}S = \int_{\cell{}} \divergence \vns^{*} \text{ }\rm{d}V, \label{eq:FV}
\end{equation}
where $\bm{n}_{f}$ is the outward normal to face $f$ of the finite volume cell $\cell{}$. The idea in \cite{Losasso;Fedkiw;Osher:06:Spatially-Adaptive-T} is to define the flux on the face of a cell to be that of the largest cell neighboring that face. For example, let us consider a finite volume cell $\cell{}$ and neighboring faces $f_{j}$, where the Hodge variable is stored at the cell centers. An illustration of such a cell, with corresponding notations, is displayed below in Fig.~\ref{fig:hodge_finite_volumes} (a) - (b). 
\begin{figure}[H]
\centering
	\begin{subfigure}[t]{0.45\textwidth}
		\centerline{
		\includegraphics[width=\textwidth]{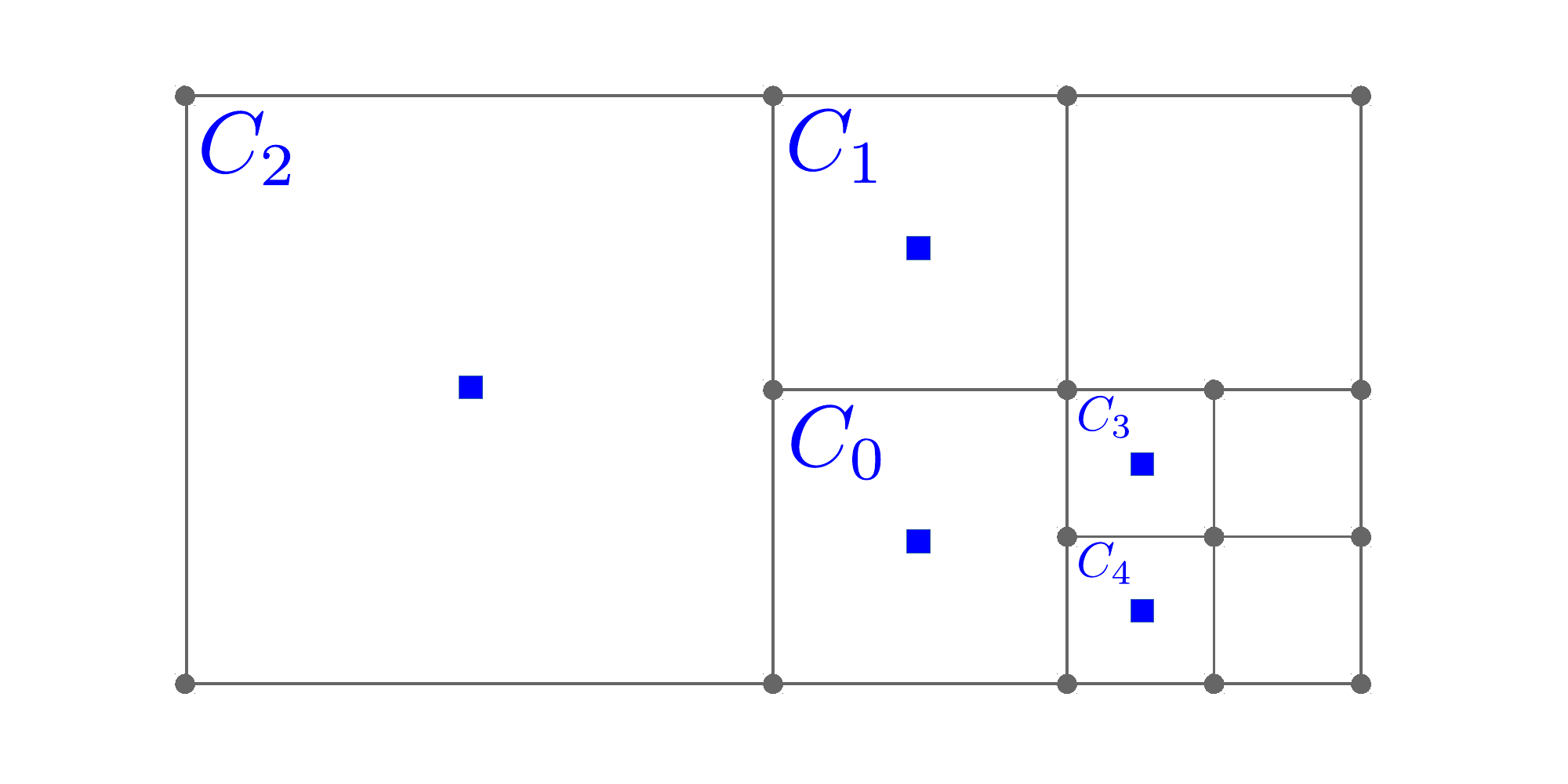} }
		\caption{}
	\end{subfigure}
	~
	\begin{subfigure}[t]{0.45\textwidth}
		\centerline{
		\includegraphics[width=\textwidth]{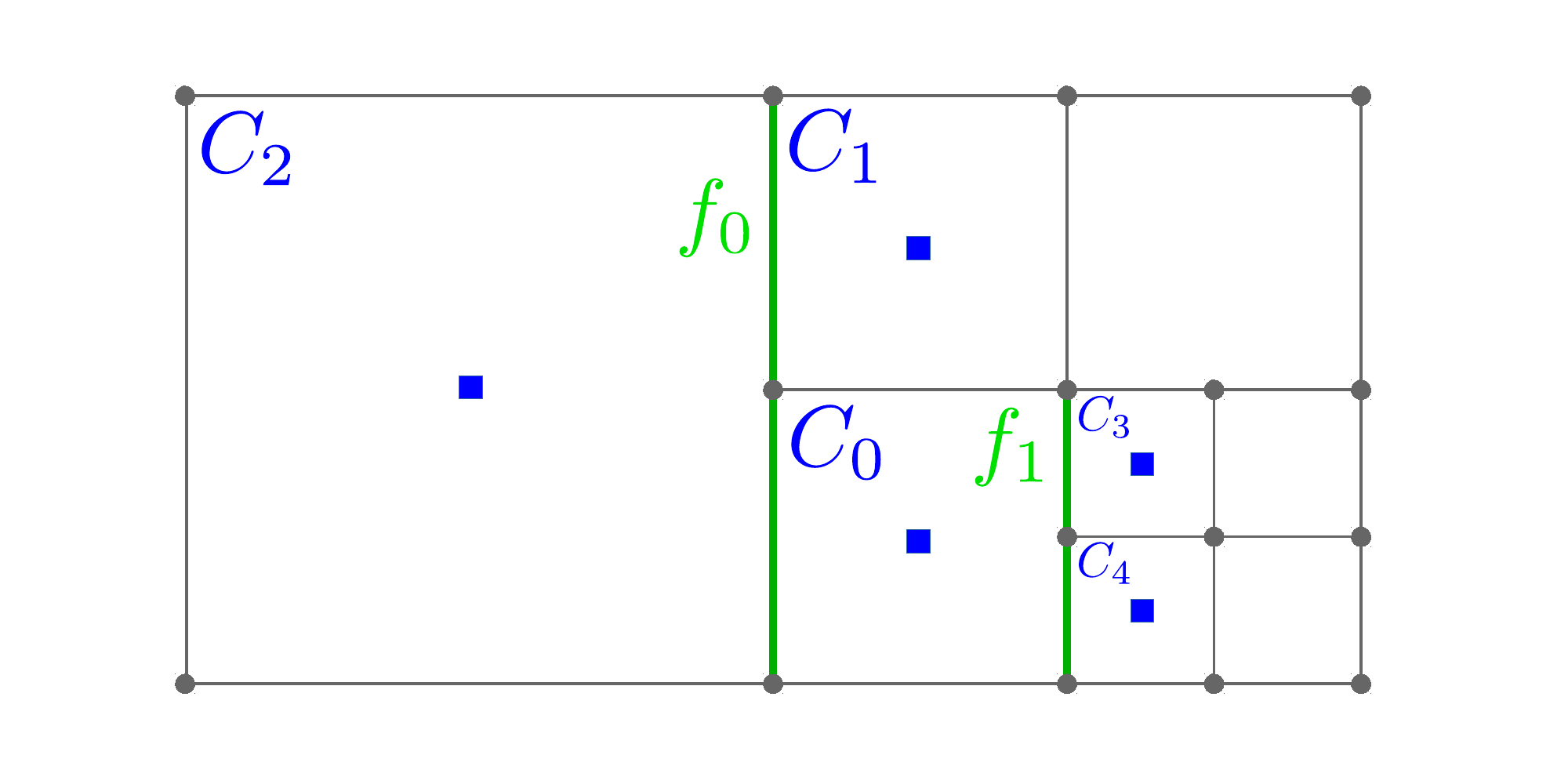} }
		\caption{}
	\end{subfigure}
	~
	\begin{subfigure}[t]{0.45\textwidth}
		\centerline{
		\includegraphics[width=\textwidth]{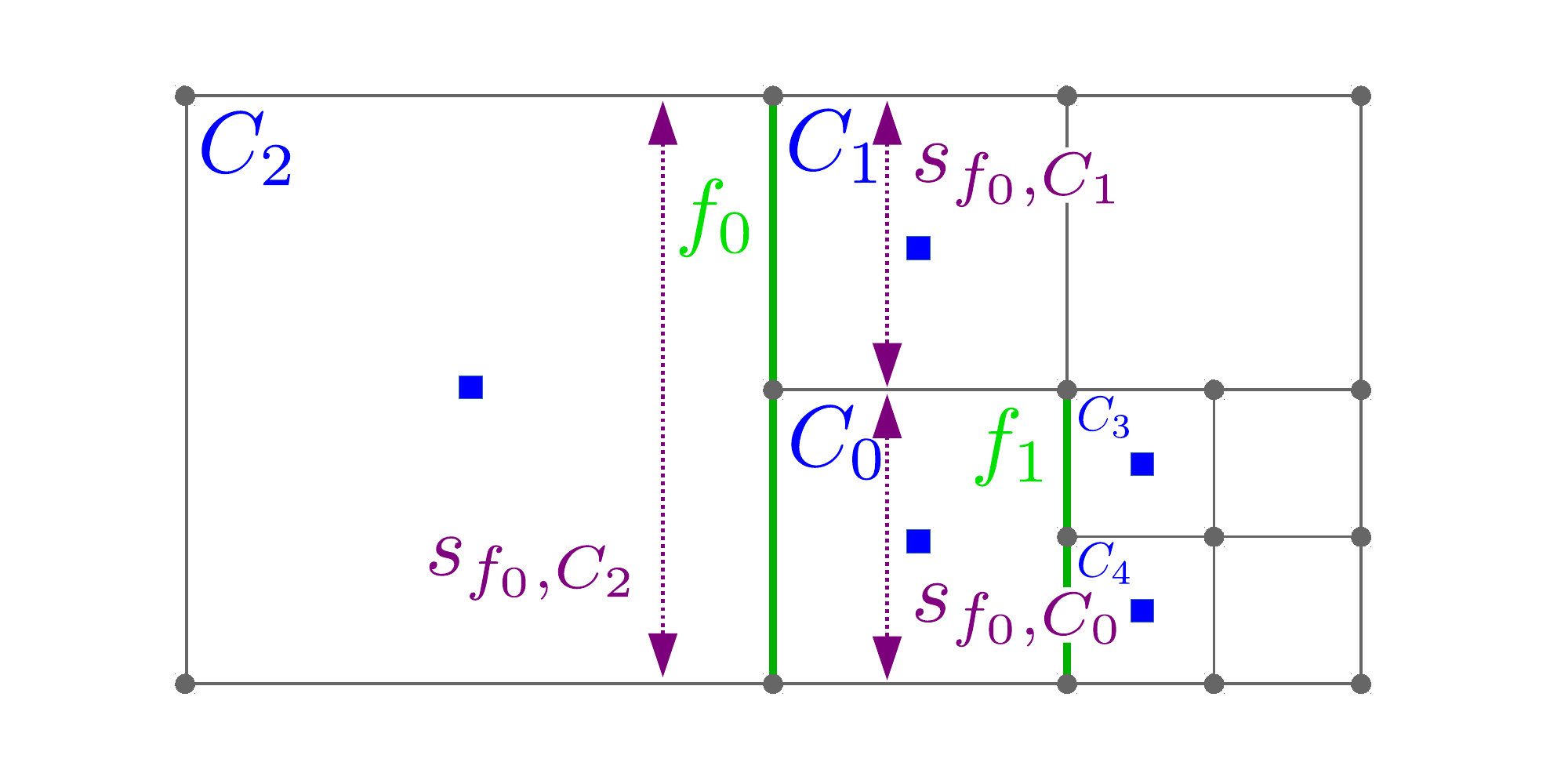} }
		\caption{}
	\end{subfigure}
	~
	\begin{subfigure}[t]{0.45\textwidth}
		\centerline{
		\includegraphics[width=\textwidth]{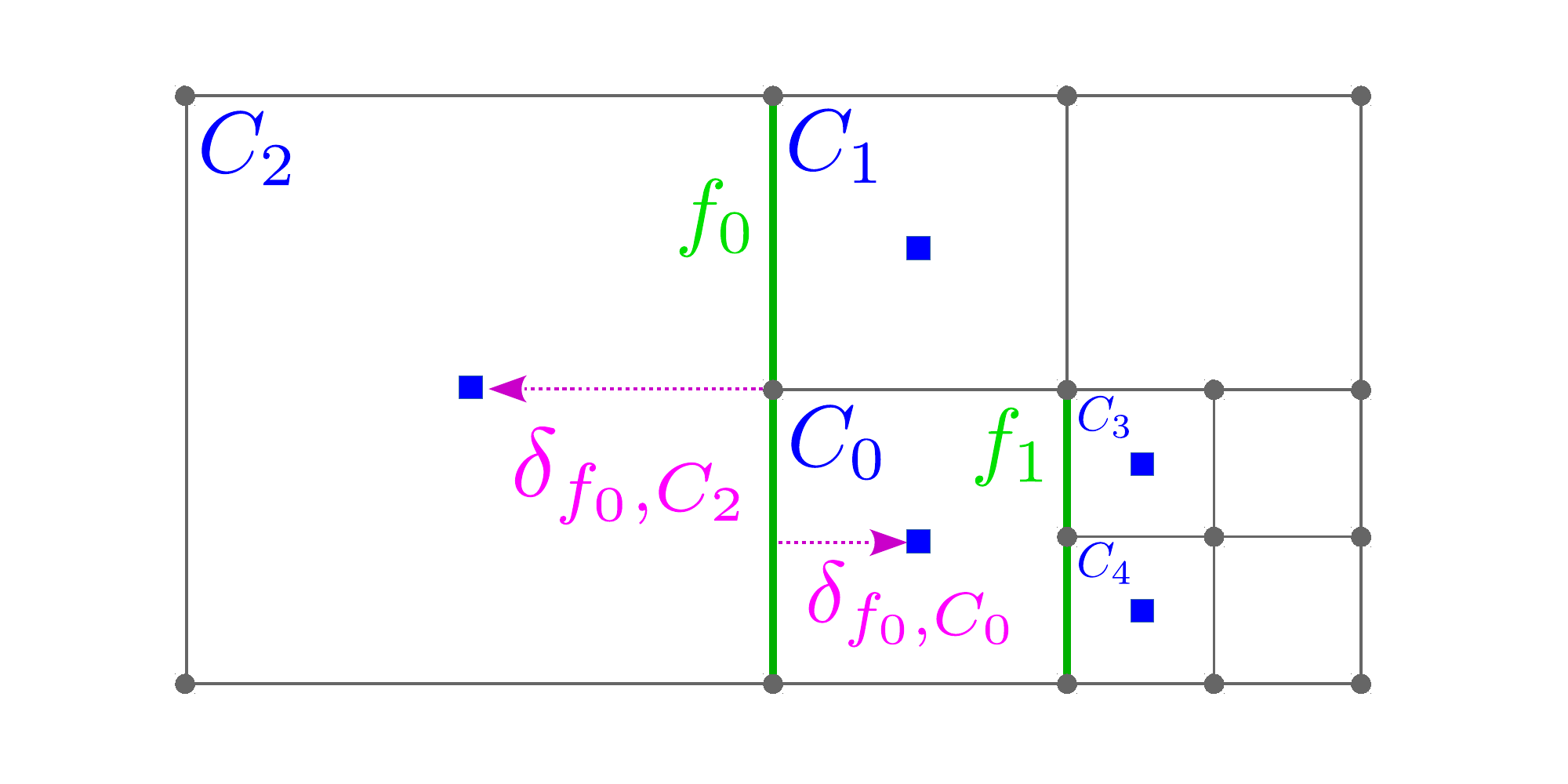} }
		\caption{}
	\end{subfigure}
	~
\caption{Example illustration of finite volume discretization notations for the Hodge variable. (a) Example labeled cells, with cell centers shown as blue squares, (b) with labeled faces displayed, (c) with edge lengths between given faces and cells displayed, (d) with signed distances between faces and cell centers displayed.  }
\label{fig:hodge_finite_volumes}
\end{figure}

For a given face $f$ at which we wish to discretize the flux of $\Phi$, we define $\cell{*}$ as the largest direct neighbor cell of the face $f$ and $N_{C}(f)$ as the set of cells in contact with $\cell{*}$ and $f$. We can then define the distances $s_{C_{i}, C_{k}}$ as the edge length connecting two cells $\cell{i}$ and $\cell{k}$, and $\delta_{f, C_{i}}$ as the length from face $f$ to the center of the cell $\cell{i}$. We can then write the average discretization distance for discretizing $\nabla \Phi \cdot \bm{n}_{f}$ across $f$ as
\begin{equation*}
\Delta = \sum_{C_i \in N_C(f)} \frac{s_{\cell{i}, \cell{*}}}{s_{\cell{*}, \cell{*}}} (\delta_{f, \cell{*}} - \delta_{f, \cell{i}}).
\end{equation*}

The discretization for the flux of $\Phi$ is then given by
\begin{equation*}
\nabla \Phi \cdot \bm{n}_f =  \sum_{\cell{i} \in N_C(f)} \left( \frac{ s_{\cell{i}, \cell{*}} }{ s_{\cell{*}, \cell{*}}} \right)  \left( \frac{\Phi_i - \Phi_{*}}{\Delta} \right)
\end{equation*}

For a given cell $C$ for which we wish to discretize the divergence of fluid velocity at the cell center, we additionally define the lengths $s_{f_j, C}$ as the length of the edge connecting cell $C$ and a given face $f_j$. We denote $N_{f}(C)$ as the set of all faces in contact with cell $C$.  

The average size of a face $f_j$ is defined as
\begin{equation*}
A_{f_{j}} = \frac{1}{2} \sum_{\cell{i} \in N_{C}(f_j)} s_{\cell{i}, \cell{*}}, 
\end{equation*}
which is analogous with the length $s_{\cell{*}, \cell{*}}$ as defined previously for a given face with corresponding cell $\cell{*}$.

The right-hand side of \eqref{eq:FV}, aka the divergence of the intermediate velocity field, can then be discretized for a given cell $C$ as

\begin{equation*}
\divergence \vns^{*} |_{C} = \frac{1}{\Delta_{C}} \sum_{f_{j} \in N_f(C)} \left( \frac{\delta_{f_j, C}}{|\delta_{f_j, C}|} \right) \left( \frac{ \sbar{f_j, C} \text{  }\vns^{*}_{f_j}}{A_{f_j}} \right) ,
\end{equation*}
where $\Delta_{C}$ is the size of cell $C$.  

An example of notations for the $x-$direction discretizations is shown above in Fig.~\ref{fig:hodge_finite_volumes}(a)-(d). For example, when considering the flux across $f_{0}$, we have $\cell{*} = \cell{2}$, $N_{C}(f_0) = \{ \cell{0}, \cell{1} \}$, and the corresponding edge lengths and signed distances shown in Fig.~\ref{fig:hodge_finite_volumes} (c) - (d). Considering the divergence of the intermediate velocity discretized for $\cell{0}$, we would have $N_{f}(C_0) = \{ f_0, f_1 \}$.

Neumann boundary conditions are applied by including the boundary condition flux term naturally into the finite volume discretization. Dirichlet boundary conditions are again applied using a similar finite difference approach as \cite{shortley1938numerical,chen2007supra}, enforcing a uniform grid near the interface.

\subsubsection{\revised{A note on the boundary conditions}}
\label{subsec:NSBCexplain}
It is worth noting that because we solve \eqref{eq:sub_momentum} in a component-wise fashion, we must correspondingly prescribe the boundary condition on fluid velocity given by \eqref{eq:finalNSInterfaceBC} in a component-wise manner. As such,  we prescribe the conditions
\begin{align} 
u|_\Gamma  &= (\vint \cdot \normal) \left(1 - \frac{\rhos}{\rhol} \right) n_x,
\label{eq:vnsxcomponentInterfaceBC} \\
v|_\Gamma &= (\vint \cdot \normal) \left(1 - \frac{\rhos}{\rhol} \right ) n_y, 
\label{eq:vnsycomponentInterfaceBC}
\end{align}
where $u$ is the $x$-component of the fluid velocity sampled on the vertical face centers, $v$ is the $y$-component of the fluid velocity sampled on the horizontal face centers, and $n_x$ and $n_y$ are the $x$ and $y$-components of the unit normal vector $\normal$, respectively. The conditions \eqref{eq:vnsxcomponentInterfaceBC} and \eqref{eq:vnsycomponentInterfaceBC} satisfy equality when substituted back into \eqref{eq:finalNSInterfaceBC} and serve to approximate the condition. 
We additionally recall that the condition which is actually imposed is on the intermediate velocity field $\vns^{*}$ rather than $\vns^{n+1}$, and so the form of the boundary condition is modified in order to account for the Hodge decomposition. Therefore, the conditions applied to the intermediate velocity field may be expressed as
\begin{align} 
v_x^{*}|_\Gamma  &= (\vint \cdot \normal) \left(1 - \frac{\rhos}{\rhol} \right) n_x + \frac{\partial \Phi}{\partial x}|_{\Gamma}, 
\label{eq:vnsxcomponentInterfaceBCHodge} \\
v_y^{*}|_\Gamma &= (\vint \cdot \normal) \left(1 - \frac{\rhos}{\rhol} \right ) n_y + \frac{\partial \Phi}{\partial y}|_{\Gamma},
\label{eq:vnsycomponentInterfaceBCHodge}
\end{align} where $\Phi$ is the Hodge variable.

The interfacial velocity $\vint$, unit normals $\normal = \left( n_x \text{    }n_y \right)^{T}$, and scalar normal velocity $\vint \cdot \normal$ are stored at the grid nodes. The unit normals are calculated using the level-set function as described in Sec.~\ref{sec:levelSetMethods}, using the node-based spatial discretizations described in Sec.~\ref{sec:quadtrees} to calculate the gradient of the level-set function. The gradient of the Hodge variable $\nabla \Phi = \left( \frac{\partial \Phi}{\partial x} \text{    } \frac{\partial \Phi}{\partial y}\right)^{T}$ is discretized at the face centers using the approach described in Sec.~\ref{sec:hodgesubsec}. 

\begin{figure}[H]
\centering
	\begin{subfigure}[t]{0.5\textwidth}
		\centering
		\includegraphics[width=\textwidth]{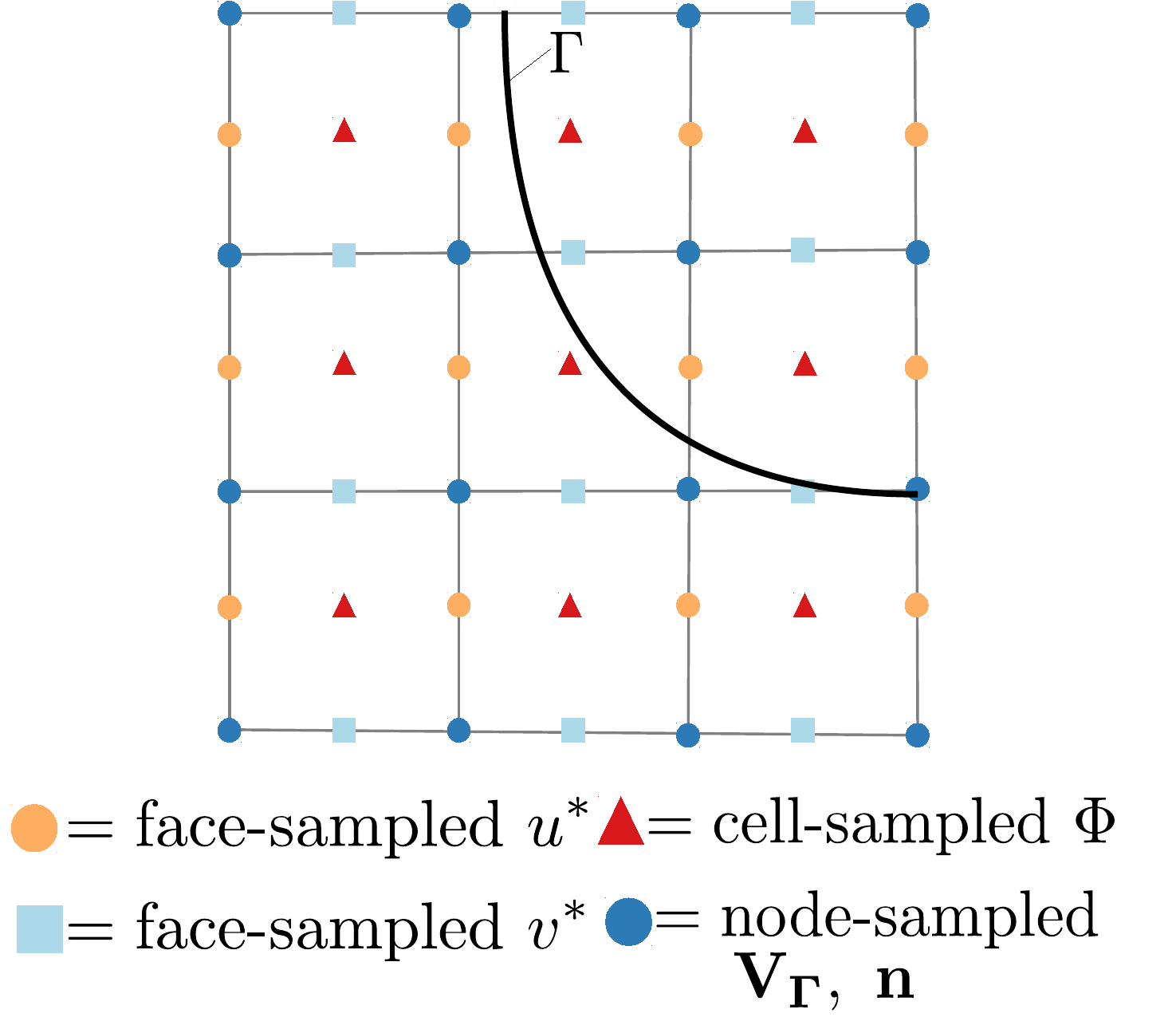} 
		\caption{}
	\end{subfigure}
	~
	\begin{subfigure}[t]{0.49\textwidth}
		\centering
		\includegraphics[width=\textwidth]{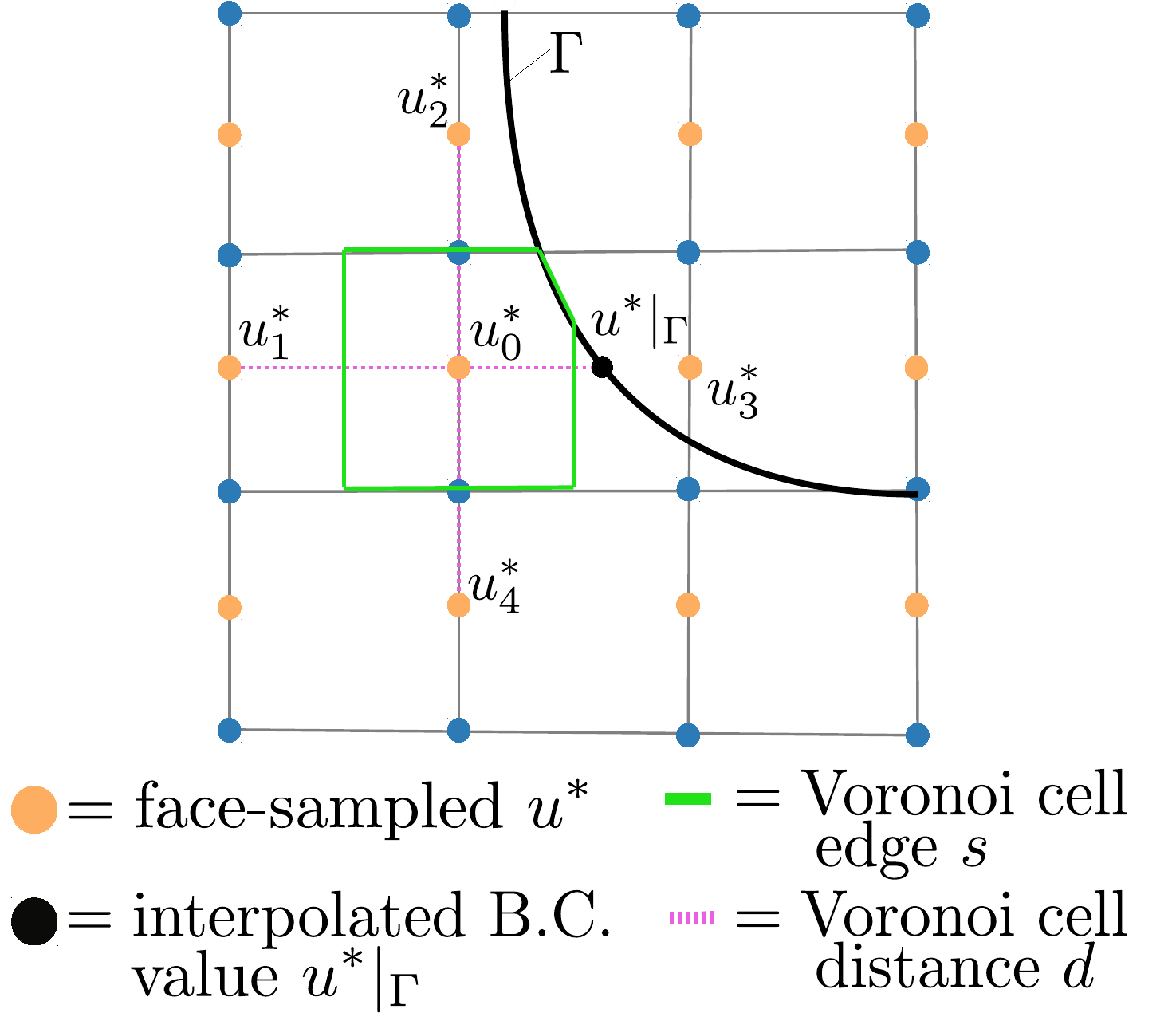} 
		\caption{}
	\end{subfigure}
	~
	\begin{subfigure}[t]{0.49\textwidth}
		\centering
		\includegraphics[width=\textwidth]{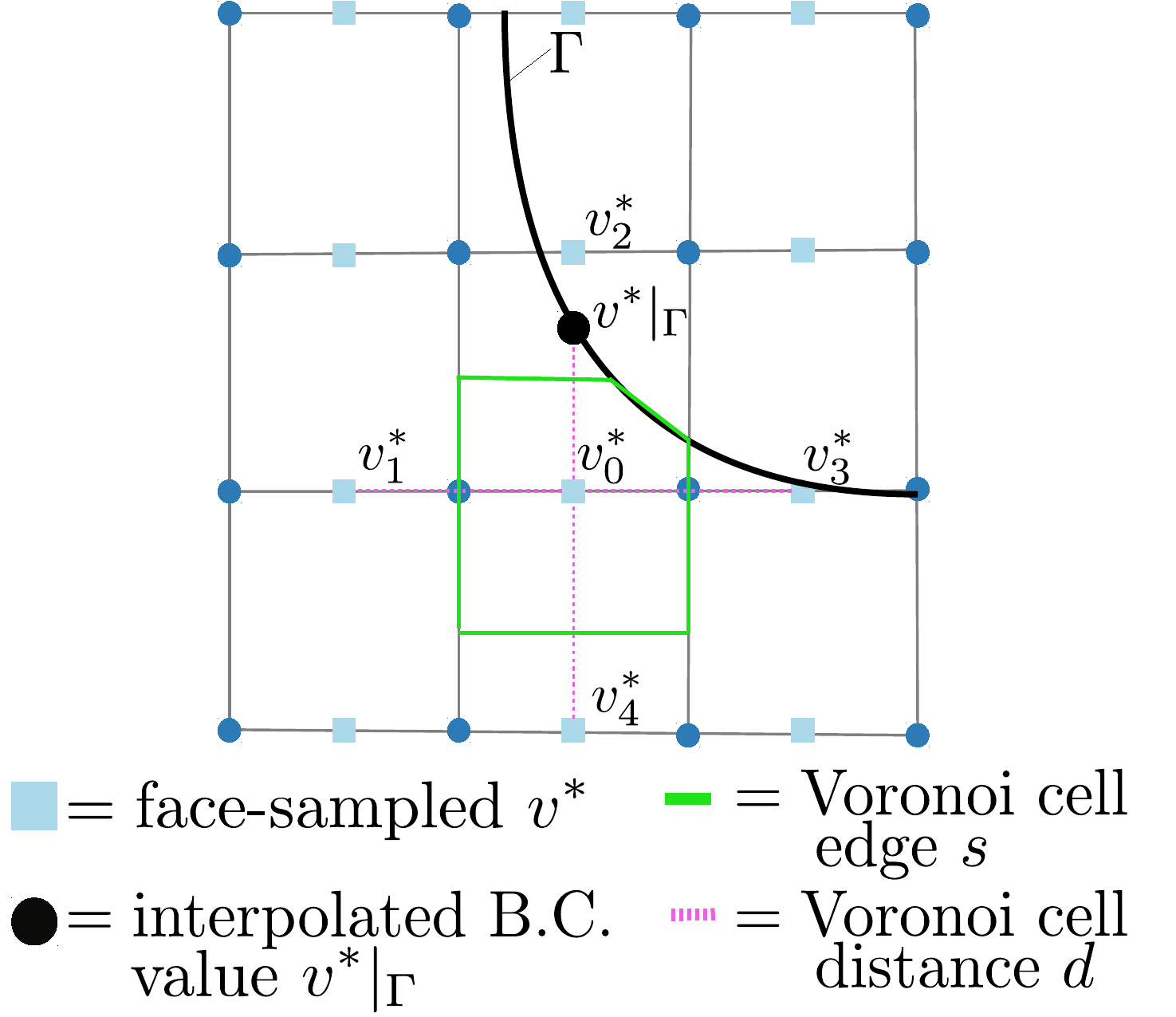} 
		\caption{}
	\end{subfigure}
	
\caption{Illustrations of the application of component-wise interfacial boundary conditions on the fluid velocity given by eqs. \eqref{eq:vnsxcomponentInterfaceBC} - \eqref{eq:vnsycomponentInterfaceBC}. (a) Data layout for relevant fields that enter the boundary condition. The interfacial velocity $\vint$ and unit normals $\normal$ are sampled at the nodes, and interpolated billinearly from the nodes to relevant locations for calculating the boundary condition values $u_\Gamma$ and $v_\Gamma$ as depicted in (b, c). (b,c) Examples of a Voronoi cells for near-interface discretizations of the (b) the $x$-component of the fluid velocity, $u$, at the vertical face centers and (c)  the $y$-component of the fluid velocity, $v$, at the horizontal face centers. }
\label{fig:ns_bc_explanation}
\end{figure}

To apply the boundary conditions, we utilize the Shortley-Weller approach \cite{shortley1938numerical, chen2007supra} similar to that described previously in Sec.~\ref{sec:poissontype}. For a given face whose neighbor lies across the interface, the boundary condition value is incorporated into the discretization instead, and the corresponding discretization distance is shortened such that reflects the distance to the interface. In this case, the boundary condition value is computed as given by either \eqref{eq:vnsxcomponentInterfaceBCHodge} or \eqref{eq:vnsycomponentInterfaceBCHodge}, where the values of $\vint \cdot \normal$, $n_x$, $n_y$, $\partial \Phi/ \partial x$, and $\partial \Phi / \partial y$ are all bilinearly interpolated to the relevant interface location required by the discretization. Illustrations of the data layout and examples of Voronoi cell discretizations close to the interface are depicted in Fig.~\ref{fig:ns_bc_explanation}.

\subsection{Extension of fields} \label{sec:extension}
It is important to define valid values for the solutions on each side of the interface in order to perform discretizations that avoid discontinuities in the solution, or simply to define values that do not exist. This is the case for example when defining valid values for the solid after we advance the interface from $t^n$ to $t^{n+1}$ or when discretizing an equation at a node that belongs to the liquid region but which requires neighboring values that are not necessarily in the liquid region. We make use of different extrapolation schemes depending on where the data is stored.

\subsubsection{PDE-based extrapolation of node-sampled fields} 
For the extension of node-sampled fields across the interface, we use the PDE-based quadratic extrapolation approach of \cite{bochkov2019multidimensional}, adapted from \cite{aslam2004partial}. Considering a scalar field $q$ that we want to extend from the region defined by $\phi\leq 0$ to the region given by $\phi>0$, we first define the unit normal, $\normal$, pointing outwards from the domain $\phi \leq 0$, and compute $q_{\normal \normal} = \nabla (\nabla q \cdot \normal) \cdot \normal$ and $q_{\normal} = \nabla q \cdot \normal$ with central differencing given in section \ref{sec:NodeBasedSpatialDiscretization}. Then, we successively solve the following set of partial differential equations for about 20 iterations each, in fictitious time $\tau$:
\begin{align*}
\frac{\partial q_{\normal \normal}}{\partial \tau} + H(\phi) (\normal \cdot \nabla q_{\normal \normal})&=0, \\ 
\frac{\partial q_{\normal}}{\partial \tau} + H(\phi) (\normal \cdot \nabla q_{\normal} - q_{\normal \normal})&=0, \\
\frac{\partial q}{\partial \tau} + H(\phi) (\normal \cdot \nabla q - q_{\normal})&=0,
\end{align*}
where the Heaviside $H(\phi)$ is defined as:
\begin{equation*}
H(\phi) = 
\begin{cases}
0, \text{ if } \phi \leq 0 \\
1, \text{ if } \phi > 0 \\
\end{cases}.
\end{equation*}

\subsubsection{Geometric extrapolation of face and cell-sampled fields} 
\begin{wrapfigure}{r}{4cm}
\centerline{
\vspace{-1cm}
\includegraphics[height = 0.3\textwidth]{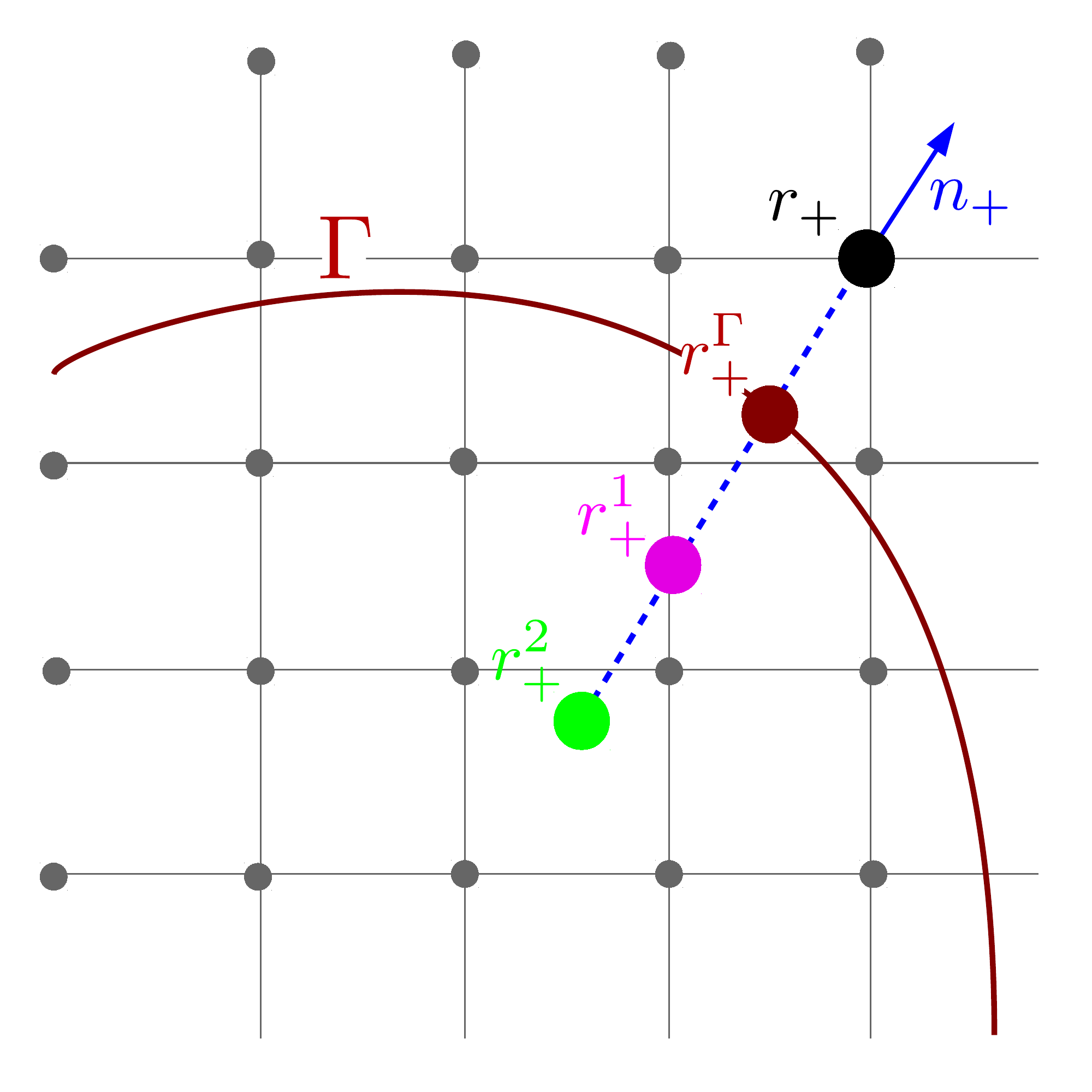} }
\vspace{1cm}
\caption{Illustration of points defined for geometric extraploation procedure.}
\label{fig:extrap}
\hspace{2cm}
\vspace{-1cm}
\end{wrapfigure}

\newcommand{\rplus}{\rVec_{+}}
For extension of face and cell-sampled fields, the data layout of the face and cell centers present a challenge in implementation compared with node-sampled fields, and thus a geometric quadratic least-squares extrapolation approach is used \cite{guittet2015stable}. Again, we consider a scalar field $q$ which we want to extend from the region $\phi \leq 0 $ to the region $\phi >0$. Let us first define a point $\rplus$ in the positive subdomain, as shown in Fig.~\ref{fig:extrap},  at which we want an extended value of $q$, denoted $q(\rplus)$. This time, we define $\normal_{+}$ as the unit normal pointing outward from the positive subdomain defined by $\phi >0$. To evaluate $q$ at this point, we build a second degree Newton polynomial along the normal direction using the interface boundary condition and two values from the known region $\phi \leq 0$.

\subsection{Interpolation methods} \label{sec:interp}
\newcommand{\xplus}{x_{+}}
\newcommand{\xminus}{x_{-}}
\newcommand{\yplus}{y_{+}}
\newcommand{\yminus}{y_{-}}

Interpolation procedures depend on the location where the data is stored for the given field of interest. For all node-sampled fields except fluid velocity components, a quadratic non-oscillatory method is used, whereas a standard quadratic approach is used for fluid velocity components, and a geometric weighted least-squares method is used for cell and face - sampled fields.

%\elyce{\underline{Addressing raphael comment:} Technically the node-sampled fluid velocity components are interpolated using a quadratic method (different from quadratic non-oscillatory), but I don't think we need to devote a section to that. }
\subsubsection{Quadratic non-oscillatory interpolation of node-sampled fields}
To perform the interpolation of node-sampled fields we utilize a quadratic non-oscillatory method, which defines the interpolant using a standard bilinear interpolation formula with added second-derivative correction terms. The second-derivatives used to apply the correction are selected using the \texttt{minmod} slope limiter, as is done in \cite{min2007second}, in order to achieve better stability in regions where the field may have sharp changes. However, we make a slight modification to the definition of the non-oscillatory second derivatives - rather than apply the \texttt{minmod} slope limiter to the second derivatives defined at all the vertices of the cell, we first interpolate the second derivatives along the faces of the cell, and then apply the \texttt{minmod} slope limiter. This modification ensures continuity of the interpolant across cells in uniform regions.

\begin{figure}[H]
	\centering
	\begin{subfigure}[t]{0.3\textwidth}
		\centering
		\includegraphics[width=\textwidth]{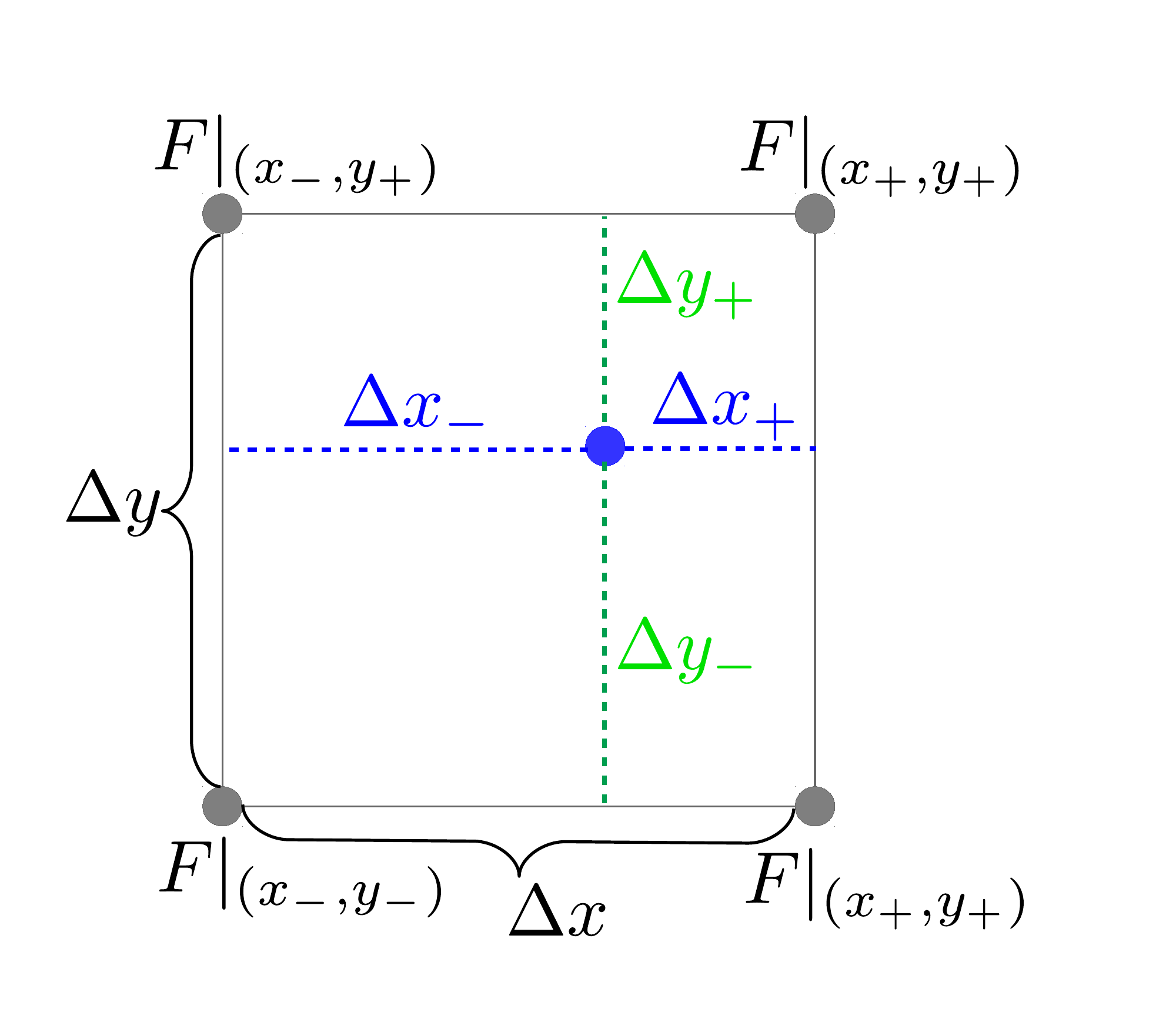} 
		\caption{}
		\label{fig:quadNO_notation}
	\end{subfigure}
	~
	\begin{subfigure}[t]{0.3\textwidth}
		\centering
		\includegraphics[width=\textwidth]{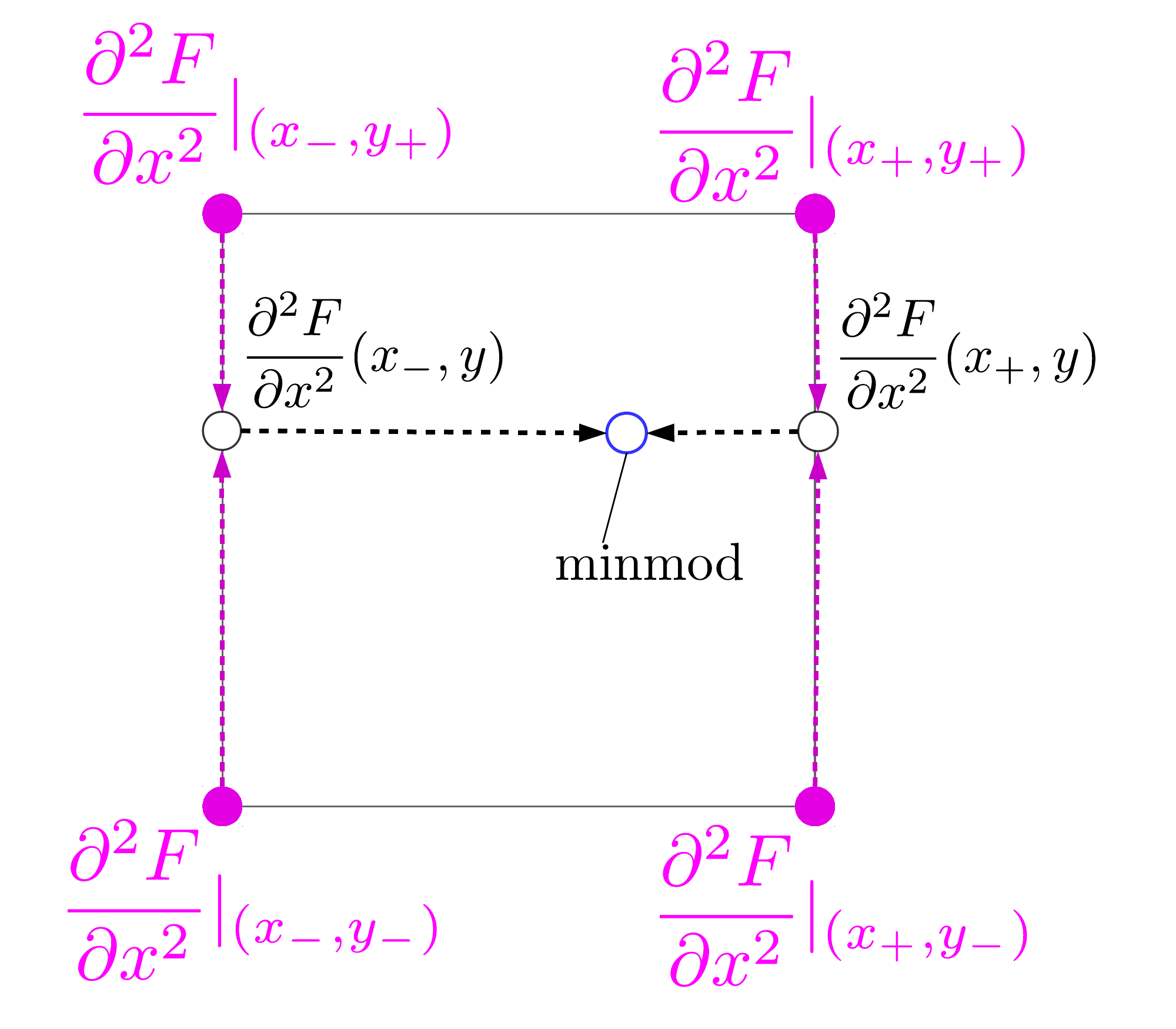} 
		\caption{}
	\end{subfigure}
	~
	\begin{subfigure}[t]{0.3\textwidth}
		\centering
		\includegraphics[width=\textwidth]{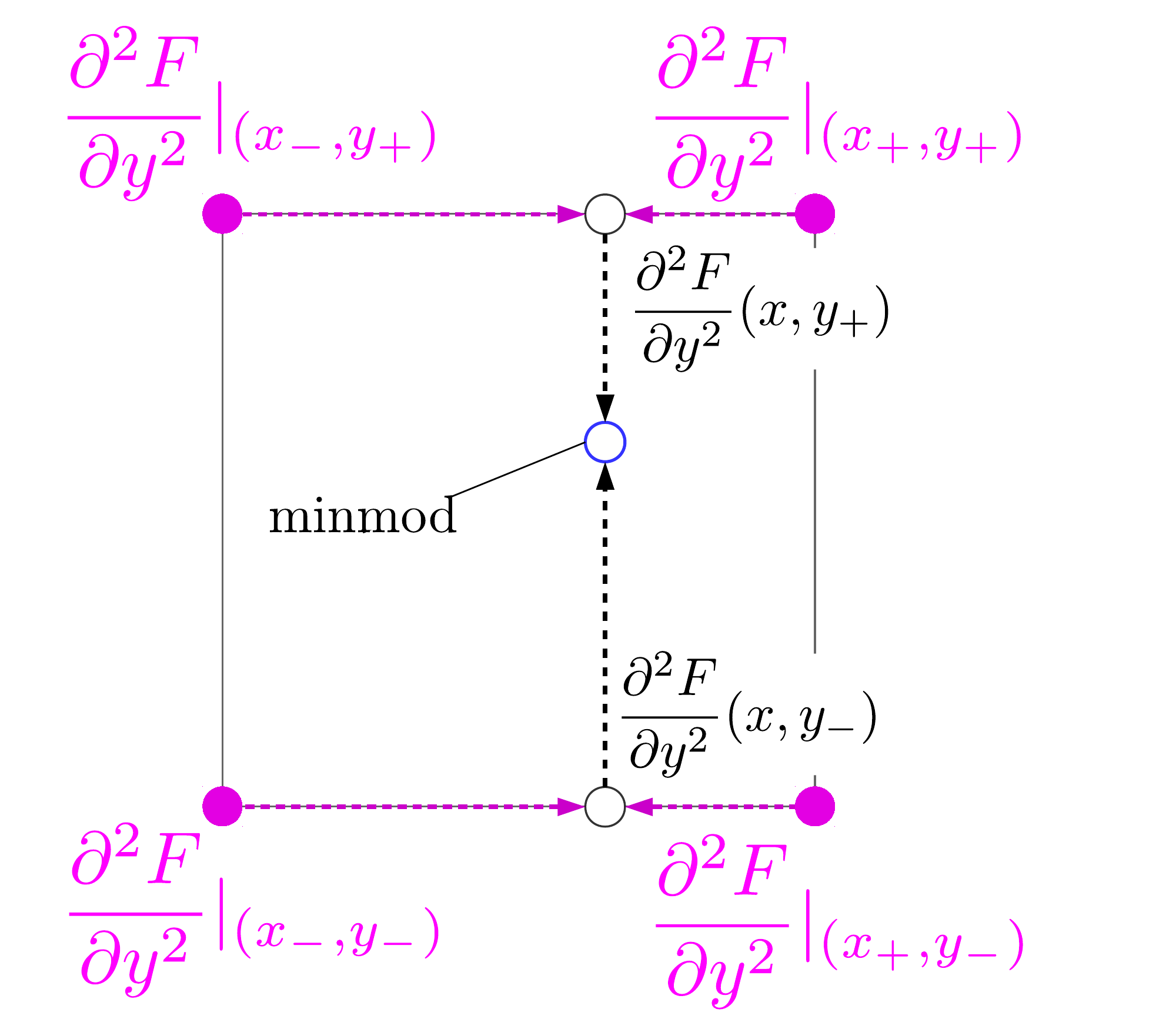} 
		\caption{}
	\end{subfigure}
	~
\caption{Illustration of quadratic non-oscillatory interpolation process with (a) grid cell with interpolation notations, (b) non-oscillatory approximation of the second-order derivative in $x$, (c) non-oscillatory approximation of the second-order derivative in $y$.}
\label{fig:quadNO_interp_process}
\end{figure}

Referring to Fig.~\ref{fig:quadNO_interp_process}, the interpolation of a scalar quantity $q(\bm{r})$ to a location $\bm{r} = (x, y)$ takes the form
\begin{gather*}
\begin{split}
q(\bm{r}) = q(\xminus,\yminus) w_{--} + q(\xplus,\yminus) w_{+-} +  q(\xminus,\yplus) w_{-+} + q(\xplus,\yplus) w_{++}  - \\
(\frac{1}{2} \Delta \xplus \Delta \xminus) q_{xx}|_{(\text{NO})} - (\frac{1}{2} \Delta \yplus \Delta \yminus) q_{yy}|_{(\text{NO})},
\end{split}
\end{gather*}  
where $q_{xx}|_{(\text{NO})}$ and $q_{yy}|_{(\text{NO})}$ are the selected ``non-oscillatory" second-order derivatives of the function $q$, and where the standard bilinear interpolation weights are given by 
\begin{align*}
\begin{split}
w_{--} &= \left(\frac{\Delta x_{+}}{\Delta x}\right) \left(\frac{\Delta y_{+}}{\Delta y}\right), \quad
w_{++} = \left(\frac{\Delta x_{-}}{\Delta x}\right) \left(\frac{\Delta y_{-}}{\Delta y}\right), \\
w_{-+} &= \left(\frac{\Delta x_{+}}{\Delta x}\right) \left(\frac{\Delta y_{-}}{\Delta y}\right), \quad
w_{+-} = \left(\frac{\Delta x_{-}}{\Delta x}\right) \left(\frac{\Delta y_{+}}{\Delta y}\right).
\end{split}
\end{align*}

To select the non-oscillatory second-order derivative values, we first interpolate the derivatives to locations on the grid cell faces that correspond to the point of interest, and then applying the \texttt{minmod} limiter. This is expressed as
\begin{align*}
\begin{split}
q_{xx}|_{NO} &= \texttt{minmod}(q_{xx}(\xminus,y), q_{xx}(\xplus,y)), \\
q_{yy}|_{NO} &= \texttt{minmod}(q_{yy}(x,\yminus), q_{yy}(x,\yplus)),
\end{split}
\end{align*}
where the second-order derivative values are interpolated along the cell faces as
\begin{align*}
\begin{split}
q_{xx}(\xminus,y) &= \frac{\Delta \yplus}{\Delta y} (q_{xx})|_{(\xminus,\yminus)} + \frac{\Delta \yminus}{\Delta y} (q_{xx})|_{(\xminus,\yplus)}, \\
q_{xx}(\xplus,y) &= \frac{\Delta \yplus}{\Delta y} (q_{xx})|_{(\xplus,\yminus)} + \frac{\Delta \yminus}{\Delta y} (q_{xx})|_{(\xplus,\yplus)}, \\
q_{yy}(x,\yminus) &= \frac{\Delta \xplus}{\Delta x} (q_{yy})|_{(\xminus,\yminus)} + \frac{\Delta \xminus}{\Delta x} (q_{yy})|_{(\xplus,\yminus)}, \\
q_{yy}(x,\yplus) &= \frac{\Delta \xplus}{\Delta x} (q_{yy})|_{(\xminus,\yplus)} + \frac{\Delta \xminus}{\Delta x} (q_{yy})|_{(\xplus,\yplus)}, 
\end{split}
\end{align*}
and the \texttt{minmod} slope limiter is given by
\begin{equation*}
\texttt{minmod}(a,b) = 
\begin{cases}
a \text{,   } |a| < |b| \text{ and } ab>0 \\
b \text{,   } |a| > |b| \text{ and } ab>0 \\
0 \text{,   } ab<0 
\end{cases}.
\end{equation*}

\subsubsection{Quadratic weighted least-squares interpolation of face and cell-sampled fields}
\newcommand{\doubleunderline}[1]{\underline{\underline{#1}}}
\newcommand{\Xmatrix}{\doubleunderline{\bm{X}}}
\newcommand{\Wmatrix}{\doubleunderline{\bm{W}}}
In the case of the Navier-Stokes fields, it may not always be possible to define a square using only face and cell centers on an adaptive grid, as it is in the case of node grid points. Therefore, it may not always be possible to write second-order accurate approximations of face and cell-sampled values, as is done for node-sampled ones. Due to this challenge in data layout, we utilize a weighted least-squares approach to interpolation of face and cell-sampled fields, as is done in \cite{guittet2015stable} and summarized below. 

\newcommand{\dd}[3]{(#1 - #1_{#2,#3})}
Considering a field $q$ sampled at the cells or the faces of the computational grid, we interpolate it at location $\bm{r}_{I} = (x_{I}, y_{I})$ by constructing a quadratic interpolant polynomial around $\bm{r}_{I}$. The procedure fetches the cell $C$ containing the point of interest $\bm{r}_{I}$ and all its second-degree neighbors (\emph{i.e.}, neighbor cells of neighbor cells). $n$ relevant data samples $\left(\bm{x}_{i}, q_{i}\right), \, i \in \left\{0, 1, \ldots, n - 1\right\}$ are accumulated in this manner by parsing this set of neighbor cells (or their appropriate faces, if interpolating face-sampled fields). The interpolation error $e_{i}$ on the $i^{\text{th}}$ sample then reads
\begin{small}
\begin{equation*}
e_{i} = \beta_0 + \beta_x \left(x_{i} - x_{I}\right) + \beta_y \left(y_{i} - y_{I}\right) + \beta_{xx} \left(x_{i} - x_{I}\right)^2 + \beta_{xy} \left(x_{i} - x_{I}\right)\left(y_{i} - y_{I}\right) + \beta_{yy} \left(y_{i} - y_{I}\right)^2 - q_{i},
\end{equation*}
\end{small}
where $\beta_0$, $\beta_x$, $\beta_y$, $\beta_{xx}$, $\beta_{yy}$, $\beta_{xy}$ are polynomial coefficients to be determined. Using matrix formalism, one has 
\begin{equation*}
\bm{e} = \Xmatrix  \bm{\beta} - \bm{q} 
\end{equation*}
where $\bm{e} = \left[e_{0} \ e_{1} \ \ldots e_{n-1}\right]^T$, $\bm{q} = \left[q_{0} \ q_{1} \ \ldots q_{n-1}\right]^T$, $\bm{\beta} = \left[\beta_{0} \ \beta_{x} \ \beta_{y} \ \beta_{xx} \ \beta_{xy} \ \beta_{yy}\right]^T$  and 
\begin{footnotesize}
\begin{equation*}
\Xmatrix = \begin{pmatrix}
1 & \left(x_{0} - x_{I}\right) & \left(y_{0} - y_{I}\right) & \left(x_{0} - x_{I}\right)^2 & \left(x_{0} - x_{I}\right)\left(y_{0} - y_{I}\right) & \left(y_{0} - y_{I}\right)^2 \\
\vdots & \vdots & \vdots & \vdots & \vdots & \vdots  \\
1 & \left(x_{n-1} - x_{I}\right) & \left(y_{n-1} - y_{I}\right) & \left(x_{n-1} - x_{I}\right)^2 & \left(x_{n-1} - x_{I}\right)\left(y_{n-1} - y_{I}\right) & \left(y_{n-1} - y_{I}\right)^2 \\
\end{pmatrix}.
\end{equation*}
\end{footnotesize}
The weighted least-square interpolation method intends to find the interpolation coefficients $ \bm{\beta}$ that minimize $\left\| \Wmatrix \bm{e}\right\|_{2}$ where the weight matrix
$$\Wmatrix = \text{diag}\left(\left[\left\|\bm{x}_{0} - \bm{x}_{I}\right\|^{-1} \ \left\|\bm{x}_{1} - \bm{x}_{I}\right\|^{-1} \ \ldots \left\|\bm{x}_{n-1} - \bm{x}_{I}\right\|^{-1}\right]\right)$$ 
attaches weight factors to interpolation errors: interpolation errors are given a weight inversely proportional to their distance to $\bm{x}_{I}$. The solution of this minimization problem is 
\begin{equation*}
\bm{\hat{\beta}}_{WLS} = (\Xmatrix^{T} \Wmatrix^{2} \Xmatrix)^{-1} \Xmatrix^{T} \Wmatrix^{2} \bm{q}
\end{equation*}
and the first component of $\bm{\hat{\beta}}_{WLS}$, i.e., $\hat{\beta}_{WLS, 0}$, is a second-order accurate interpolation of $q$ at the desired point $\bm{r}_{I}$.

For each interpolation, a $6\times6$ symmetric positive definite system must be inverted in two dimensions, which we do using a Cholesky decomposition. While this approach parallelizes easily, it is relatively costly and is used sparingly. For example, velocity field values are interpolated from the faces to the nodes once at the end of each time step, so that the semi-Lagrangian backtrace calculations and their corresponding interpolations can be performed using node-sampled values. By doing so, we perform only one weighted least-squares interpolation per grid point, rather than multiple (as would be required for the backtrace calculations).

\color{black}

\section{Convergence test}\label{sec:numerical_verification}
\begin{figure}[H]
\centering
	\hspace{-15mm}
	\begin{subfigure}[t]{0.33\textwidth}
		\centering
		\includegraphics[height=0.28\textheight]{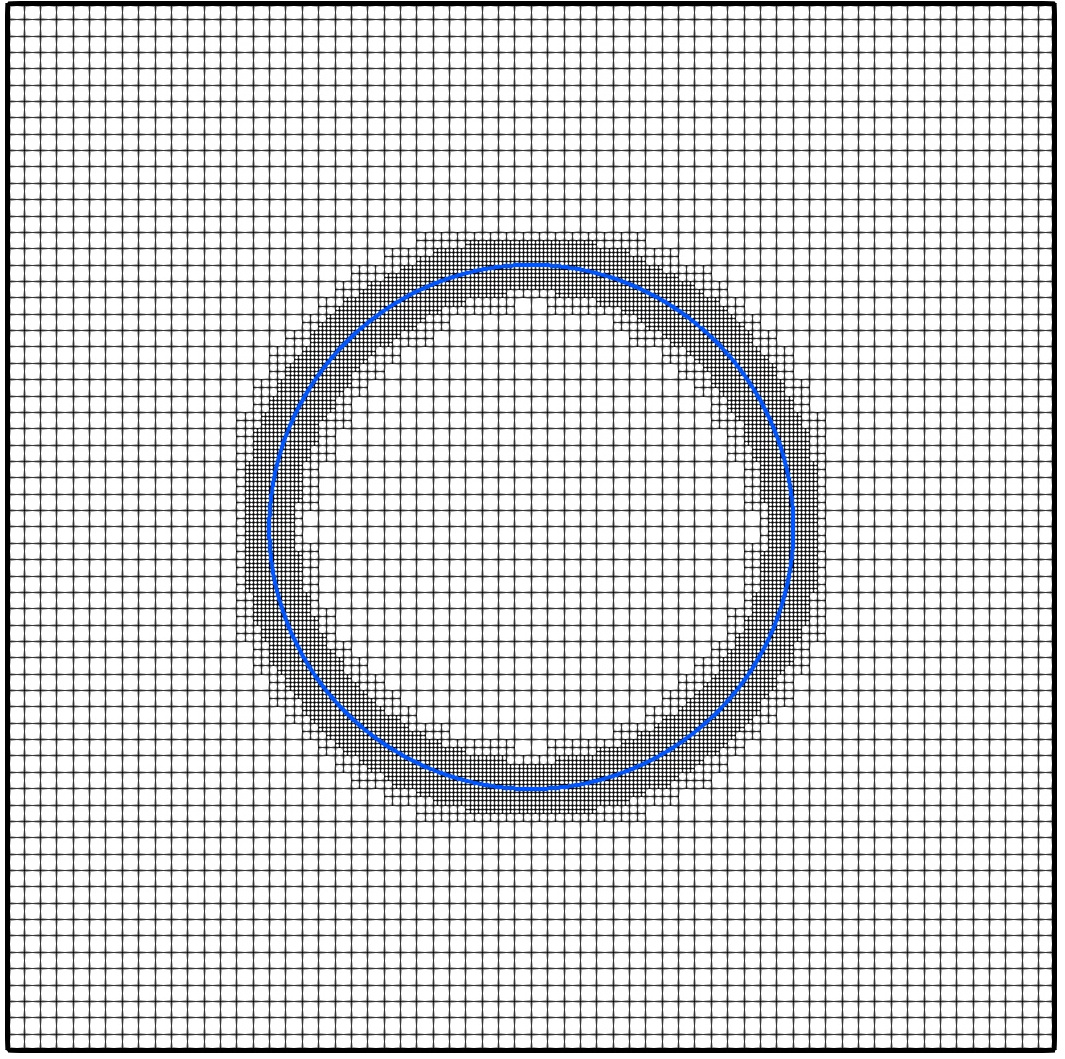} 
		\caption{}
	\end{subfigure}
	\hspace{3mm}
	~
	\begin{subfigure}[t]{0.33\textwidth}
		\centering
		\includegraphics[height=0.28\textheight]{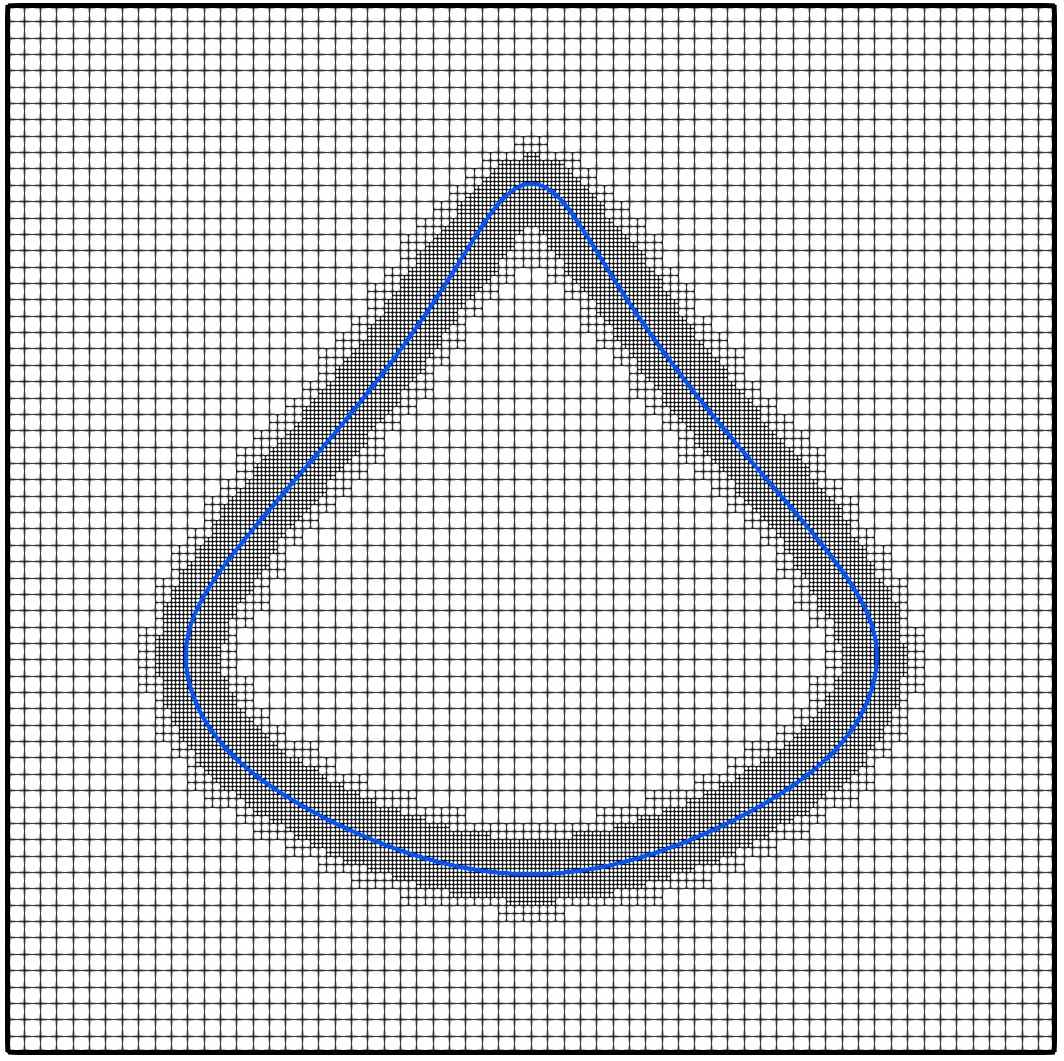} 
		\caption{}
	\end{subfigure}
	\hspace{3mm}
	~
	\begin{subfigure}[t]{0.33\textwidth}
		\centering
		\includegraphics[height=0.28\textheight]{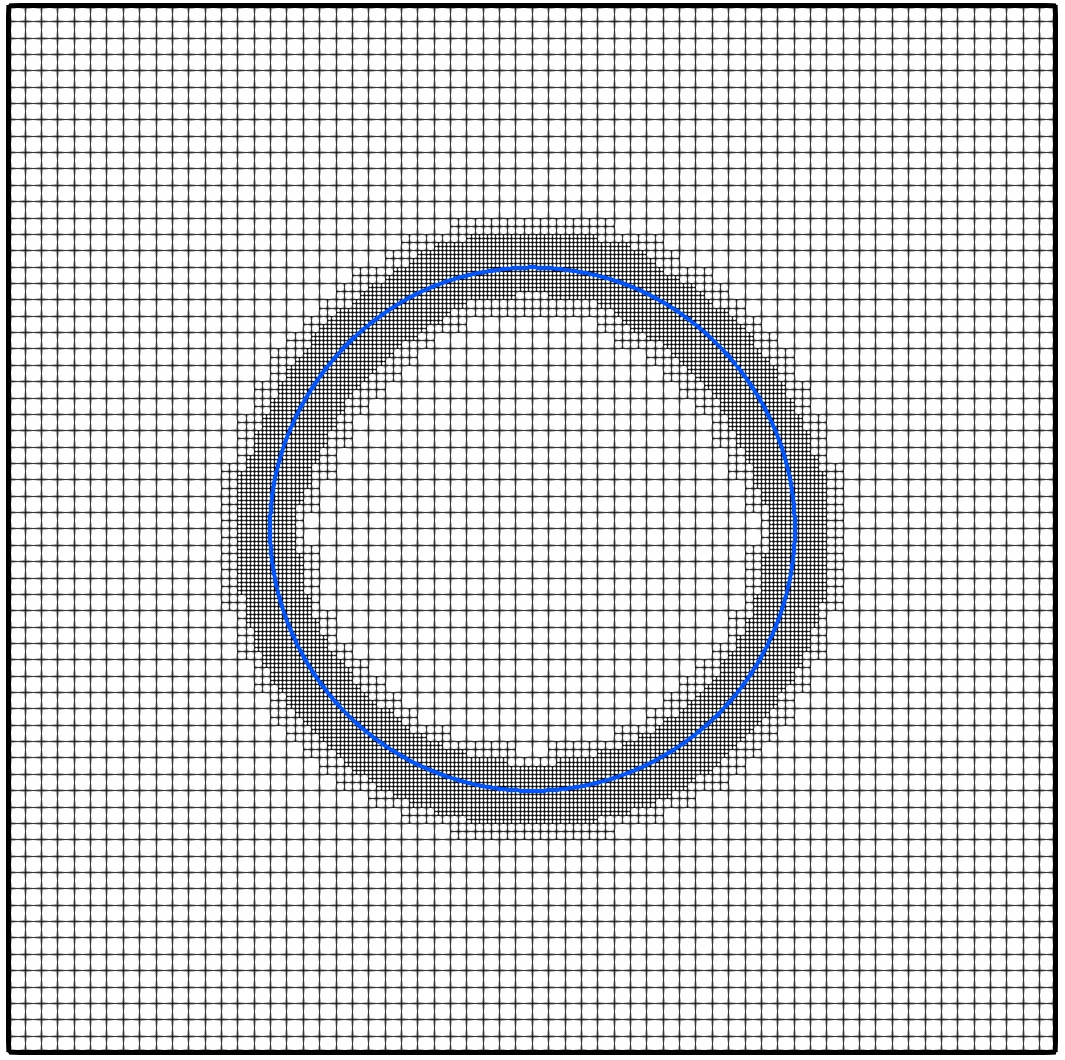} 
		\caption{}
	\end{subfigure}
	\hspace{-10mm}
	~
\caption{\revised{Time evolution of $\phi$ under verification test, (a) at initial state, (b) at half-way time, (c) at final time, for 5/7 grid.}}
\label{fig:verification_test_lsf_evolution}
\end{figure}

For the given system of PDEs, there is no known analytical solution. Therefore, we investigate the convergence behavior of the solver by considering a numerical verification test for which we select analytical solutions for $T_l$, $T_s$, $\vns$, $\press$, and $\vint$, and augment the system of equations with the appropriate synthetic forcing terms to produce such solutions. We note that for the verification test, all non-dimensional groups and material parameters are set equal to 1. Convergence behavior is then evaluated for these fields based on time-averaged \Irevised{$L_{\infty}$} norm of the spatial error for each field. To test the convergence of the level set function $\phi$, we select the interfacial velocity field such that it deforms the interface sufficiently, and then switch the sign of the velocity field at $t = t_{\text{final}}/2$, so that $\phi$ should recover its initial state at $t = t_{\text{final}}$  (as is shown in Fig.~\ref{fig:verification_test_lsf_evolution}). The error in $\phi$ is thus evaluated by the difference between $\phi_{\text{initial}}$ and $\phi_{\text{final}}$. 

We consider the domain $\Omega = [-\pi, \pi] \times [-\pi, \pi]$, which is initially split into the subdomains $\liqDomain$, $\solDomain$ by the level-set function $\phi$ such that 
\begin{equation*}
\begin{cases}
\phi = R - \sqrt{x^2 + y^2},\\
\displaystyle R = \frac{\pi}{2}, \\
\liqDomain = (x,y) \in \Omega | \phi < 0, \\
\solDomain = (x,y) \in \Omega | \phi > 0, \\
\Gamma = (x,y) \in \Omega | \phi = 0 ,
\end{cases}
\end{equation*}

We select the following solutions for fluid velocity and pressure
\begin{subequations}
\begin{equation*}
v_{x} = \cos{t} \sin{x} \cos{y},
\end{equation*}
\begin{equation*}
v_{y} = -\cos{t} \cos{x} \sin{y},
\end{equation*}
\begin{equation*}
P = 0,
\end{equation*}
\end{subequations}
and temperature fields 
\begin{subequations}
\begin{equation*}
T_l = \sin{x} \sin{y} \left( x + \cos{x}\cos{y}\cos{t} \right),
\end{equation*}
\begin{equation*}
T_s = \cos{x}\cos{y} \left( \cos{t}\sin{x}\sin{y} - 1 \right),
\end{equation*}
\end{subequations}
which produce an interfacial velocity of
\begin{equation*}
\vint \cdot \normal = \left(V_{\Gamma, x} n_x + V_{\Gamma,y} n_y \right),
\end{equation*}
where $V_{\Gamma, x}$ and $V_{\Gamma, y}$ are given by
\begin{subequations}
\begin{equation*}
V_{\Gamma,x} = \cos{y}\sin{x} - \sin{x} \sin{y} - x \cos{x} \sin{y}, 
\end{equation*}
\begin{equation*}
V_{\Gamma,y} = \cos{x}\sin{y} - x \cos{y} \sin{x}.
\end{equation*}
\end{subequations}

\begin{figure}[H]
\centerline{
	\includegraphics[width=0.65\textwidth]{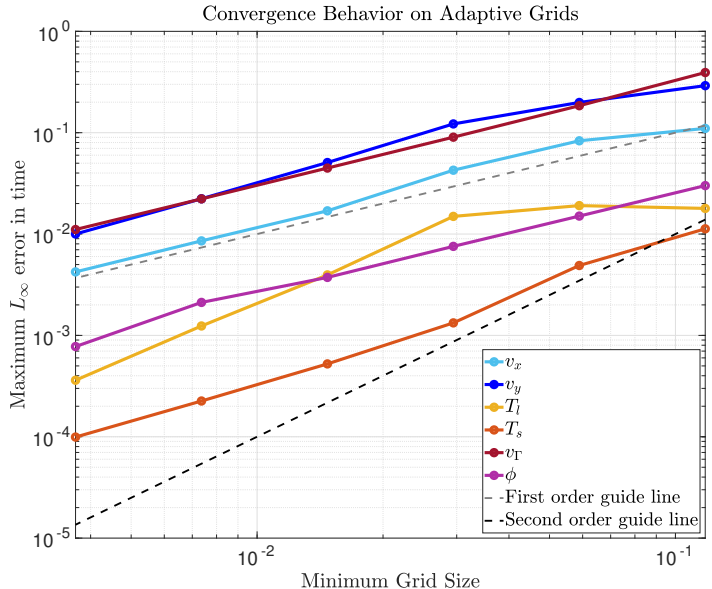} }
	\caption{Convergence results for numerical verification test.}
	\label{fig:verification_test_1}
\end{figure}

Simulations are run for $t_{0} = 0$ to $t_{f} = \pi/3$ at 6 different grid level configurations. A test for grid level 5/7 takes 94 seconds on a Dell 7810 tower with 6 3.60 GHz processes. Convergence results are shown in Fig.~\ref{fig:verification_test_1} and Table \ref{tab:verif1_convergence}.

Convergence rates between first and second order are achieved for $v_x$, $v_y$, $T_l$, and $T_s$, which is reasonable based on the combination of numerical methods utilized. As the computation of $\vint$ depends on the gradient of the scalar fields $T_l$, $T_s$, we expect $\vint \cdot \normal$ and therefore the level-set function $\phi$ defining the interface to converge at approximately first order, which they do.

\begin{table}[H]
{ \footnotesize{
\centerline{
\begin{tabular}{|c|c|c|c|c|c|c|c|}
\hline
$l_{\text{min}}/l_{\text{max}}$& $N$ & $v_{x}$ & $v_{y}$  & $T_l$ & $T_s$ & $\vint \cdot \normal$ & $\phi$  \\
\hline 
 3/5 & 59 & \num{  1.14e-01 } &  \num{  3.59e-01 } &  \num{  1.72e-02 } &  \num{  1.03e-02 } &  \num{  3.92e-01 } &  \num{  2.61e-02 }    \\  
  \hline  
 4/6 & 96 & \num{  9.05e-02 } &  \num{  2.54e-01 } &  \num{  1.90e-02 } &  \num{  4.07e-03 } &  \num{  1.89e-01 } &  \num{  1.05e-02 }    \\  
  \hline  
 5/7 & 125 & \num{  4.55e-02 } &  \num{  1.24e-01 } &  \num{  1.48e-02 } &  \num{  8.83e-04 } &  \num{  8.97e-02 } &  \num{  5.01e-03 }    \\  
  \hline  
 6/8 & 181 & \num{  1.70e-02 } &  \num{  5.17e-02 } &  \num{  3.92e-03 } &  \num{  1.95e-04 } &  \num{  4.48e-02 } &  \num{  2.25e-03 }    \\  
  \hline  
 7/9 & 313 & \num{  8.56e-03 } &  \num{  2.23e-02 } &  \num{  1.23e-03 } &  \num{  5.46e-05 } &  \num{  2.22e-02 } &  \num{  1.19e-03 }    \\  
  \hline  
 8/10 & 572 & \num{  4.23e-03 } &  \num{  1.00e-02 } &  \num{  3.59e-04 } &  \num{  1.31e-05 } &  \num{  1.11e-02 } &  \num{  7.74e-04 }    \\  
  \hline  
  \hline 
\hline
\textbf{Convergence rate} & - & 1.01 & 1.07 & 1.19 & 1.97 & 1.03 & 1.03 \\
\hline 
\end{tabular}}
\caption{Convergence results for verification test. Displayed is maximum $L_{\infty}$ error in time for each field at various grid levels, and corresponding convergence rates. $N$ is the average number of nodes per spatial direction.}
\label{tab:verif1_convergence}
}}
\end{table}

\section{Ice growth on a cooled cylinder in crossflow -- Quantitative benchmark} \label{sec:okada_compare}
\newcommand{\cylDomain}{\Omega_{\text{cyl}}}
\newcommand{\cylTemp}{T_{\text{cyl}}}

Our solver is validated using the configuration of ice growth on a cooled cylinder in a crossflow, as studied experimentally by Okada \etal \cite{okada1978freezing} and Cheng \etal \cite{cheng1981experimental} 
by comparing the time evolution of interface shape with experimental results. Additionally, we compare the local heat transfer behavior at the forward stagnation point with an empirical \revised{correlation} developed by \citea{perkins1964local} \cite{perkins1964local}. The physical system for the problem is shown below in Fig.~\ref{fig:okada_problem_setup}. We consider the problem divided into three subdomains: (i) the liquid water $\liqDomain$, (ii) the solid ice $\solDomain$, and (iii) the cooled cylinder $\cylDomain$ for which $\cylTemp $ is assumed constant. The physical properties of ice and water are chosen to match those of \cite{okada1978freezing} to yield the ratios of physical quantities and non-dimensional numbers as given by Table \ref{tab:okada_properties}. The characteristic temperature difference for this problem is given by $\Delta T = T_{\infty} - T_{\text{cyl}}$, where $T_\infty$ is the freestream fluid temperature. 

\begin{figure} [H]
\centering
\includegraphics[width=0.75\textwidth]{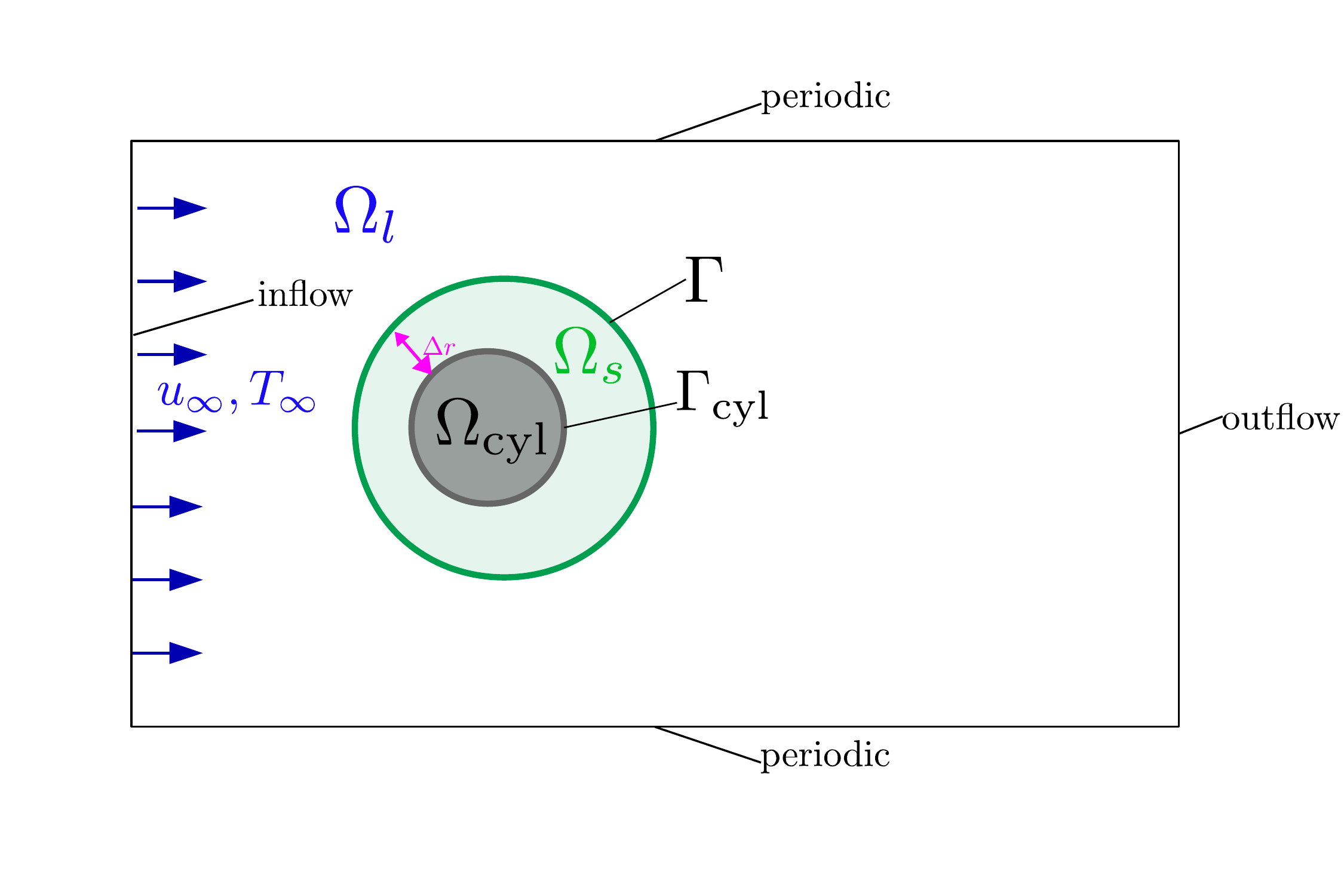}
\caption{Illustration of problem setup for ice growth on cooled cylinder in a crossflow (not drawn to scale)}
\label{fig:okada_problem_setup} 
\end{figure}

\begin{table}[H]
\centering
\begin{tabular}{|c|c|}
\hline 
Prandtl number, $Pr$ & $13$ \\
\hline
Ratio of densities, $\rho_l / \rho_s$ & $1.09$ \\ 
\hline 
Ratio of thermal diffusivities, $\alpha_l / \alpha_s$ & $0.11$ \\ 
\hline 

\begin{tabular}{l} 
Ratio of surface tension and\\
characteristic length, $\sigma / d$ 
\end{tabular} & $\num{1.2e-8}$  \\
\hline
Ratio of thermal conductivities, $k_l / k_s$ & $0.25$ \\ 
\hline  
Freestream fluid temperature, $T_{\infty}$ & $ 2.5 {}^{\circ} C$\\
\hline
Cylinder diameter, $d$ & $35$ mm \\
\hline
\end{tabular}
\caption{Parameters used in the ice growth on cooled cylinder problem, chosen to match conditions from \cite{okada1978freezing}.} 
\label{tab:okada_properties}
\end{table}

For ice shape comparison, the simulations are run for the cases (i) $Re= 201 $, $St = 0.069$ ($T_{\text{cyl}} = {-2.5}^{\circ} C$) and (ii) $Re = 506$, $St = 0.081$ ($T_{\text{cyl}} = {-7.5}^{\circ} C$) in order to match the experimental conditions given by Okada \etal \cite{okada1978freezing}. In the regime we are considering, the momentum and thermal boundary layers $\delta_m$ and $\delta_T$ are such that $\delta_m \sim \delta_T$. Thus, the maximum level of refinement is chosen such that there are a minimum of $10 - 15$ grid cells through the boundary layer thickness for the $Re = 201$ case, and $6 - 12$ for the $Re = 506$ case. Since $\delta_T \sim \delta_m \sim (r_{cyl} + r_{ice})/ \sqrt{Re_d}$, the boundary layer thickness grows as the thickness of the ice increases, and the thinnest boundary layer we encounter is at the beginning of the simulations. In order to provide an initial ice domain for the solver, the radius of the ice layer is initialized as a cylinder with $r_{ice} = 1.10 $ $r_{cyl}$. The initial size is chosen to be as small as possible for the grid resolution used, whilst still resolving the ice layer by a minimum of 6 grid cells in order to ensure that the initial ice domain is well-defined. This corresponds to an initial ice thickness of $ \sim 1 $ mm.

\begin{figure}[H]
	\centering
	\begin{subfigure}[t]{0.75\textwidth}
		\centerline{
		\includegraphics[width=1.3\textwidth]{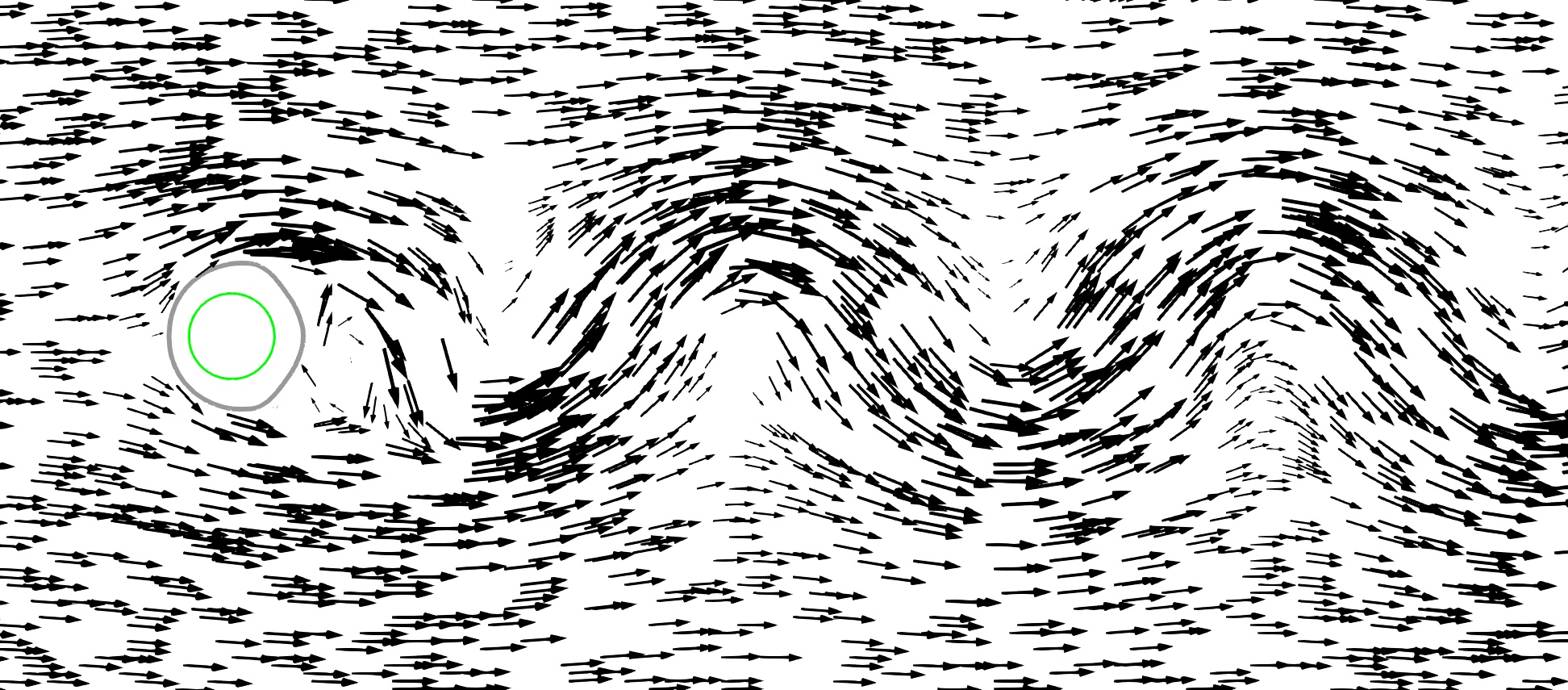}}
		\caption{}
	\end{subfigure}
	\begin{subfigure}[t]{0.75\textwidth}
		\centerline{
		\includegraphics[width=1.3\textwidth]{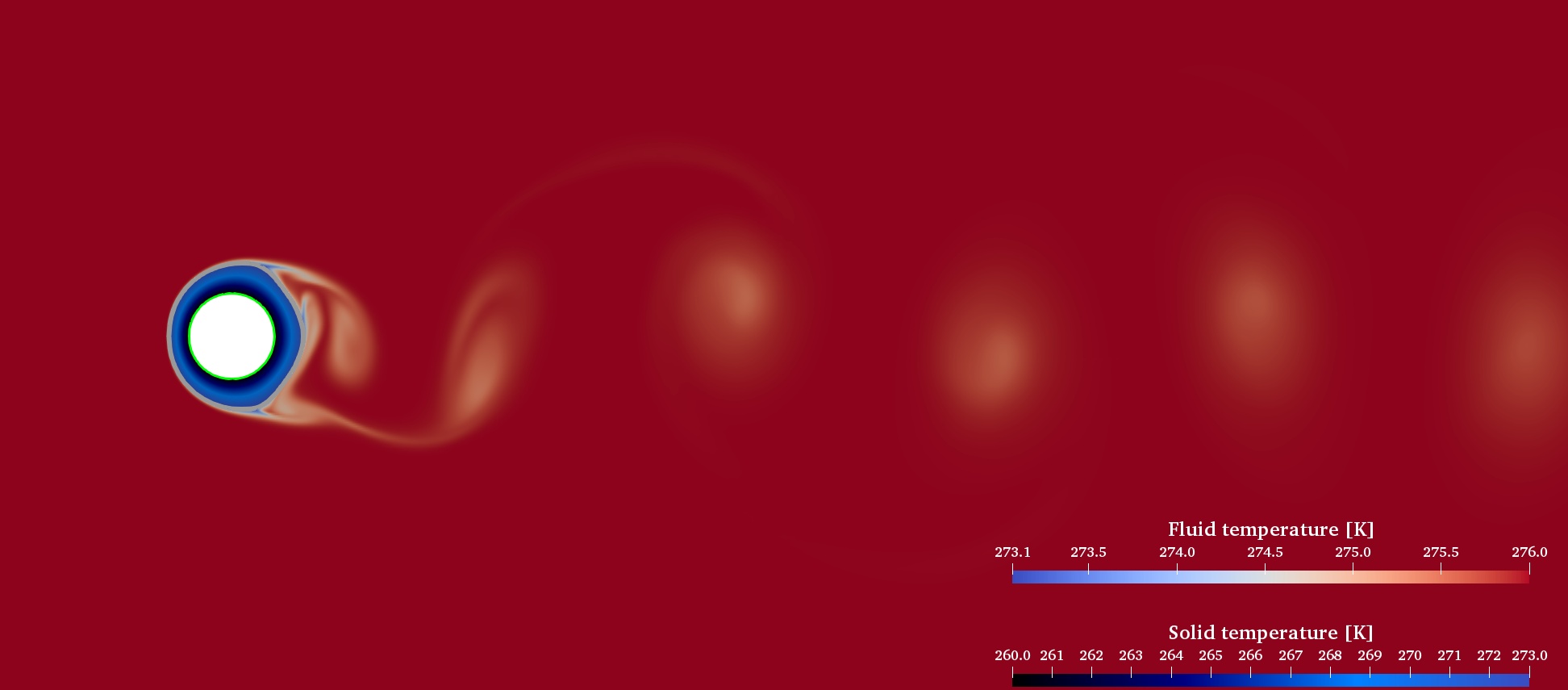}}
		\caption{}
	\end{subfigure}
	\caption{(a) Velocity vectors and (b) temperature fields for the simulation of ice growth on cooled cylinder at $ Re = 506 $ and $St = 0.081$, with the cylinder interface plotted in green and the ice interface plotted in gray.}
	\label{fig:OkadaSimulation}
\end{figure}

Examples of velocity vectors and temperature fields for a snapshot in time are displayed in Fig.~\ref{fig:OkadaSimulation}(a) and Fig.~\ref{fig:OkadaSimulation}(b), respectively. We recapture the von-Karman vortex street in the flow field, and pockets of cooler fluid temperature can be seen in the wake behind the ice.  The ice-water interface evolves to a non-uniform shape, \revised{due to} the complex flow profile and resulting heat transport around the ice. \revised{The details of this shape evolution will be further discussed in Sec. \ref{sec:investigating}. }

\subsection{Ice Shape}
\begin{figure}[H]
	\centering
	\begin{subfigure}[c]{0.47\textwidth}
		\centerline{\includegraphics[height=\textwidth]{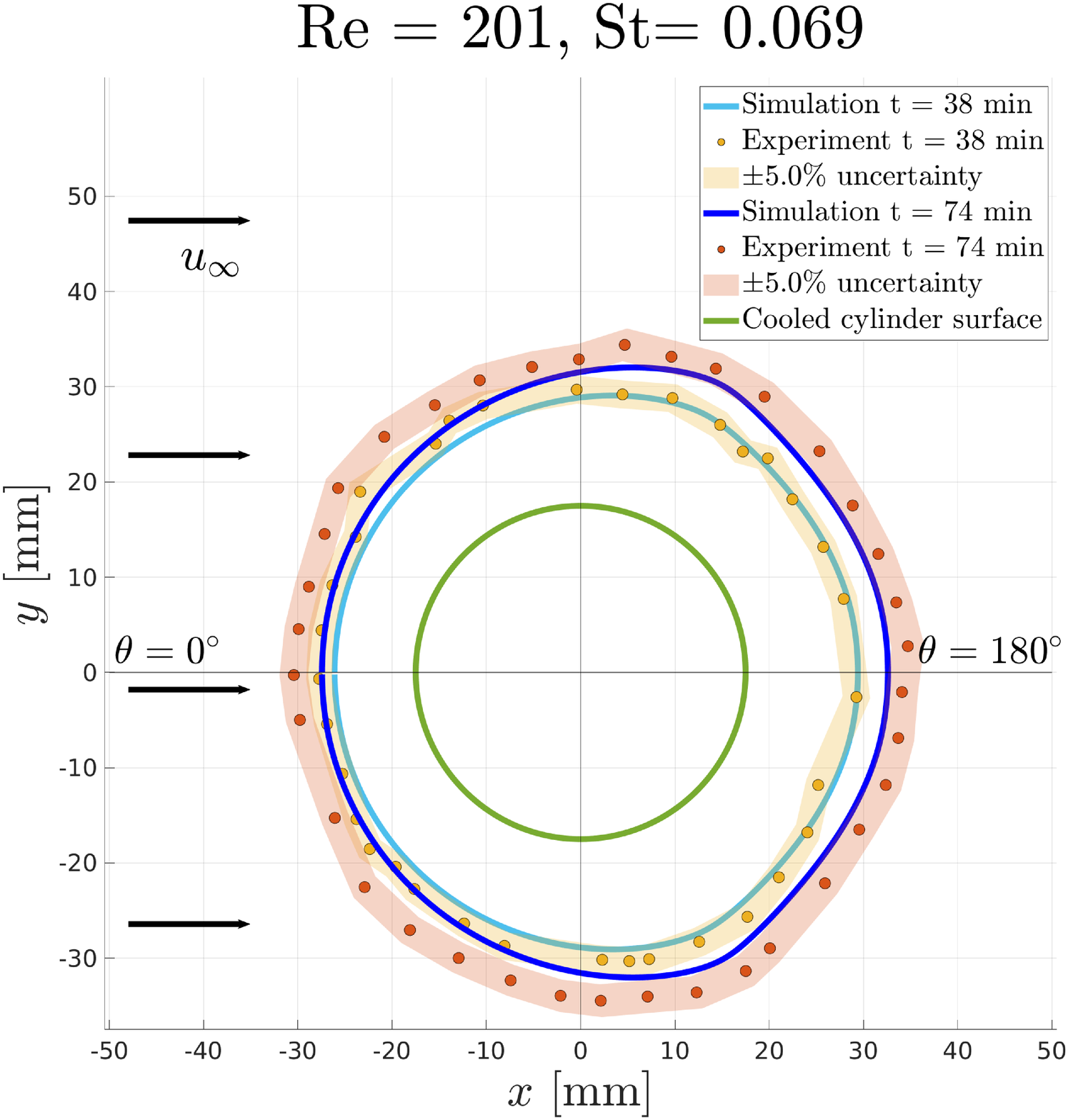}}
		\caption{}
	\end{subfigure}
	\hspace{6mm}
	\begin{subfigure}[c]{0.47\textwidth}
		\centerline{\includegraphics[height=\textwidth]{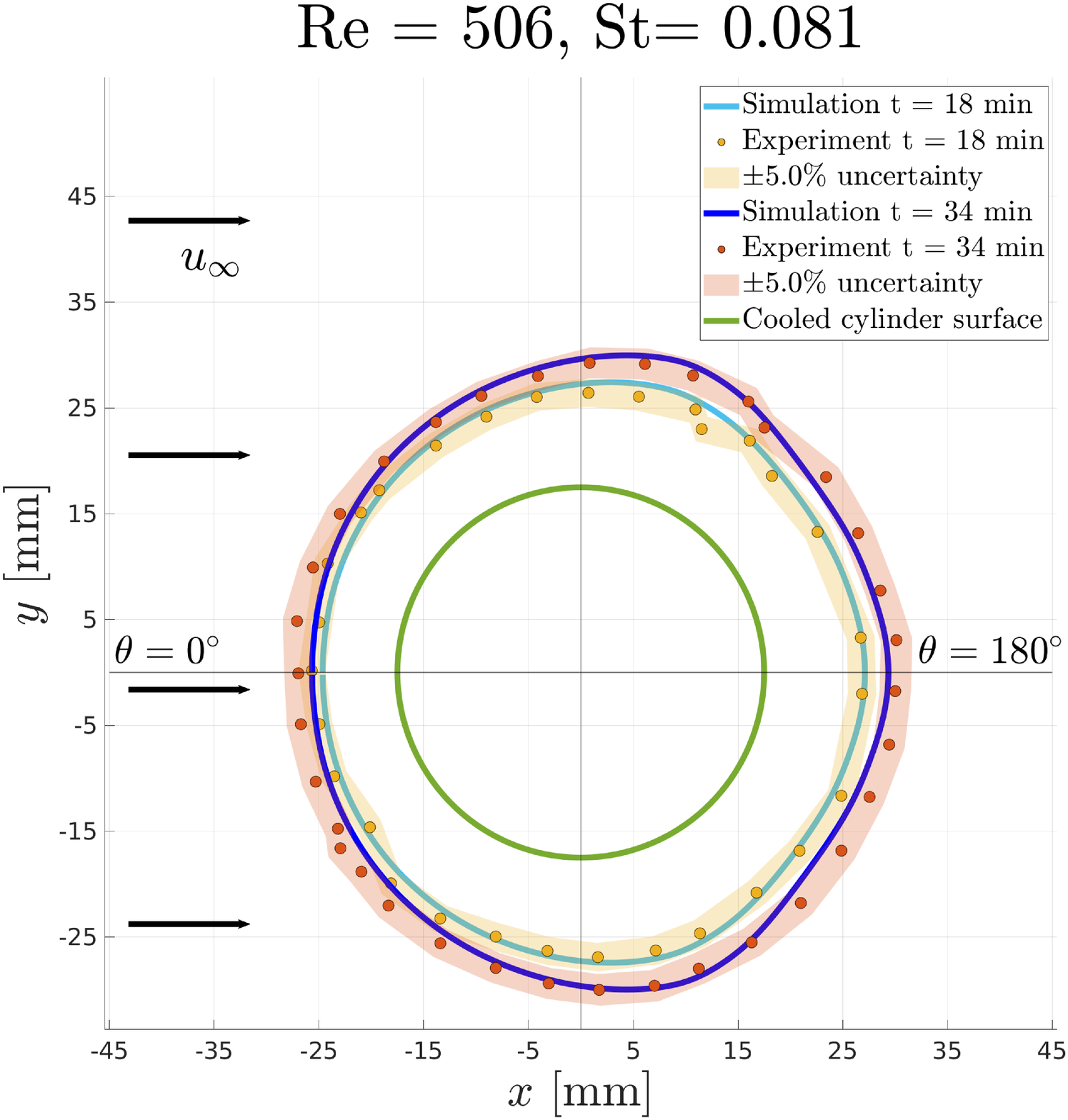}}
		\caption{}
	\end{subfigure}
	\begin{subfigure}[c]{0.47\textwidth}
		\centerline{\includegraphics[height=\textwidth]{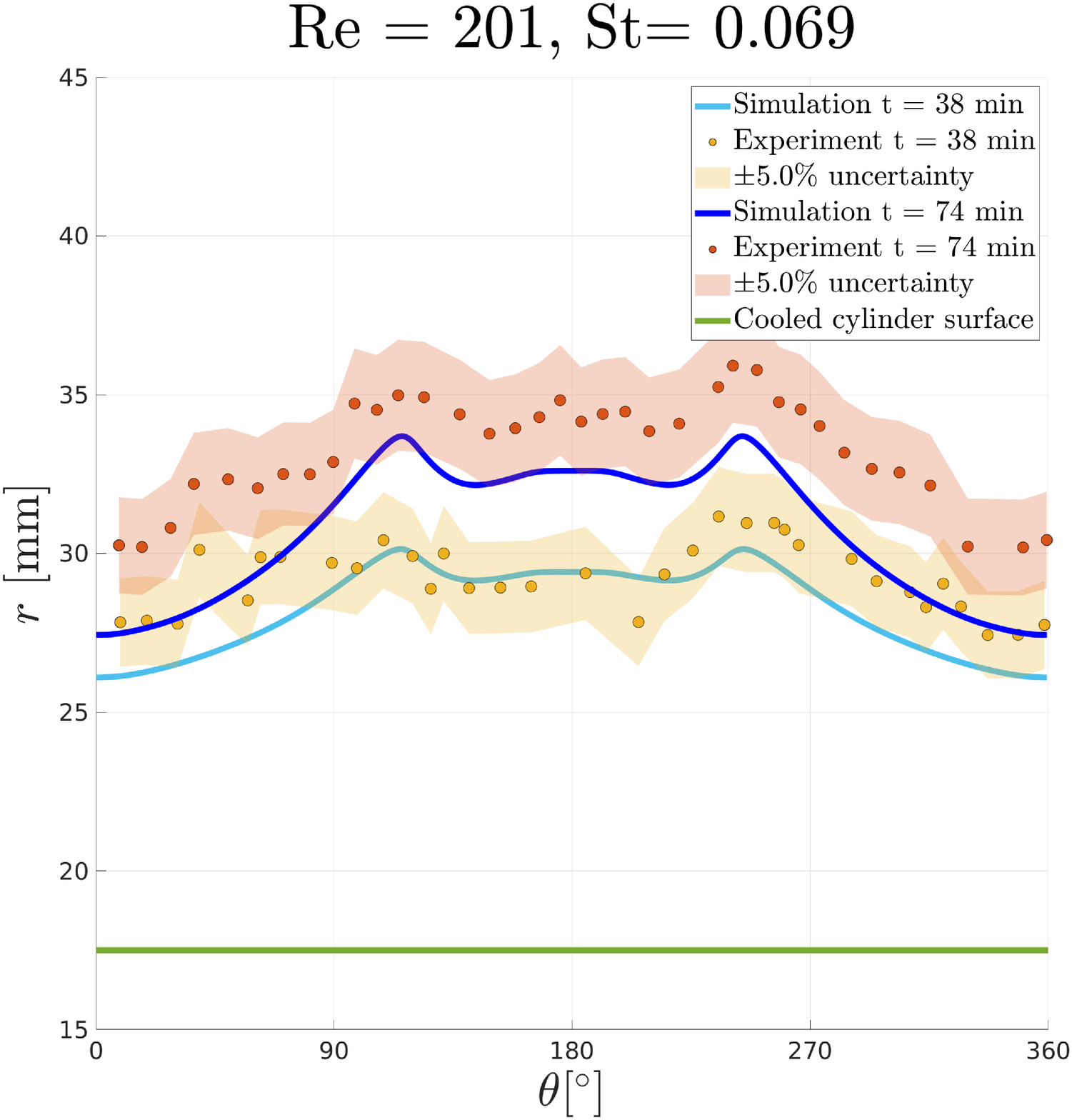}}
		\caption{}
	\end{subfigure}
	\hspace{6mm}
	\begin{subfigure}[c]{0.47\textwidth}
		\centerline{\includegraphics[height=\textwidth]{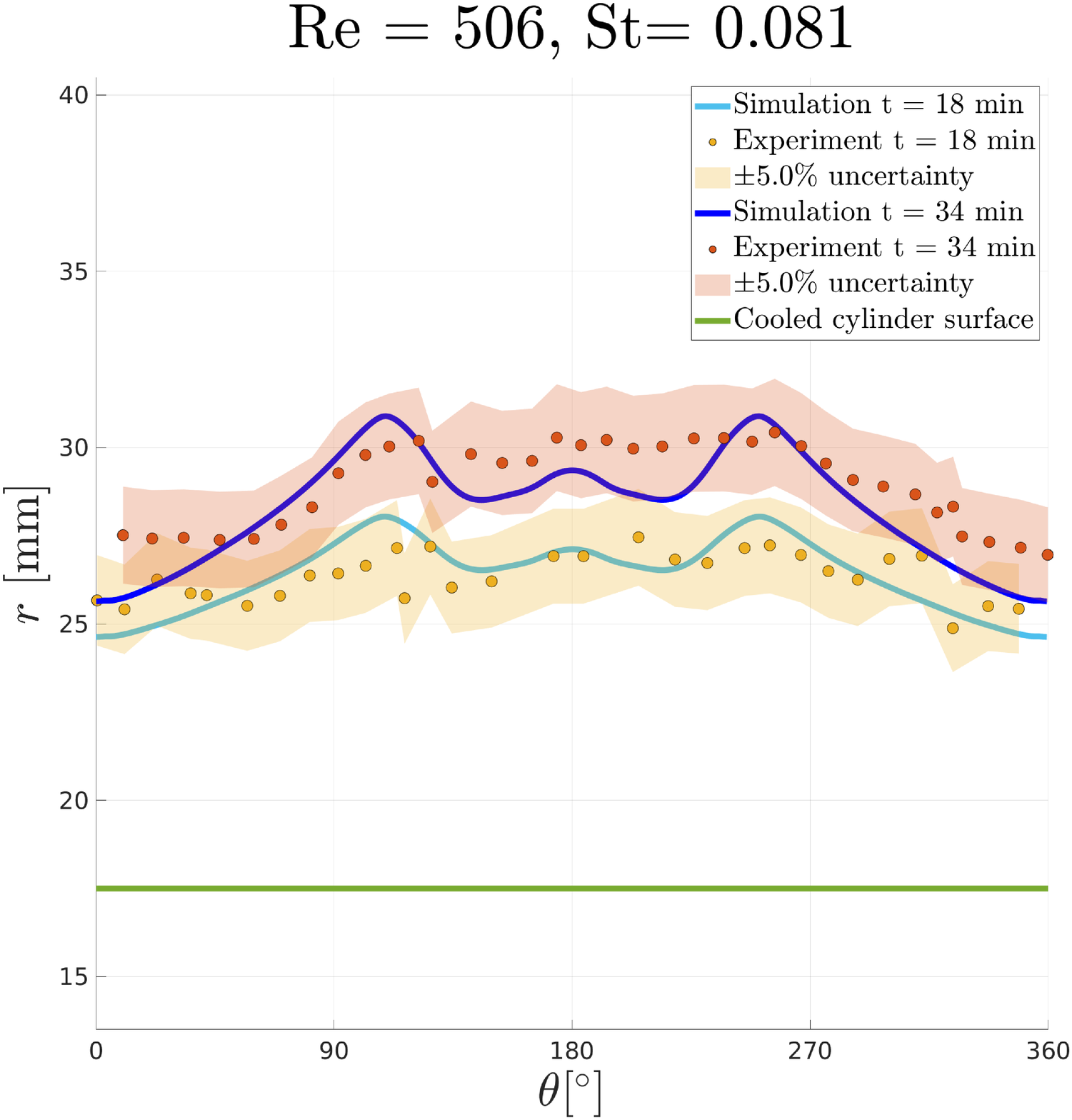}}
		\caption{}
	\end{subfigure}
	\caption{Ice interface contour comparison between simulations and experiments of \cite{okada1978freezing} in (a) - (b) $x-y$ coordinates and (c)-(d) in $r-\theta$ coordinates. The green line represents the cooled cylinder surface, the blue lines represent simulation results, the orange and red dots represent experimental results, and the orange and red shaded areas represent the estimated experimental uncertainties of $\pm 5 \%$. The coloring corresponds to different times, as denoted in the figure legends. }
	\label{fig:OkadaResults}
\end{figure}

The ice interface at different times is plotted against the results of \cite{okada1978freezing} in Figs.~\ref{fig:OkadaResults}(a)-(d). Note that in the plotted results, the angle over the cylinder $\theta$  is taken to be $0^{\circ}$ at the front of the cylinder  and $180 ^{\circ}$ at the back. The general shape that evolves is consistent with the characteristic shape observed in \cite{okada1978freezing} -- with a relatively circular front shape, a sharp transition point around $\theta \approx 120 ^\circ $, and a more blunted back shape. As can be seen in the contour plots shown in Figs.~\ref{fig:OkadaResults}(a)-(d), we find good qualitative agreement of overall interface shape between simulation and experimental results for both cases at varying times. 
When comparing interface evolution quantitatively, we note that in \cite{okada1978freezing}, the interface between the ice and the water was measured using a vernier caliper. An estimate of the experimental uncertainty of such measurements could be around 0.5 - 1 mm, corresponding to an error of about 3-5 \%. Additionally, the initial conditions and boundary conditions (\emph{i.e.} constant temperature in the cooled cylinder) are more challenging to control experimentally than in numerical simulations. For example, \cite{okada1978freezing} notes a $T_{cyl}$ fluctuation of up to $\approx \pm 2.5 ^\circ C$. Within these experimental uncertainties, Figs.~\ref{fig:OkadaResults}(a)-(d) demonstrate that the numerical approach reproduces quantitatively the time evolution of the ice shape. Because the ice shape is governed by the flow field around the interface, this result also validates accurate calculation of the flow dynamics.

\subsection{Local heat transfer at the forward stagnation point}

\begin{figure}[H]
\centerline{
\includegraphics[height=0.5\textwidth]{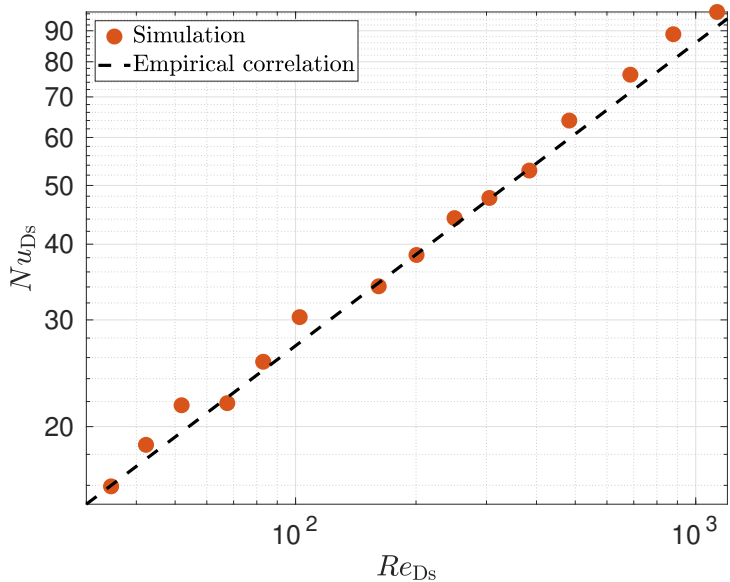}}
\caption{Simulation results for the Nusselt number at the forwards stagnation point plotted with the empirical correlation given by equation \eqref{eq:NuDs_corr} \cite{perkins1964local}.}
\label{fig:stagnationNu}
\end{figure}

In addition to ice shape evolution, we also compare local heat transfer behavior produced by the solver with experimental results. Since the ice shape at the front is cylindrical, we can compare the local heat transfer at the forward stagnation point with local heat transfer for a circular cylinder under forced convection, as is done in Cheng \etal \cite{cheng1981experimental}. \citea{perkins1964local} \cite{perkins1964local} provide the empirical correlation for the Nusselt number at the stagnation point as given by
\begin{equation}
Nu_{D_s} = 1.08 \text{ } Re_{D_s}^{0.5} \text{ } Pr^{0.36}, 
\label{eq:NuDs_corr}
\end{equation}
where $Re_{D_s}$ is the Reynolds number at the stagnation point given by $Re_{D_s} = \rho_l D_{s} u_\infty / \mu_l$.
Results are obtained from the simulations of ice growth on a cooled cylinder in crossflow which are described later in Sec.~\ref{sec:investigating} and compared with the empirical correlation described in equation \eqref{eq:NuDs_corr}. We calculate the local Nusselt number at the stagnation point as $Nu_{D_s} = h_s D_s/k_l$, where $D_s$ is the diameter of the ice at the forward stagnation point, and $h_s$ is the local heat transfer coefficient at the forward stagnation point defined as $h_s = (k_l \partial T_l / \partial n)|_{\theta = 0 ^\circ}/ (T_\Gamma - T_\infty)$. Simulation results are plotted with the empirical correlation in Fig.~\ref{fig:stagnationNu}, where the $Nu_{D_s}$ and $Re_{D_s}$ obtained from simulations are those taken at the final time ($t = 40 $ minutes) of each simulation.  As demonstrated in Fig.~\ref{fig:stagnationNu}, we find very good agreement between the simulation results and the empirical correlation for the heat transfer. 

\section{Shape dynamics and heat transfer for varying $Re$ and $St$ numbers} \label{sec:investigating}
Now that the numerical method has been quantitatively validated, we discuss in more detail the shape dynamics and heat transfer results for a large range of $Re$ and $St$ numbers to investigate their effects on the resulting interface morphologies and heat transfer, and how these might be related. We use the same problem configuration as described in Sec.~\ref{sec:okada_compare}, and perfom simulations for Reynolds numbers $Re = $ $[20, 40, 100, 200, 500]$ and Stefan numbers $St = [0.07, 0.13, 0.26]$. The boundary layer is resolved by $15-20$ grid points for $Re = 20, 40, 100$ and $10-12$ grid points for $Re = 200, 500$. It is also worth noting that for $Re = $ $100$, $200$, $500$ cases in which vortex shedding emerges, we introduce a slight perturbation in the initial condition of the velocity field as is done by  Laroussi \etal \cite{laroussi2014triggering} in order to reduce the amount of simulation time required to arrive at the steady state. This takes the form $u_{0} = U_{\infty}(1 + 0.25 \sin{(2\pi \beta y / H)}) $, where $\beta$ is $0.25$, and $H$ is the height of the computational domain. In these cases, the interfacial velocity is constrained to zero for a prescribed startup time in order to allow the flow to develop, and we define the start time as the time when the interface is first allowed to move.

\subsection{Ice shape}

\begin{figure}[H]
\centering
	\begin{subfigure}[c]{0.85\textwidth}
		\centerline{\includegraphics[width=1.2\textwidth]{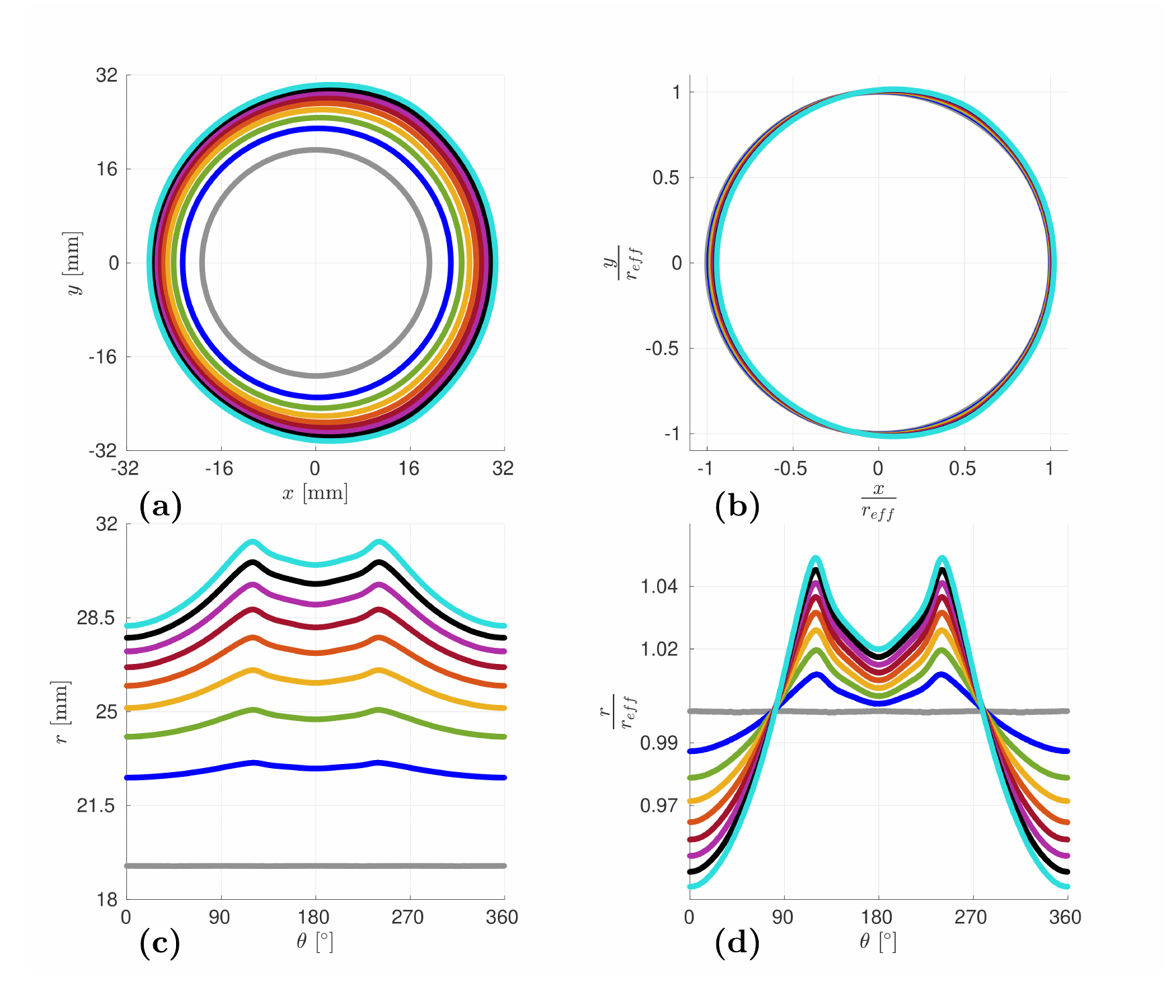} } 
	\end{subfigure}
	\hspace{2mm}
	\begin{subfigure}[c]{0.10\textwidth}
		\centerline{\includegraphics[width=1.2\textwidth]{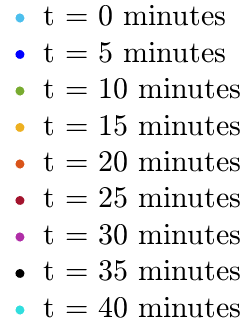}} 
	\end{subfigure}
	~
\caption{Example of ice evolution for $Re = 100$, $St = 0.07$. The interface position at each time is plotted as (a) contour data in $xy$ coordinates, (b) shape factor in $xy$ coordinates, (c) contour data in polar coordinates, (d) shape factor in polar coordinates.}
\label{fig:IceEvolutionExample}
\end{figure}

We consider the time evolution of the ice-water interface, and in particular the role that $Re$ and $St$ play in the emerging shape. We define a \emph{shape factor} of the ice as the ice contour data normalized by the effective radius at the given time, where the effective radius is defined as $r_{ \rm{eff}} = \sqrt{A_{\rm{ice}}/\pi }$,  and $A_{\rm{ice}}$ is the cross-sectional area of the ice and cylinder. Fig.~\ref{fig:IceEvolutionExample} demonstrates such an evolution in time with dimensional contour plots on the left hand side, and shape factor plots on the right hand side. 

\begin{figure}[H]
\centering
	\begin{subfigure}[c]{0.85\textwidth}
		\centerline{\includegraphics[width=0.75\textheight]{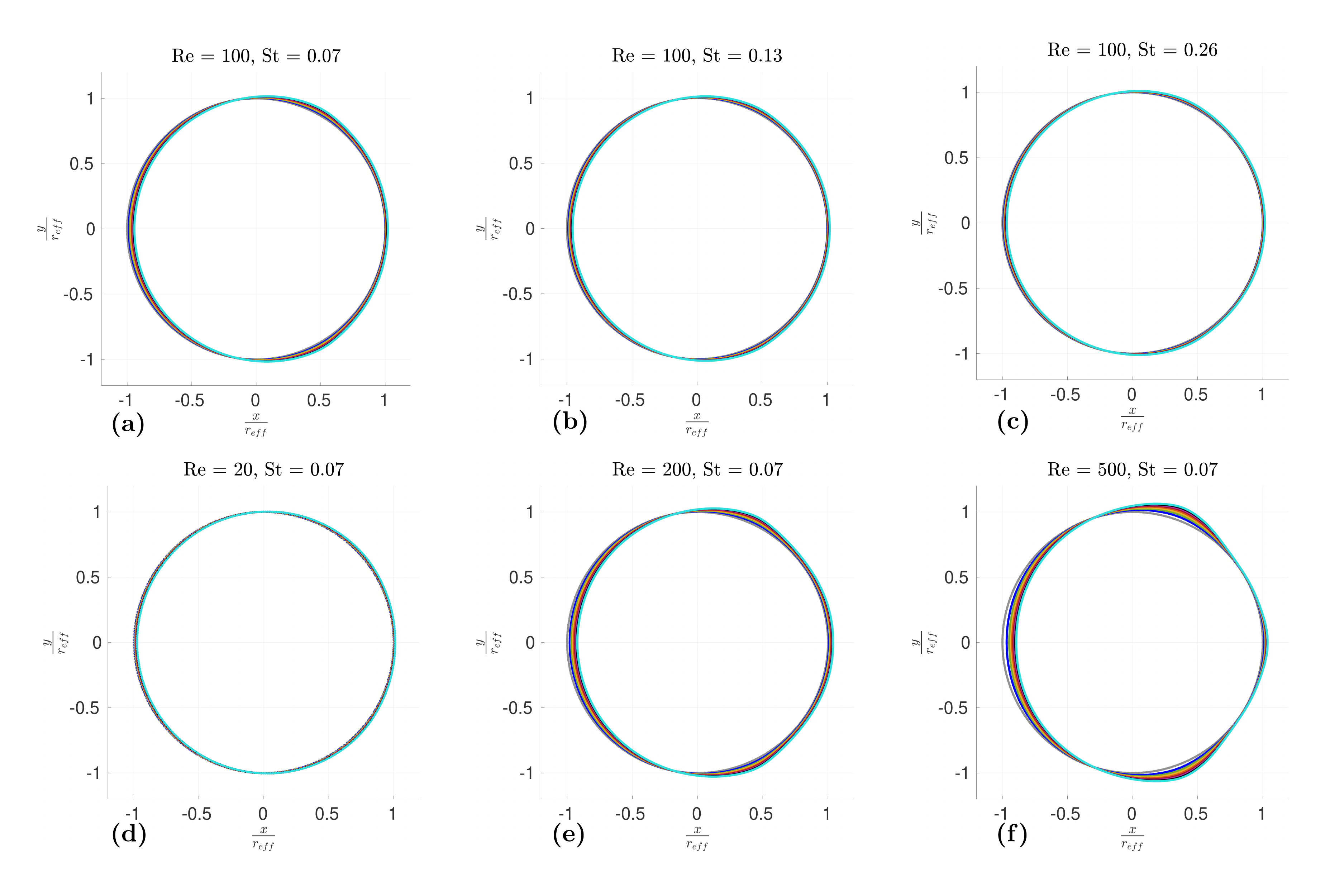} } 
	\end{subfigure}
	\hspace{6mm}
	\vspace{3mm}
	\begin{subfigure}[c]{0.10\textwidth}
		\centerline{\includegraphics[width=0.085\textheight]{time_legend.png}} 
	\end{subfigure}
	~
\caption{Time evolution of the ice shape factor for varying $Re$ and $St$, with $x / r_{\text{eff}}$ on the $x$-axis and $y / r_{\text{eff}}$ on the $y$-axis. (a)-(c) Influence of increasing $St$ with constant $Re$. (d)-(f) Influence of increasing $Re$ with constant $St$.}
\label{fig:VarReStContoursXY}
\end{figure}

\begin{figure}[H]
\centering
	\begin{subfigure}[c]{0.85\textwidth}
		\centerline{\includegraphics[width=0.75\textheight]{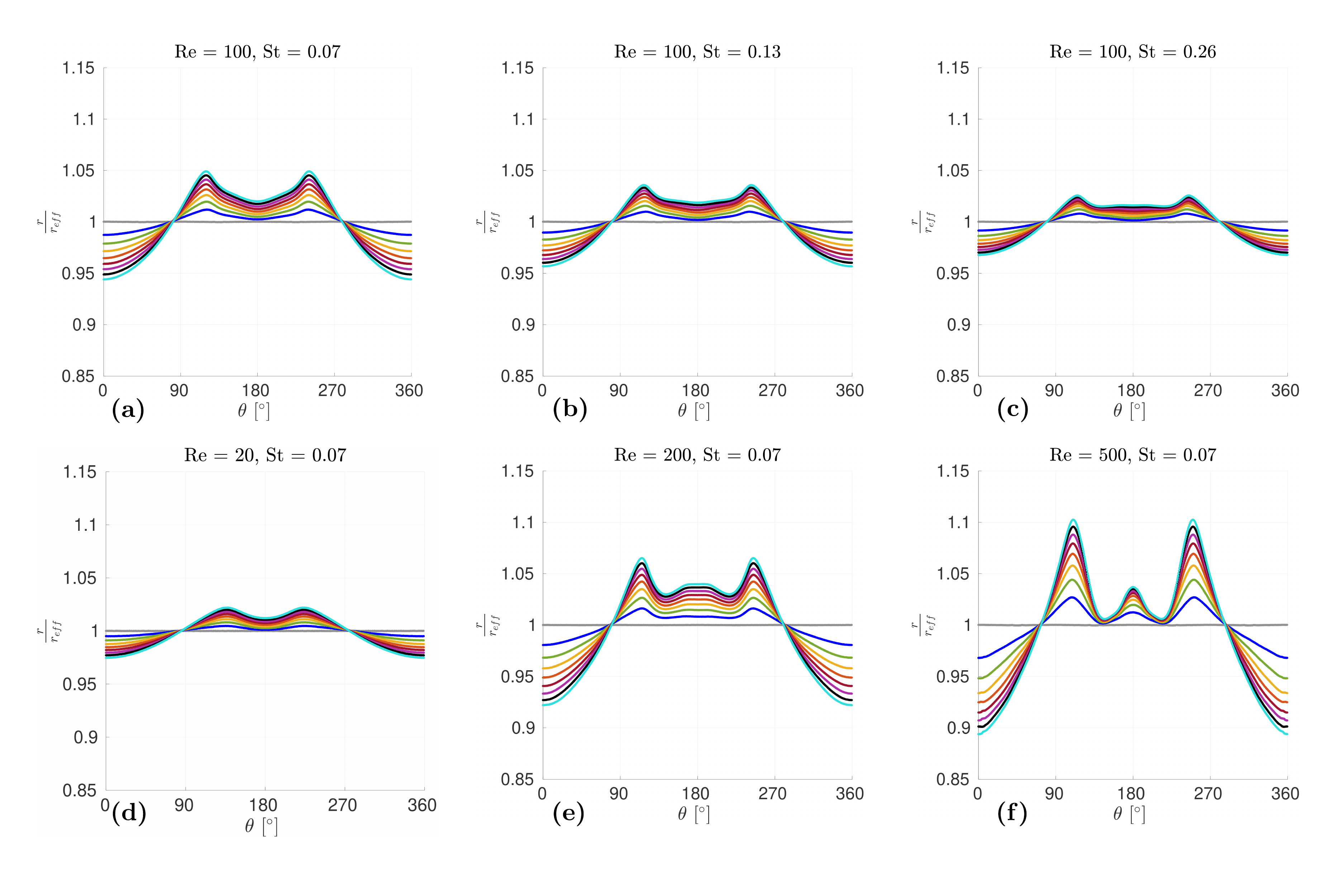} } 
	\end{subfigure}
	\hspace{6mm}
	\vspace{3mm}
	\begin{subfigure}[c]{0.10\textwidth}
		\centerline{\includegraphics[width=0.085\textheight]{time_legend.png}} 
	\end{subfigure}
	~
\caption{Time evolution of the ice shape factor for varying $Re$ and $St$, with $\theta$ on the $x$-axis and $r / r_{\text{eff}}$ on the $y$-axis. (a)-(c) Influence of increasing $St$ with constant $Re$. (d)-(f) Influence of increasing $Re$ with constant $St$.}
\label{fig:VarReStContoursRTheta}
\end{figure}

The influence of $Re$ and $St$ on the evolution of the ice interface and the shape factor is reported in Figs.~\ref{fig:VarReStContoursXY} - \ref{fig:VarReStContoursRTheta}.

\begin{figure}[H]
\centering
	\begin{subfigure}[c]{0.33\textwidth}
		\centerline{\includegraphics[height=0.2\textheight]{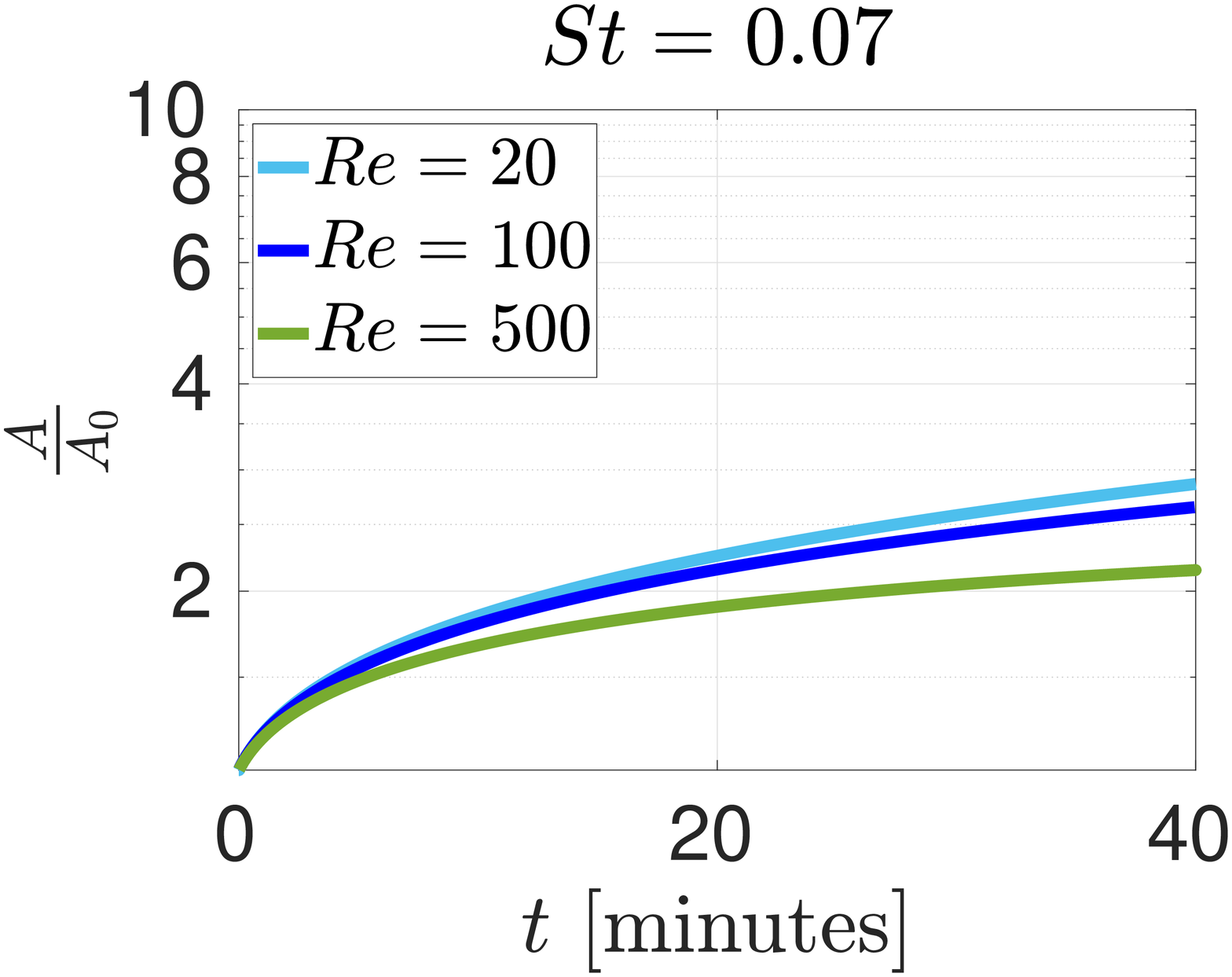}}
	\end{subfigure}
	\hspace{-3mm}
		\begin{subfigure}[c]{0.33\textwidth}
		\centerline{\includegraphics[height=0.2\textheight]{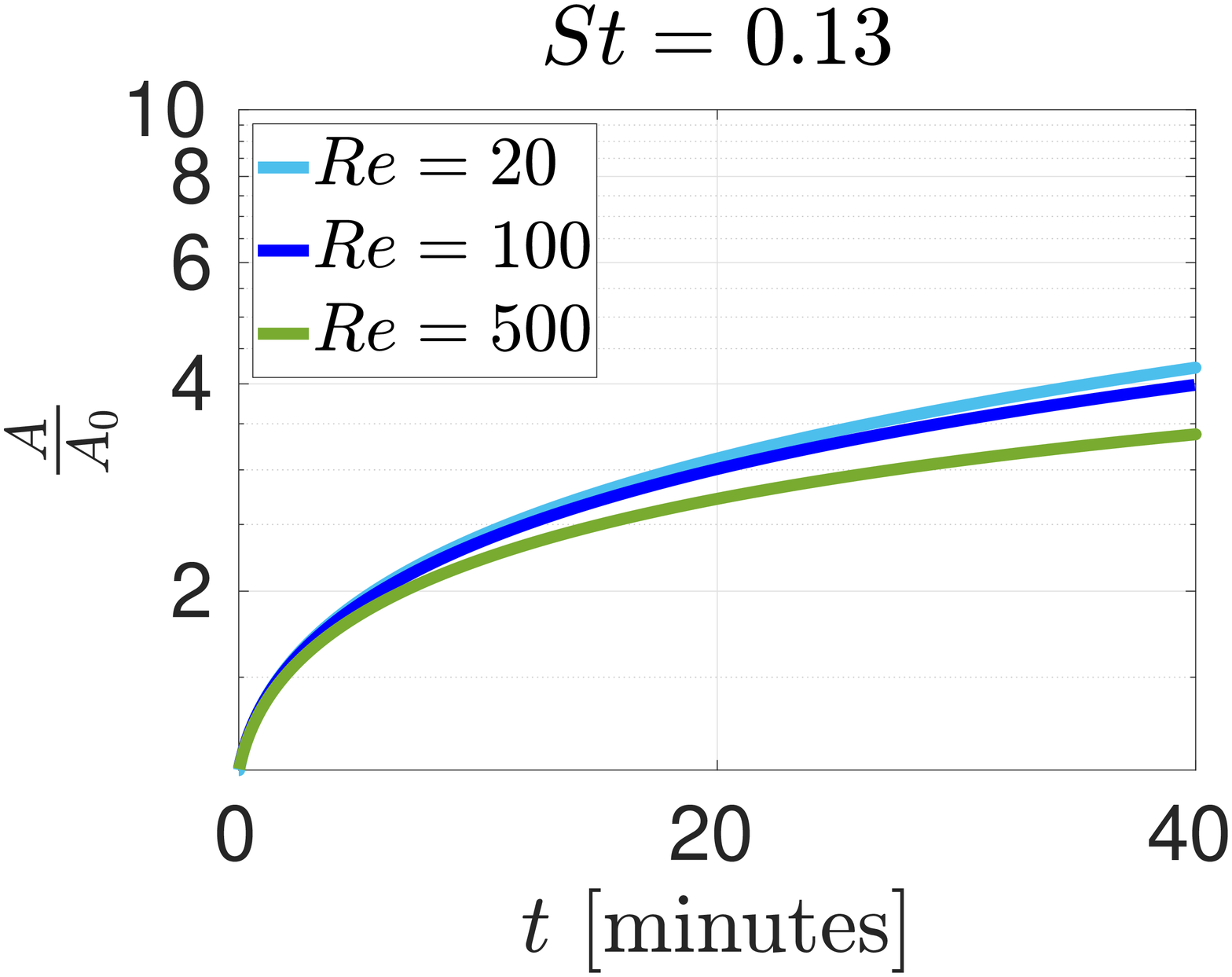}}
	\end{subfigure}
	\hspace{-3mm}
		\begin{subfigure}[c]{0.33\textwidth}
		\centerline{\includegraphics[height=0.2\textheight]{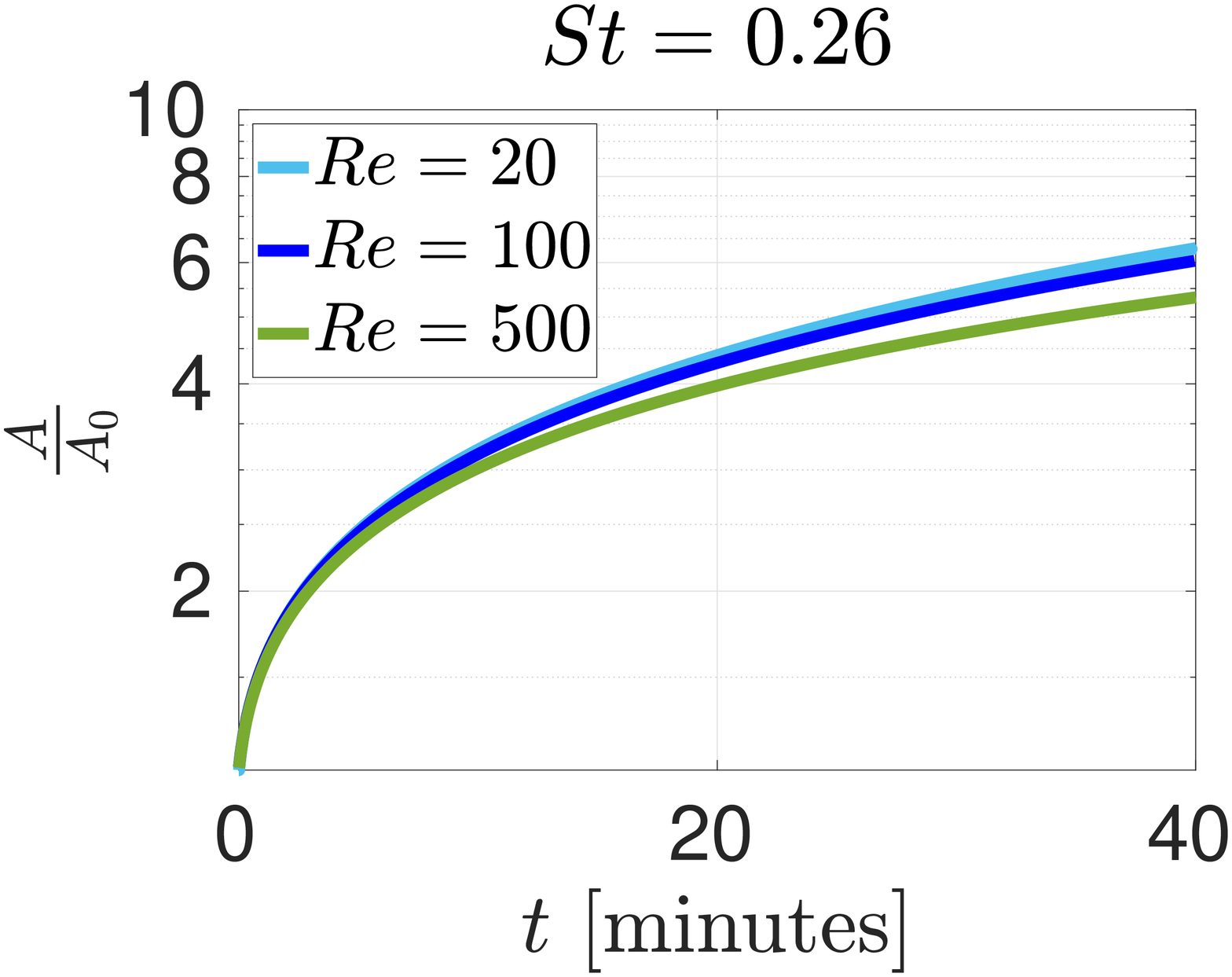}}
	\end{subfigure}
	\caption{Cross-sectional area evolution of ice over time for varying $Re$ and $St$, where $A_0$ is the initial area.
	\label{fig:areaevo}}
\end{figure}

The balance of diffusive and convective transport of heat is seen to play a significant role in the evolution of the ice shape factor and control the rear symmetry of the ice shape, as well as influence the evolution of ice area, as depicted in Fig.~\ref{fig:areaevo}.  Though not displayed above, the cases of no fluid flow ($Re = 0$) maintain a purely cylindrical shape at all times. In general, a lower value of the ratio $St/Re$ leads to a higher shape factor -- an ice shape with regions of greater nonuniform curvature, and vice versa. This is likely related to the relative strength of recirculation effects in the wake of the object compared with the Stefan number driven diffusion, as well as the size and role of the boundary layer, which we dicsuss in more detail in the following section.

\subsection{Local heat transfer around the ice and its role in interface morphology}
\newcommand{\solidflux}{k_s \partial T_s / \partial n}
\newcommand{\liquidflux}{k_l \partial T_l / \partial n}

The evolution of the ice shape has a unique relationship with heat transfer that can be difficult to predict; this is because the heat transfer along the surface develops to be highly dependent on the shape of the interface at any given time \cite{epstein1983complex, hao2002heat, wang2021how}.  Accurate numerical simulation provides an advantage in studying the coupling between ice growth, heat transfer, and fluid flow, as heat transfer behavior at the interface is readily available from simulation data. Taking a closer look at the heat transfer behavior at the ice-water interface can give us more insight into the roles that advection and diffusion play in determining the emerging interface morphology, especially in regimes where these effects are of comparable influence. We recall that the interfacial velocity is governed by a jump in temperature flux across the interface, \emph{i.e.} $V_{\Gamma} \sim (\solidflux - \liquidflux)|_\Gamma$. In the case we consider, it is reasonable to assume that the magnitude of the solid heat flux $\solidflux$ is governed mainly by the diffusion of heat in the ice due to the cooled cylinder, captured by the $St$ number. Larger $St$ number implies larger temperature differences across the ice, and therefore larger solid temperature gradients. Meanwhile, the magnitude of the fluid heat flux $\liquidflux$ is governed mainly by the advection of heat due to flow transport around the ice's surface. This flow transport is characterized by the $Re$ number, which will govern flow transport properties such as boundary layer thickness and recirculation effects that ultimately determine the fluid temperature profile and thus fluid temperature gradient near the interface. It is also important to note not only the average effects of diffusion in the solid and advection in the fluid, but the local effects that vary as a function of location along the interface, which may yield helpful explanations in understanding the shape factor profile of the ice-water interface with varying $\theta$ around the interface. For example, we report in Figs.~\ref{fig:VarReStContoursXY} - \ref{fig:VarReStContoursRTheta} the variation in shape factor for different trends in $Re$ and $St$. In general, increasing $St$ tends to result in a more uniformly cylindrical shape of the ice, while increasing $Re$ results in more locations of sharp change in shape along the interface. Additionally, for an increasing $Re$, ice shape factor minima and maxima varied in both size and azimuthal location along the interface. These trends can be better understood by examining the heat fluxes for both solid and fluid along the interface, and how they vary with $Re$ and $St$.

\begin{figure}[H]
	\centerline{\includegraphics[height=0.35\textwidth]{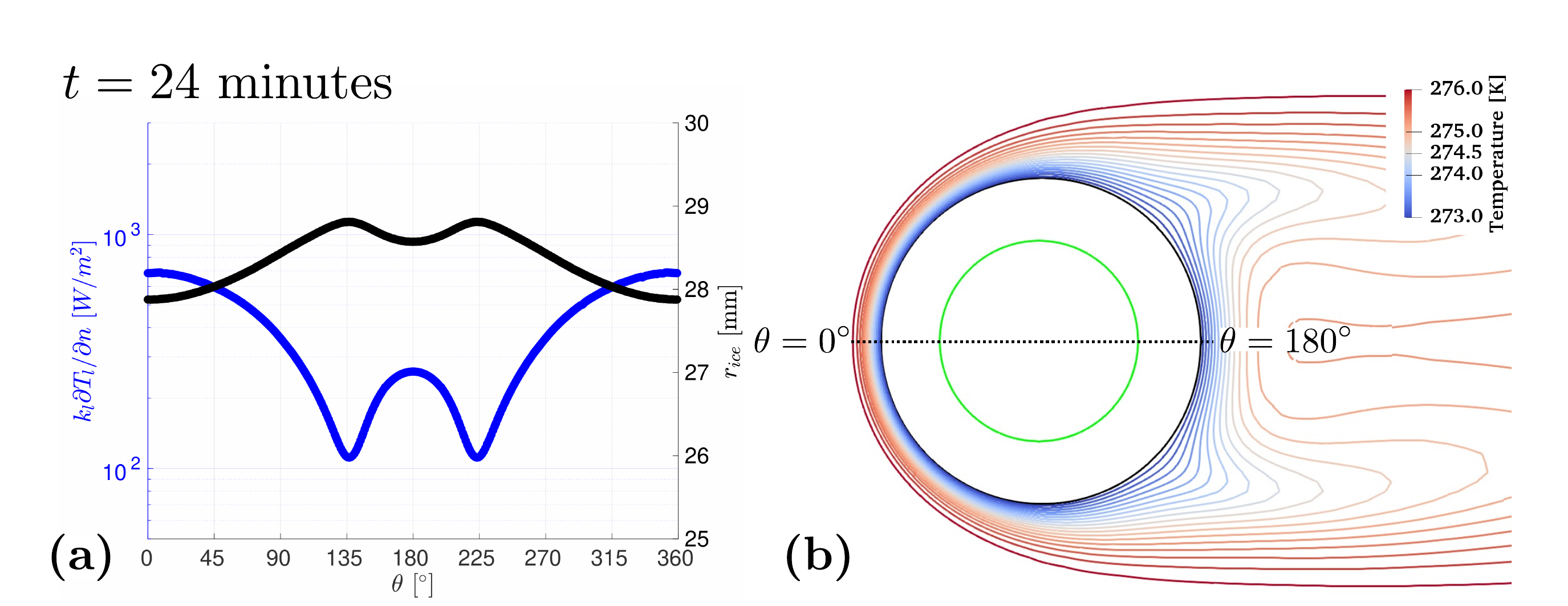} }
	\caption{For $Re = 20$, $St = 0.07$, (a) local fluid heat flux at the interface and ice radius and (b) fluid temperature contours at  $t = 24$ minutes. In this case, the fluid heat flux distribution is symmetrical in $\theta$ as correlated with the geometry of the ice shape and maintains the same shape in time -- there is no notable time variation in the shape of the distribution within the timescales considered.}
	\label{fig:heatfluxesRe20dT10}
\end{figure}

\begin{figure}[H]
	\centerline{\includegraphics[height=0.7\textwidth]{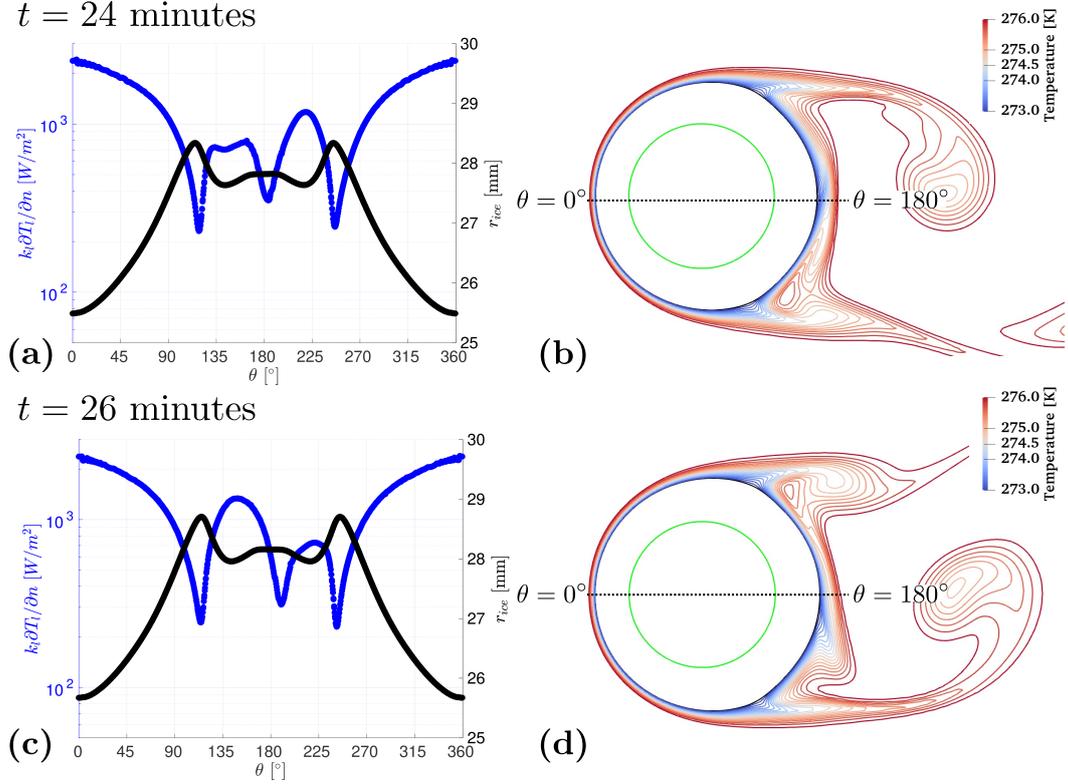} }
	\caption{For $Re = 200$, $St = 0.07$, local fluid heat at the interface and ice radius at (a) $t = 24$ minutes and (c) $t = 26$ minutes, and fluid temperature contours at (b) $t = 24$ minutes and (d) $t = 26$ minutes. In this case the fluid heat flux distribution is asymmetrical in $\theta$ as correlated with pockets of warmer fluid recirculated by vortices, and this distribution varies in time depending on the location of the vortices. }
	\label{fig:heatfluxesRe200dT10}
\end{figure}

First, we examine the effect of $Re$ by considering the magnitude of the local fluid heat flux $\liquidflux$, ice radius and temperature contours in the fluid around the interface for two different Reynolds numbers ($Re = 20$ and $Re = 200$) and the same Stefan number ($St = 0.07$), as illustrated in Figs.~\ref{fig:heatfluxesRe20dT10}-\ref{fig:heatfluxesRe200dT10}. The role that $Re$ plays is best explained by two factors -- the effect of the boundary layer (both thickness and separation point), and the effect of recirculation zones in the wake of the cylinder.

The role of the boundary layer separation can be understood by looking at the fluid heat flux behavior around the interface, and how it correlates with the boundary layer separation and the evolution of the interface shape. In Figs. \ref{fig:heatfluxesRe20dT10}-\ref{fig:heatfluxesRe200dT10}, we see that for both stable flows (\emph{i.e.} no periodic vortex shedding) shown in Fig.~\ref{fig:heatfluxesRe20dT10} and unstable flows shown in Fig.~\ref{fig:heatfluxesRe200dT10}, one can note pockets of cool fluid around the interface at  $\theta \approx 135 ^\circ, 225 ^\circ$ for $Re = 20$ and $\theta \approx 115 ^\circ, 245 ^\circ$ for $Re = 200$ on the temperature contour plots, which correspond to local minima in heat transfer rate and local maxima in the ice radius. These points correspond to the locations of boundary layer separation, where the flow detaches from the ice surface and thus leaves a quiescent region of flow directly behind the separation point, resulting in regions of cool fluid. Because the fluid temperature is already at or close to the interface temperature, at these locations there is minimal heat transfer and the fluid solidifies more quickly, thus creating a larger ice radius. At these locations, a self-reinforcing behavior begins to emerge between boundary layer separation and fluid heat flux. The boundary layer separation causes a local minima in fluid heat flux and corresponding sharp change in interface shape, which in turn reinforces boundary layer separation at the location. This relationship between the heat flux and boundary layer separation was observed experimentally by Cheng \etal  \cite{cheng1981experimental} for freezing/melting. \revised{Huang \etal \cite{mac2015shape} experimentally studied a body dissolving in crossflow (a concentration-driven problem) and  noted an analogous phenomena, providing similar reasoning for role of the concentration gradient and boundary layer separation point in the emerging shape.} 

Additionally, we note that increasing $Re$ leads to a general increase in fluid heat flux at the interface. This effect is also likely related to the boundary layer. Recall that in this regime, the thermal boundary layer $\delta_T$ is of the same order as the momentum boundary layer $\delta_m$ ($\delta_T \sim \delta_m$). Because $\delta_m \sim 1/\sqrt{Re}$, increasing $Re$ results in a thinner $\delta_T$, thus producing larger fluid temperature gradients near the interface, as  the fluid temperature must transition from $T_\Gamma$ at the interface to the free-stream value $T_\infty$ in a much narrower region. 

Meanwhile, the role of recirculation effects can be examined by noting both similarities and differences between the $Re=20$ and $Re=200$ cases. For both cases, warmer pockets of fluid are transported via recirculation effects to the region spanned by $\theta \approx 140 ^\circ - 220 ^\circ $, resulting in local maxima in heat transfer. The ice solidifies much more slowly in this region due to the warmer fluid temperatures, giving way to local minima in ice radius. However, these recirculation effects become more pronounced in cases where vortex shedding begins to arise, as can be noted by some interesting differences between the $Re = 20$ case compared with the $Re = 200$ case. For example, when the flow in the wake of the cylinder is stable in the $Re = 20$ case (\emph{i.e.} no periodic shedding of vortices), the local heat transfer is correlated with the geometry of the ice, but keeps the same shape as a function of time (Fig.~\ref{fig:heatfluxesRe20dT10}). In contrast, for the $Re = 200$ case shown in Fig.~\ref{fig:heatfluxesRe200dT10}, one can note the asymmetries associated with which side of the cylinder the vortex shedding has transferred more heat to at a given time. At the $t = 24$ minutes, there is a discernible pocket of warm temperature around $\theta \approx 220 ^\circ$ on the temperature contour plot (Fig.~\ref{fig:heatfluxesRe200dT10}a) with corresponding greater heat transfer compared with the heat transfer at $\theta \approx 140 ^\circ$ (Fig.~\ref{fig:heatfluxesRe200dT10}b). A similar phenomena can be noticed at $t= 26$ minutes, with the pocket of warm fluid on the opposite side now (Fig.~\ref{fig:heatfluxesRe200dT10}c,d). Because the vortex shedding is oscillatory in nature and takes place at a much faster time scale than the interface evolution, this asymmetrical effect likely evens out over time to produce a symmetrical shape in ice contour. The effects of recirculation are more pronounced at higher Reynolds numbers in the considered flow regimes, which facilitate higher rates of heat transport to the back of the ice, leading to larger temperature gradients and resulting heat fluxes.

\revised{The shape dynamics of a body that is melting/dissolving in flow has been studied theoretically by Huang \etal \cite{mac2015shape} and \citea{moore2017shapes} \cite{moore2017shapes}, utilizing the general driving insight that a dissolving body with a uniform flux (and therefore interface velocity) along the surface will preserve its shape while continuing to shrink in size. They employ conformal mapping approaches in conjunction with free-streamline theory (FST) \cite{hureau1996ideal, alben2004flexibility, moore2013self} to find analytical solutions for the terminal frontal shape, which is found to be cylindrical. The analytical solutions do not apply past the point of boundary layer separation, but a flattened shape at the back has been observed experimentally by \cite{mac2015shape}. The work of Huang \etal \cite{mac2015shape} and analytical solutions derived occur at a much higher Reynolds regime than we consider, and consider the case of a body solely dissolving (rather than our case which features elements of both growth and shrinkage), so strict quantitative agreement of shape may not be expected. Nonetheless, the qualitative similarity of the emerging shape in our results and that acknowledged by \cite{mac2015shape, moore2017shapes} is evident.}

\begin{figure}[H]
	\centerline{\includegraphics[height=0.8\textwidth]{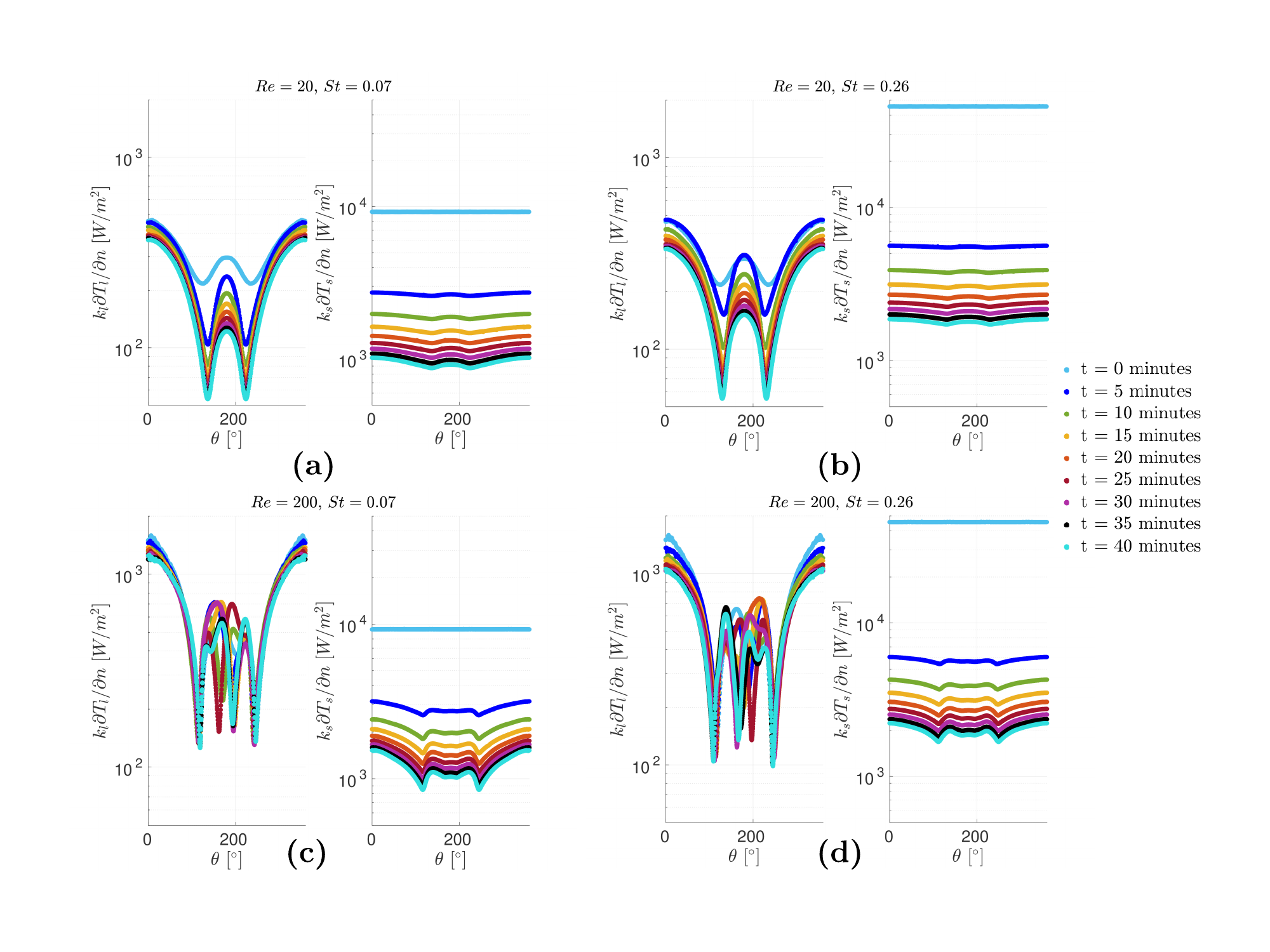} }
	\caption{The effect of varying $Re$ and $St$ on the magnitudes of the fluid heat flux $\liquidflux$ and the solid heat flux $\solidflux$ at the ice-water interface, as shown for $Re = 20, 200$ and $St = 0.07, 0.26$. The left panel of each plot shows the time evolution of the fluid heat flux $\liquidflux$ in time, while the right panel of each plot shows the time evolution of the solid heat flux $\solidflux$ in time. Plots viewed from left to right demonstrate the effect of increasing $St$ for constant $Re$. Plots viewed from top to bottom demonstrate the effect of increasing $Re$ for constant $St$. Colors represent different times. }
	\label{fig:tiled_heat_flux_evos}
\end{figure}

Keeping in mind the effect of recirculation and boundary layer separation, we can now take a closer look at the influence of the Stefan number $St$ on the magnitude of the solid heat flux and the shape of the ice. Because the cooled cylinder is held at a constant temperature (no variation with $\theta$), one would expect that varying the $St$ will produce uniform changes in the solid heat flux $\solidflux$ across the ice's surface. This is highlighted in Fig.~\ref{fig:tiled_heat_flux_evos}, which reports the time evolution of the fluid heat flux $\liquidflux$ and the solid heat flux $\solidflux$ at the ice-water interface for $Re = 20$, $200$ and $St = $ $0.07$, $0.26$. The solid heat flux at the interface is fairly uniform for all cases, aside from some slight nonuniformities which likely develop as the interface develops more nonuniformities in shape due to fluid flow effects discussed previously. Additionally, increasing $St$ results in a uniform increase in solid heat flux at the interface. This is in contrast to the effect of increasing $Re$, which results in not only an increase in average fluid heat flux at the interface, but also in nonuniform changes to the fluid heat flux distribution (in $\theta$) as $Re$ is changed. 

Therefore, the relationship between $St$ and $Re$ also plays a role in the emerging shape as it influences the local evolution of $V_\Gamma$. This can be thought of in terms of uniform solid conductive effects competing with nonuniform fluid convective effects. Regimes with higher $St$ numbers lead to a uniformly higher contribution from the $\solidflux$ term and thus more uniform cylindrical shapes. In contrast, regimes with relatively higher $Re$ result in higher and more nonuniform contributions from the $\liquidflux$ term, yielding interface shapes with more locations of nonuniformity and higher degrees of nonuniformity (\emph{i.e.} higher shape factor). This is of course limited to the cases where the cooled cylinder temperature $T_{\text{cyl}}$ is uniform in $\theta$, and it would be interesting to examine the competition of these effects for a case where $T_{\text{cyl}}$ is nonuniform. For example, one might be able to exert control over the heat fluxes and emerging shape by accounting for the discussed roles of $St$ and $Re$ and applying carefully selected temperature distributions to $T_{\text{cyl}}$.

\subsection{Average heat transfer around the ice-water interface}
\begin{figure}[H]
\centering
	\begin{subfigure}[c]{0.45\textwidth}
		\centerline{\includegraphics[height=0.2\textheight]{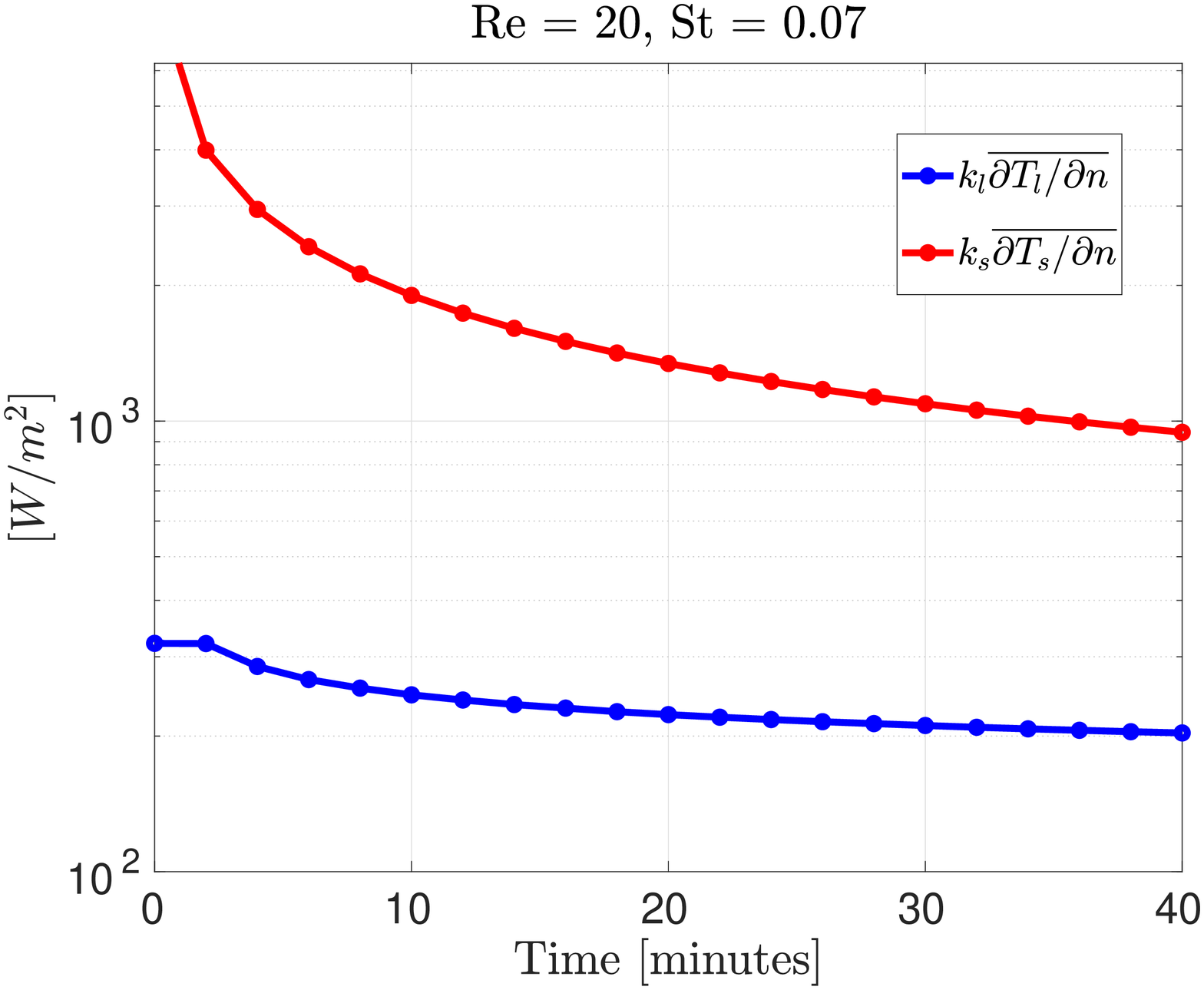}}
		\caption{}
	\end{subfigure}
	\hspace{-3mm}
	\begin{subfigure}[c]{0.45\textwidth}
		\centerline{\includegraphics[height=0.2\textheight]{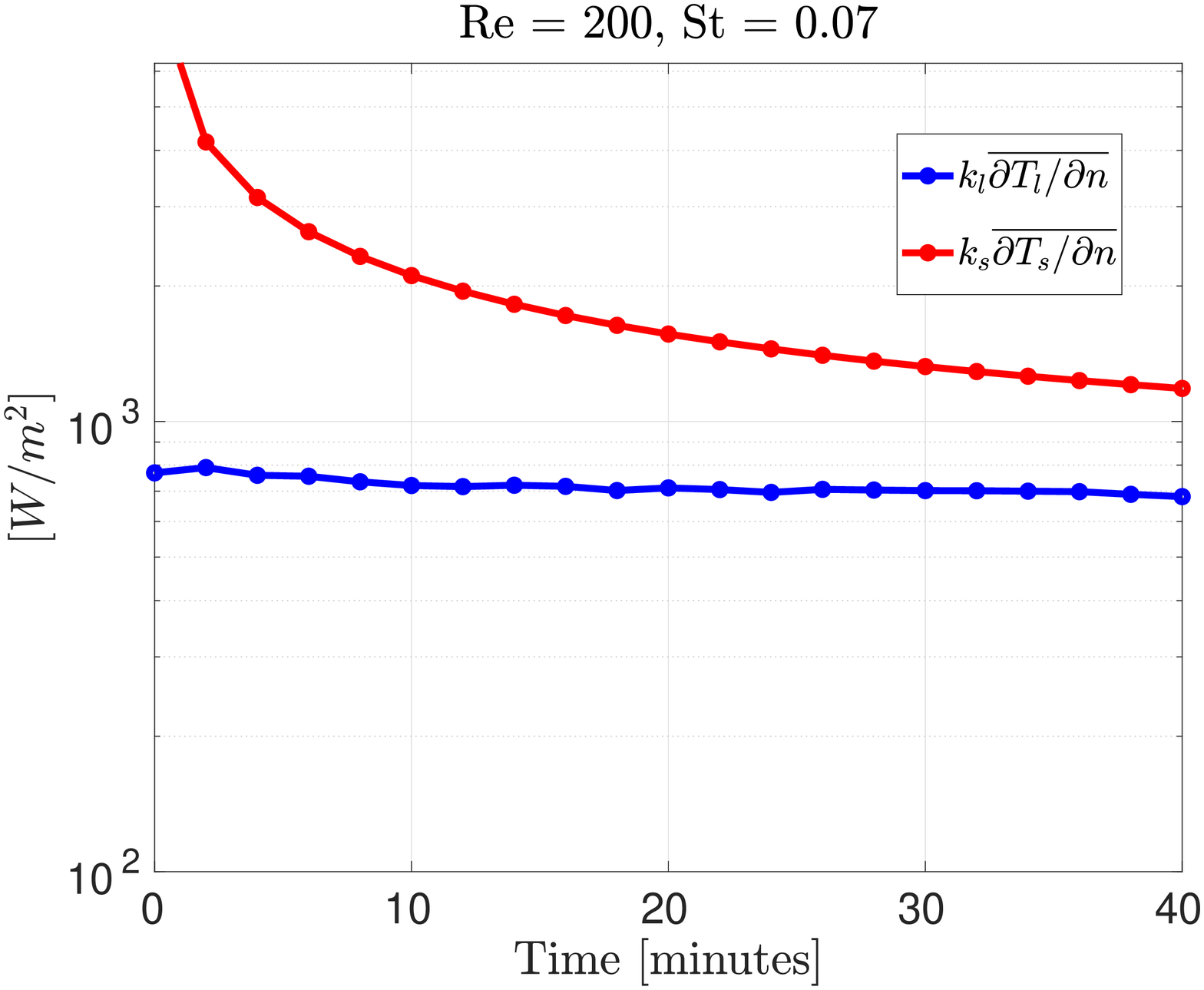}}
		\caption{}
	\end{subfigure}
	\hspace{-3mm}
	\begin{subfigure}[c]{0.45\textwidth}
		\centerline{\includegraphics[height=0.2\textheight]{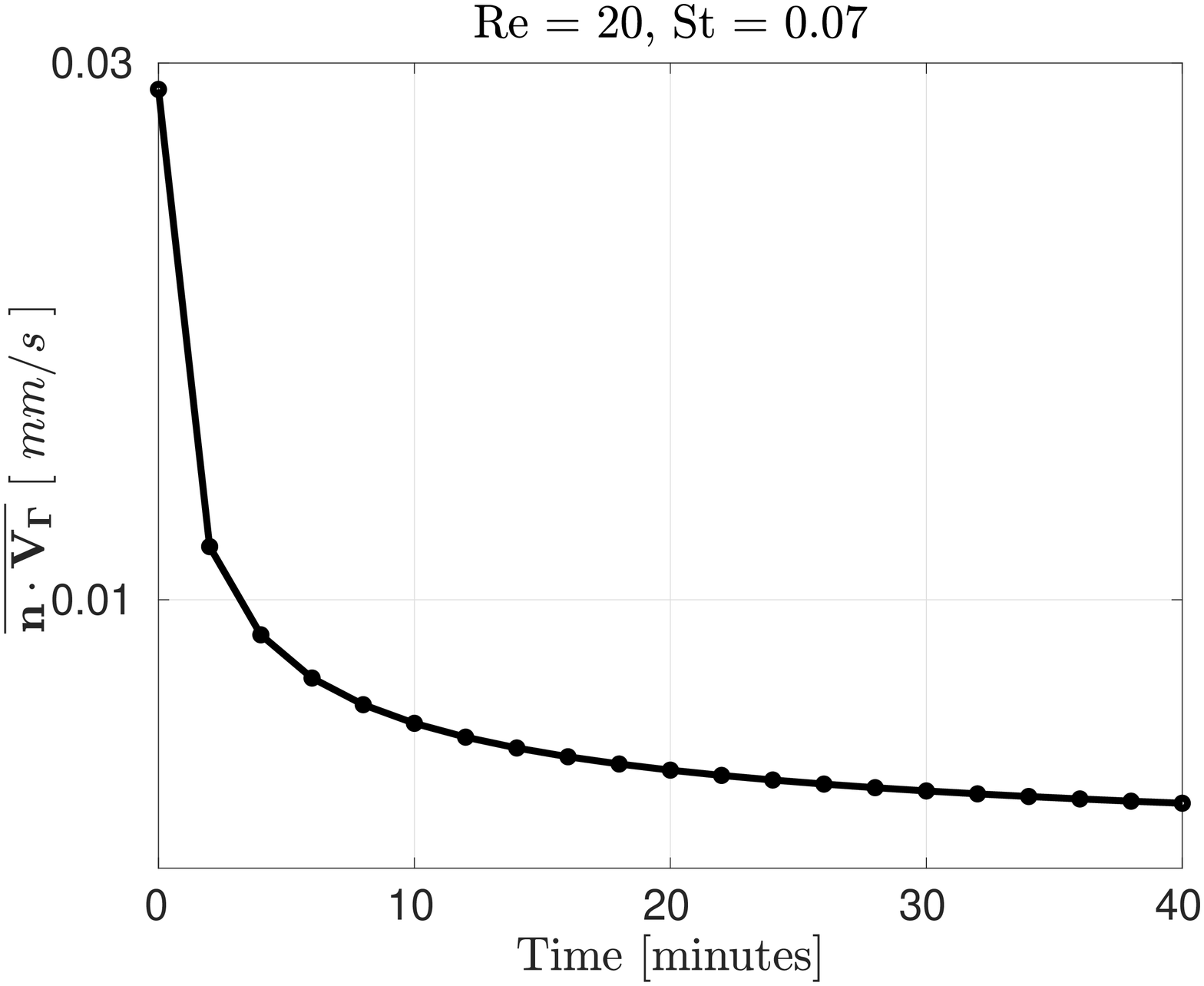}}
		\caption{}
	\end{subfigure}
	\hspace{-3mm}
	\begin{subfigure}[c]{0.45\textwidth}
		\centerline{\includegraphics[height=0.2\textheight]{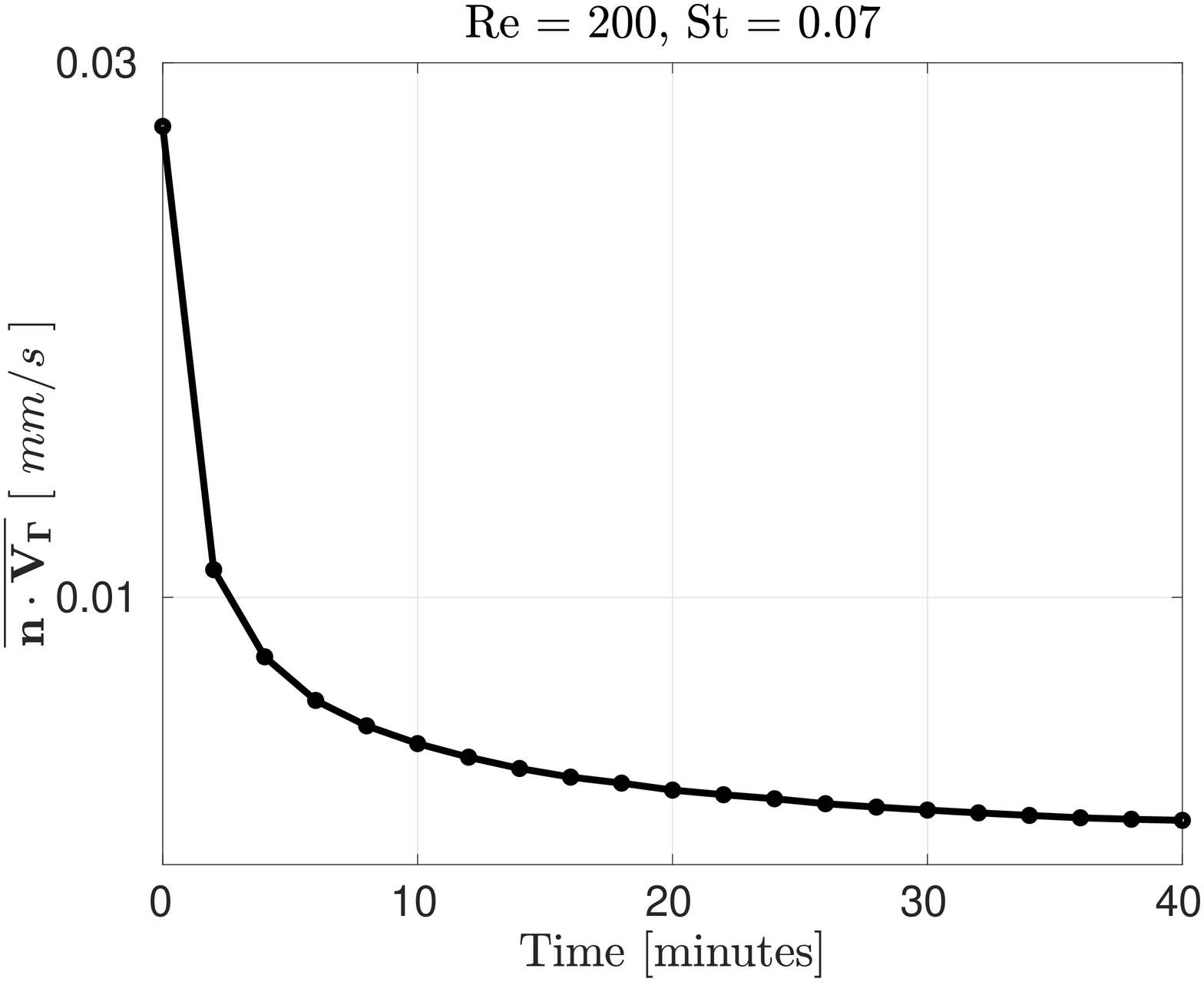}}
		\caption{}
	\end{subfigure}
	
	\caption{Time evolution of heat fluxes at the interface and interfacial velocity for $Re = 20$ and $Re = 200$ with $St = 0.07$. (a)-(b) Solid (red) and fluid (blue) heat fluxes at the interface. (c)-(d) Average interfacial velocity.  Although the interfacial velocities approach very near to zero, especially relative to flow velocities $u_{\infty} \sim 1 - 10 $ mm/s, the difference in fluid and solid heat fluxes can remain significant, and it may require much longer timescales in order to reach a true steady state, corresponding to a vanishing interfacial velocity. }
	\label{fig:heatfluxevo}
\end{figure}

During the formation of ice around the cooled cylinder, a particular quantity of interest is the Nusselt number $Nu_d$, which describes the ratio of convective to conductive heat transfer in the fluid. The local Nusselt number is defined as $Nu_d = h k_l / d$, where $d$ is the diameter of the cylinder and $h$ is the local heat transfer coefficient defined as $h = (k_l \partial T_l / \partial n)/ (T_\Gamma - T_\infty)$. Accurate calculation of the Nusselt number neccessitates measurement of the local fluid temperature gradients at the interface, which remains an experimental challenge. Previous experimental evaluations of the Nusselt number for this problem have relied on making a steady state assumption that if the interfacial velocity is close to zero ($V_\Gamma \sim 0$), $k_l \partial T_l / \partial n = k_s \partial T_s / \partial n$, and therefore the solid heat flux $\solidflux$, which is much easier to measure, can be used to calculate the Nusselt number instead. However, Fig.~\ref{fig:heatfluxevo} highlights that this assumption may not be the most reliable one, as the difference between solid and fluid heat fluxes at the interface can still be quite large for very small interfacial velocities and the steady state may not be a condition that is easily reached because of the long time scales, or checked because very small $V_\Gamma$ are still not near the steady state. This is a case where the numerical tool can provide new insight, as the fluid heat fluxes at the interface are readily available from simulation data, and we need not rely on any assumptions to compute $Nu_d$. 

\begin{figure}[H]
\centering
	\begin{subfigure}[c]{0.45\textwidth}
		\centerline{\includegraphics[height=0.25\textheight]{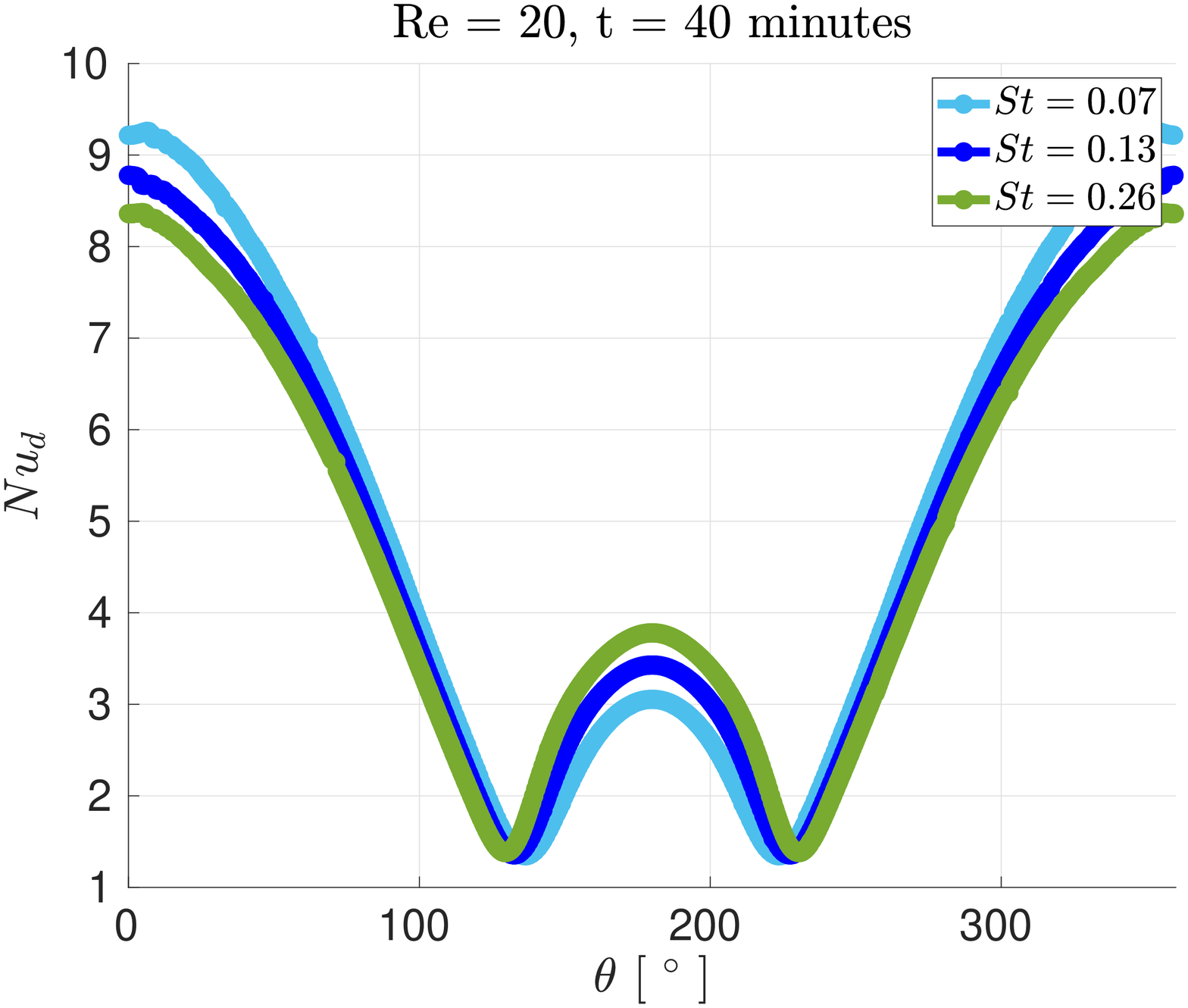}}
		\caption{}
	\end{subfigure}
	\hspace{-3mm}
	\begin{subfigure}[c]{0.45\textwidth}
		\centerline{\includegraphics[height=0.25\textheight]{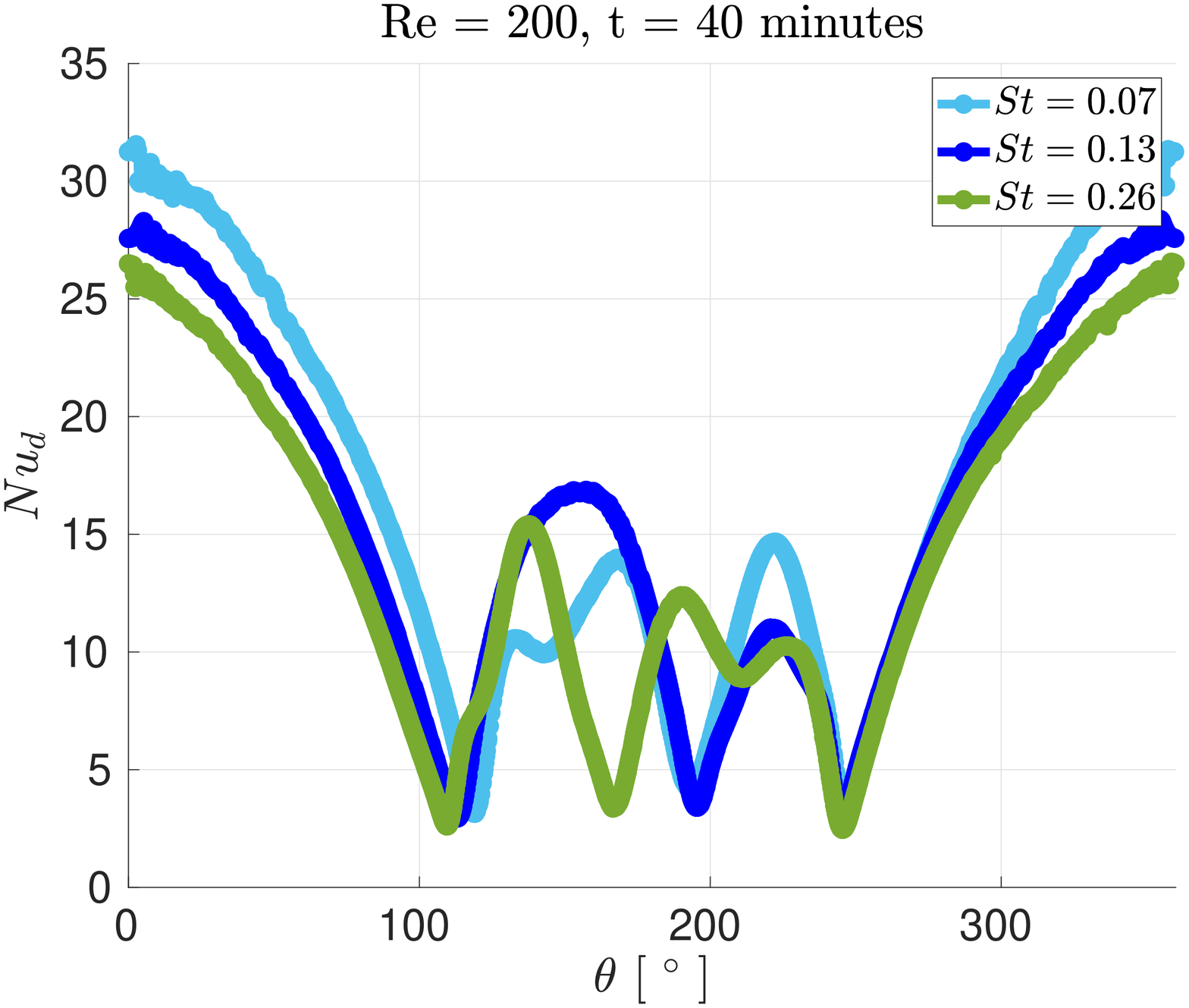}}
		\caption{}
	\end{subfigure}
		\begin{subfigure}[c]{0.45\textwidth}
		\centerline{\includegraphics[height=0.25\textheight]{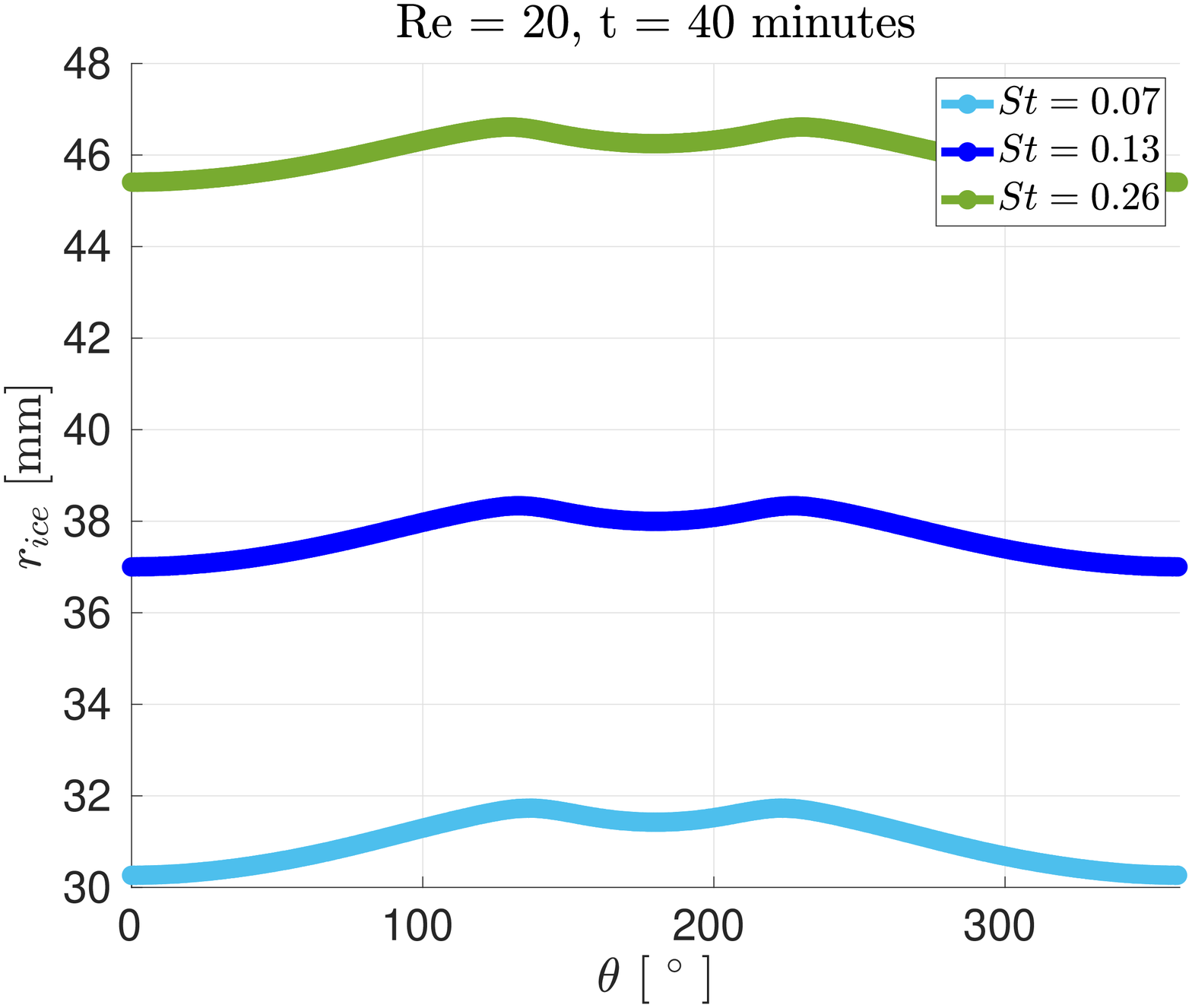}}
		\caption{}
	\end{subfigure}
	\hspace{-3mm}
	\begin{subfigure}[c]{0.45\textwidth}
		\centerline{\includegraphics[height=0.25\textheight]{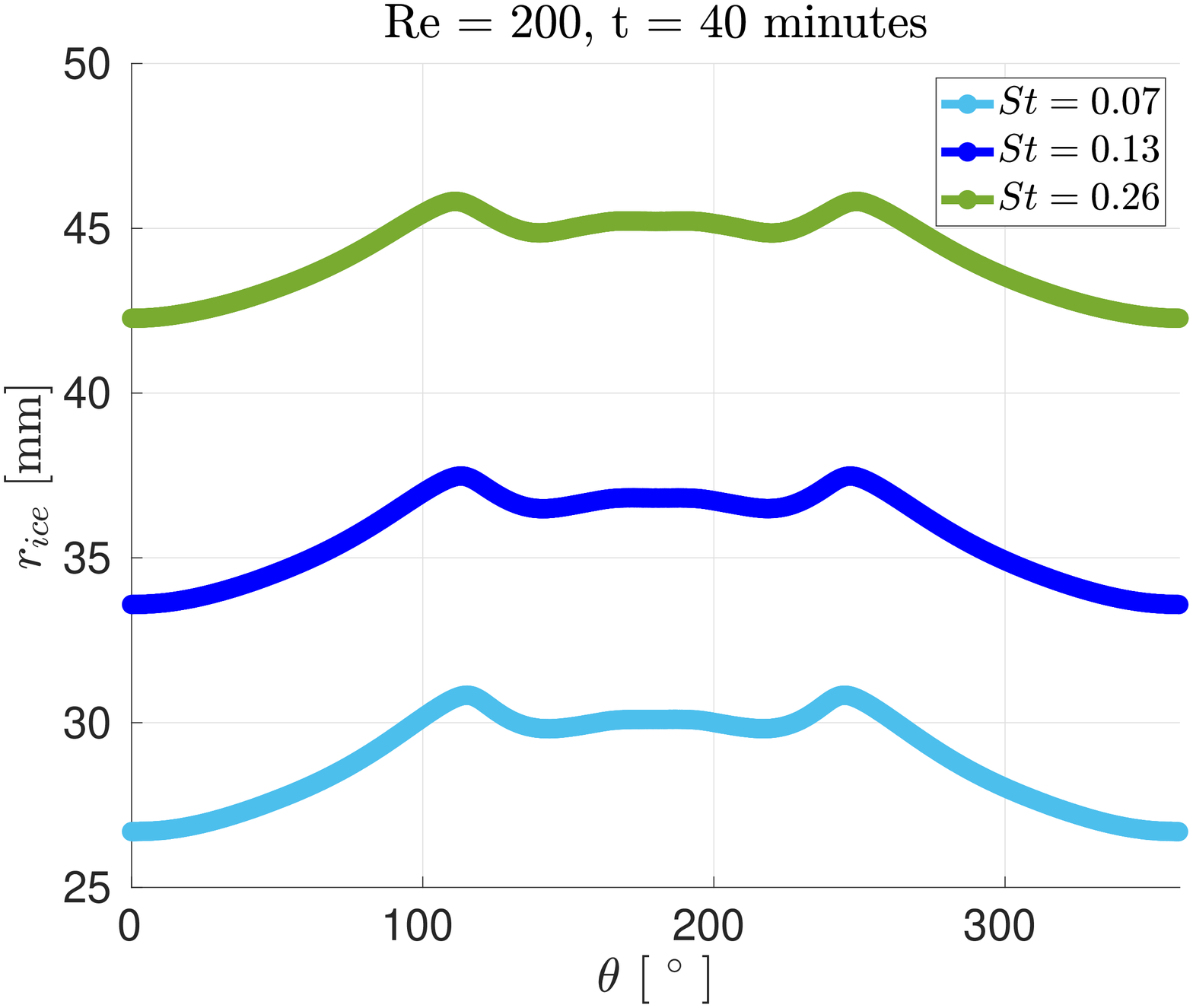}}
		\caption{}
	\end{subfigure}
	\caption{Local $Nu_d$ distributions at $t = 40$ minutes for varying $St$ number for (a) $Re = 20$ and (b)$Re = 200$, and $r_{ice}$ distributions at $t = 40$ minutes for varying $St$ number for (c) $Re = 20$ and (d)$Re = 200$.}
	\label{fig:final_time_Nusselts}
\end{figure}

Cheng \etal  \cite{cheng1981experimental} proposed an empirical correlation to describe the average Nusselt number at the ice's surface, $\overline{Nu_d}$, for a range of $Re$ and cooling temperature ratios (which can be considered an analog to the Stefan number), where $\overline{Nu_d}$ was found to increase with both $Re$ and $St$. We recover a similar trend as \cite{cheng1981experimental} in terms of $Re_d$ -- the average Nusselt number $\overline{Nu_d}$ scales with $\sim Re_d^{0.5}$. However, in terms of $St$ (or cooling temperature ratio), we obtain an opposite trend. We see that as $\Delta T$ and thus $St$ increases, the average Nusselt decreases, while \cite{cheng1981experimental} reports the opposite. This discrepancy may be explained by the use of the steady state assumption by \cite{cheng1981experimental} in computing Nusselt number at the ice interface as discussed previously, and by noting again the trends observed for local fluid heat flux in Fig.~\ref{fig:heatfluxevo}. As seen in Fig.~\ref{fig:heatfluxevo}, increasing $St$ (and correspondingly, $\Delta T$) does indeed result in an increase in the \emph{solid} heat flux at the interface; however, the same trend is not apparent for the fluid heat flux. If a steady state assumption was made and the solid heat flux was used to compute the Nusselt number instead of the fluid heat flux, it follows that one might arrive at such a trend as reported in \cite{cheng1981experimental}. Access to the fluid heat fluxes via simulation data allows us to directly capture the trends in fluid heat flux, and therefore provide a more accurate relation for changes in $\overline{Nu_d}$ with respect to $St$. However, one must then consider why an increase in $St$ results in a decrease, albeit slight, in the fluid heat flux. One possible explanation might be reached by considering how $St$ affects the thickness of the ice, and in turn, the resulting flow dynamics. For example, we have noted that a larger value of $St$ results in a larger magntidue of the solid heat flux $\solidflux$, which in turn leads to a larger interfacial velocity $V_\Gamma$ and results in a larger ice thickness for the same given time compared with a lower $St$ value. If the thickness of the ice is larger, then the effective radius of the body $r_{ice}$ seen by the flow, corresponding to the sum of the radius of the cylinder $r_{cyl}$ and the ice thickness, will be larger for a given time. Recalling that the thermal boundary layer is of the same order as the momentum boundary layer ($\delta_T \sim \delta_m $) and the momentum boundary layer scales with the radius of the body as $\delta_m \sim (r_{ice} ) / \sqrt{Re_d}$, one may note that this increase in ice thickness will result in a larger boundary layer. A thicker thermal boundary layer will also imply a thicker region over which the fluid temperature must go from the interfacial value $T_{\Gamma}$ to the freestream temperature $T_{\infty}$, thus leading to smaller gradients in fluid temperature. This effect results in a lower value of fluid heat flux $\liquidflux$, and consequently a lower local Nusselt number $Nu_d$, in all the regions where the boundary layer remains attached (\emph{i.e.} $\theta \approx 0 - 135, 225 - 360 $), as is illustrated in Fig.~\ref{fig:final_time_Nusselts}, which in turn leads to a lower values of average Nusselt number for increasing values of $St$.

\begin{figure}[H]
\centerline{\includegraphics[height=0.25\textheight]{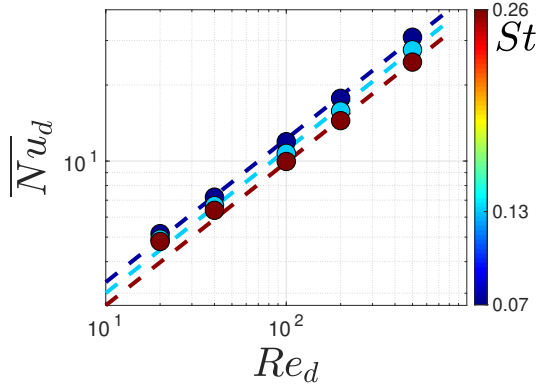}}
\caption{Simulation data and proposed correlation given by equation \eqref{eq:mynusseltcorrelation} are plotted for varying $Re$ and $St$. The circles represent simulation data, and the dotted lines represent the proposed correlation at different values of $St$. The colors correspond to different values of $St$. }
\label{fig:mynusselt}
\end{figure}

Therefore, our numerical results allow us to present a new correlation using the average $\overline{Nu_d}$ at the final time of the simulations, $t = 40$ minutes. We report that for all the cases except $Re = 500$, the average Nusselt number $Nu_d$ at the final simulation time is not changing by more than $\approx 1 \%$ over a span of $2$ minutes, and for the $Re = 500$ cases it is not changing by more than $\approx 5 \%$ over a span of $2$ minutes. Our proposed correlation is given by
\begin{equation}
\overline{Nu_d} = 0.59 \text{ } Re^{0.57} \text{ } St^{- 0.16}, 
\label{eq:mynusseltcorrelation}
\end{equation}
where the average Nusselt number over the ice's surface is computed as $\overline{Nu_{d}} = \frac{1}{2 \pi} \int_0^{2 \pi} {Nu}_{d}$ d$\theta$. The simulation data is plotted along with the proposed correlation in Fig.~\ref{fig:mynusselt}.

\section{\revised{Melting porous media}}
\label{sec:porousmedia}
Lastly, we briefly explore an example to highlight the capability of the method applied to multi-body problems of higher complexity. We examine a case of $30$ cylindrical solid bodies melting in a crossflow to represent a porous media, as is similarly done by de Anna \etal \cite{de2017prediction}.  We consider a square domain of size $[0, 2]$ $\times$ $[0, 2]$ mm, with solid bodies (referred to as ``grains'') of varying initial size such that the initial porosity of the domain is $60 \%$, and examine the case where the solid is ice and the fluid is water. The characteristic length scale for the problem is taken to be $2$ mm, representing the length of the domain. We prescribe a free-stream temperature $T_\infty = 275 $ $K$ at the inlet, and a constant pressure drop $\Delta P = 10$ $Pa$ across the domain in the $x$-wise direction from left to right. On the top and bottom walls, we apply a no-slip condition on fluid velocity, a homogeneous Neumann condition on pressure, and a homogeneous Neumann condition on the temperature. We prescribe a homogeneous Neumann condition on both temperature and velocity at the right-most wall. We take the initial temperature of the ice $T_0$ to be $272$ $K$, and the initial temperature of the water to be $273$ $K$.

It is additionally worth noting that because we specify a pressure drop rather than a free-stream velocity, the characteristic fluid velocity and therefore Reynolds number of the problem is initially unknown, and thus we require a different forumlation of the governing equations that does not include the Reynolds number. Rather than non-dimensionalizing velocity by a characteristic fluid velocity, we choose to nondimensionalize velocity using the thermal diffusivity as $\hat{\vns} = \vns / (\alphal / l)$ instead (as is done in \cite{zabaras2004stabilized}), where $l$ is the characteristic length scale of the problem. 
This yields the following change to the dimensionless system of equations given in Eqs.\eqref{eq:finalIncompressible} - \eqref{eq:finalNSInterfaceBC}: 
\begin{itemize}
\item \eqref{eq:finalNS} becomes $\ddt{\vns} + \advection{\vns}{\vns} = Pr \nabla^{2}\vns - \nabla P $ 
\item \eqref{eq:finalHeatLiq} becomes $\ddt{\temperaturel} + \advection{\vns}{\temperaturel} = \laplace{\temperaturel}$
\item \eqref{eq:finalHeatSol} becomes $\ddt{\temperatures} = \frac{\alpha_s}{\alpha_l} \laplace{\temperatures} $
\item \eqref{eq:finalStefan} becomes $\normal \cdot \vint = St \frac{\alphas}{\alphal}  \left(\nabla \temperatures -  \frac{k_l}{k_s} \nabla \temperaturel \right) \cdot \normal$
\item Equations \eqref{eq:finalIncompressible}, \eqref{eq:finalGibbsThomson}, \eqref{eq:finalNSInterfaceBC} remain unchanged
\end{itemize}
where $Pr$ is the Prandtl number given by $Pr = \mu_l /( \rhol \alphal)$. 

\begin{figure}[H]
\centering
	\begin{subfigure}[c]{0.33\textwidth}
		\centerline{\includegraphics[height=0.25\textheight]{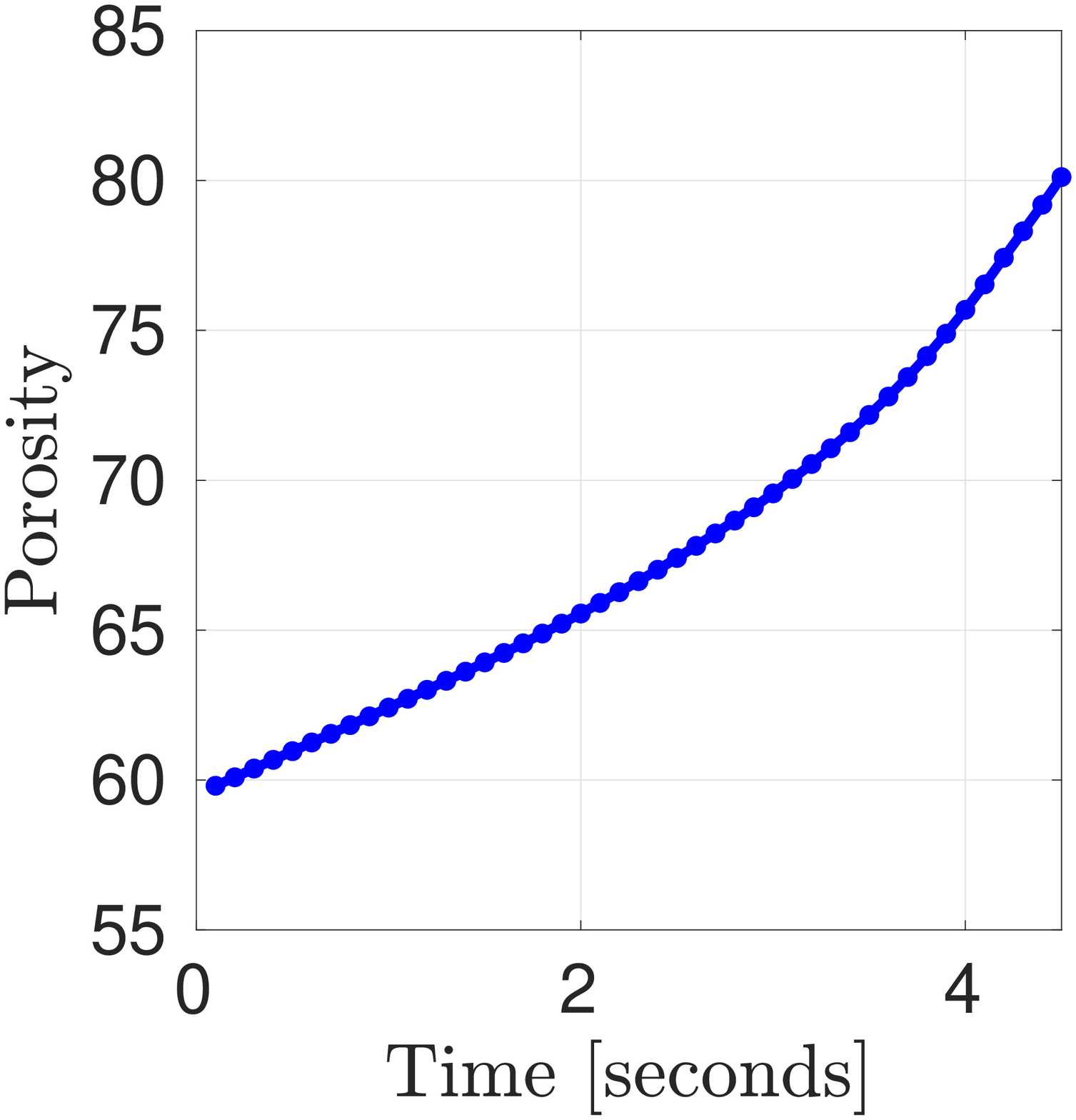}}
		\caption{}
	\end{subfigure}
	\hspace{-3mm}
	\begin{subfigure}[c]{0.33\textwidth}
		\centerline{\includegraphics[height=0.25\textheight]{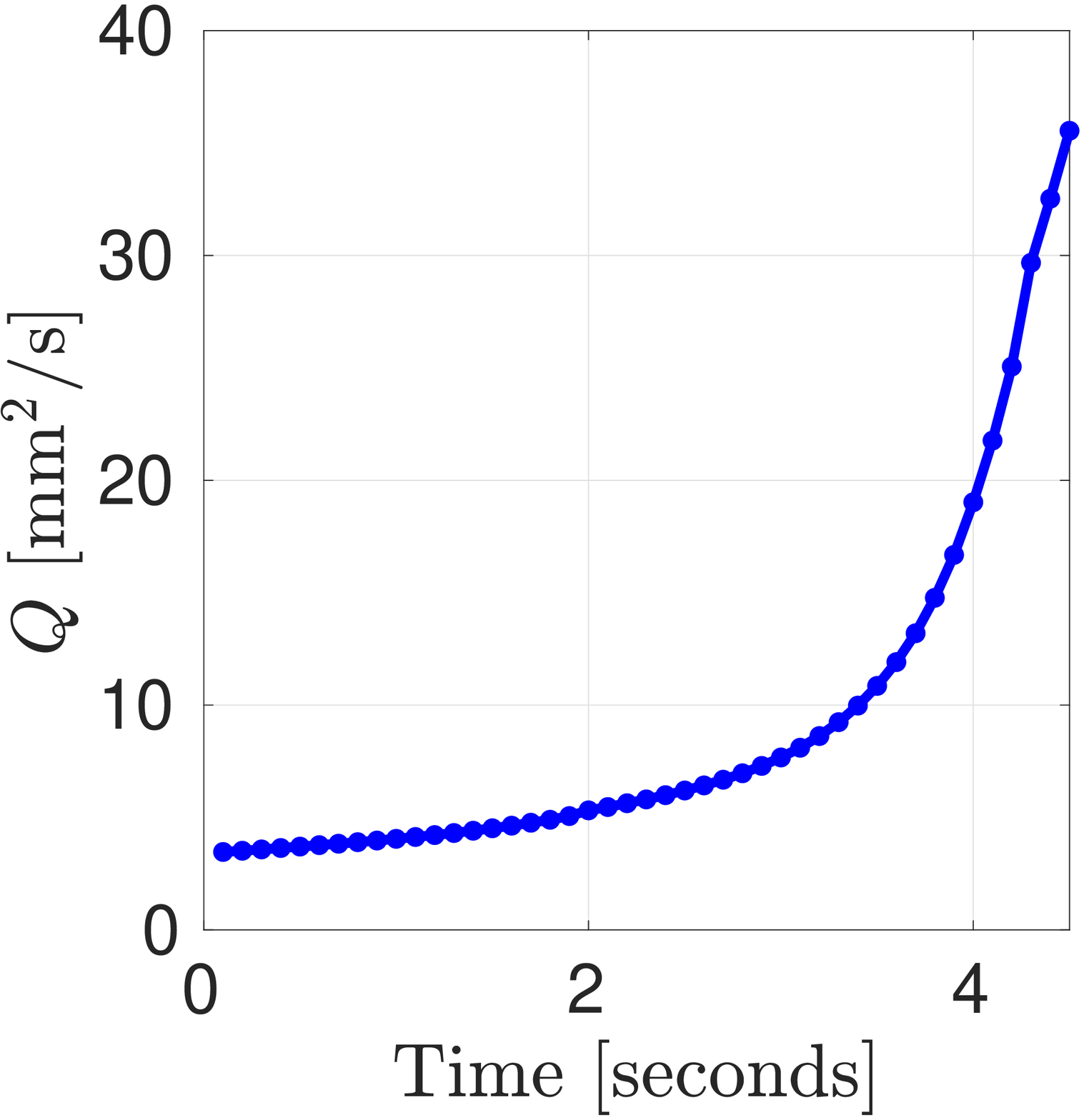}}
		\caption{}
	\end{subfigure}

	\caption{Time evolution of (a) porosity and (b) flow rate across the outlet of the domain.  }
	\label{fig:PoroAndQ}
\end{figure}

\begin{figure}[H]
\centering
\vspace{-10mm}
\centerline{\includegraphics[width=\textwidth]{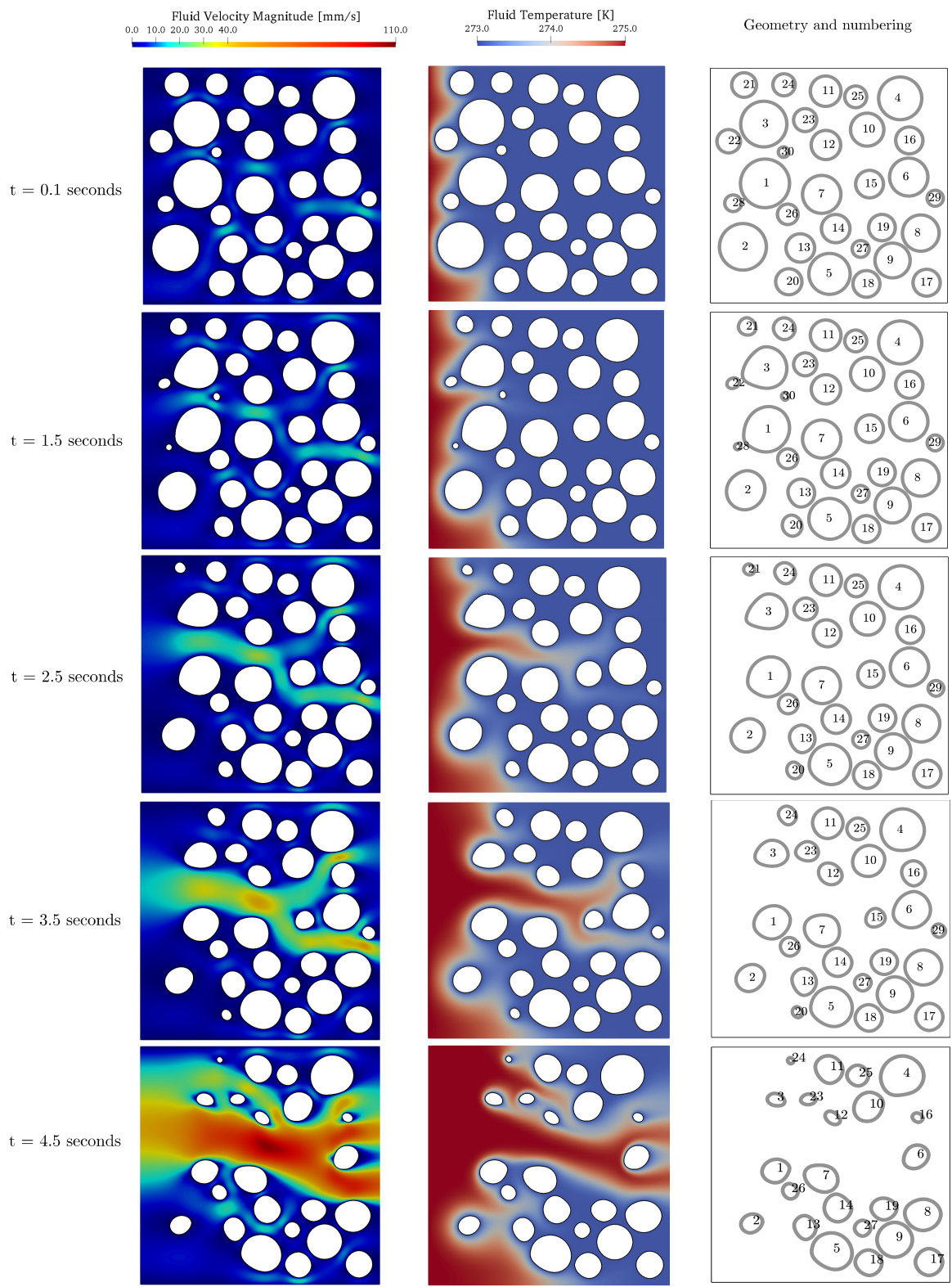}}
\caption{30 bodies melting in flow, highlighting the (left) fluid velocity magnitude, (middle) fluid temperature field, and (right) geometry and numbering at $t = 0.1, 1.5, 2.5, 3.5 $ and $4.5$ seconds.}
\label{fig:PoroEvoInTime}
\end{figure}

The simulation is carried out from $t = 0$ to $t = 4.5 $ seconds. The time evolution of porosity and flow rate across the outlet are plotted in Fig.~\ref{fig:PoroAndQ}. Snapshots in time of the fluid velocity magnitude, fluid temperature field, and body geometry for reference are shown in Fig. \ref{fig:PoroEvoInTime}. We compute a resulting Reynolds number as $Re = \rhol \bar{u} \bar{d}/ \mul$, where $\bar{u}$ is the average fluid velocity across the outlet, and $\bar{d}$ is the average body diameter for a given timestep. The resulting Reynolds number is $Re = 0.4$ at the start of the simulation, and reaches $Re \approx 3.5$ by the end.

\begin{figure}[H]
		\centerline{\includegraphics[height=0.4\textheight]{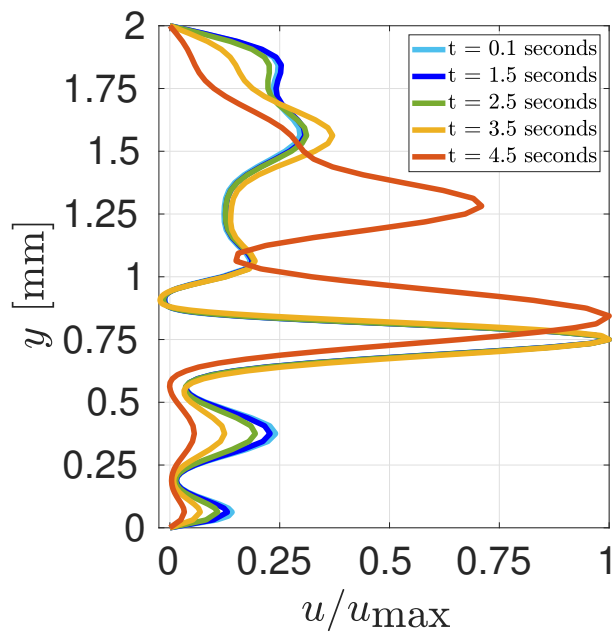}}
		\caption{Fluid velocity profile $u/u_{\textrm{max}}$ across the outlet at $t = $ $0.1$ s, $1.5$ s, $2.5$ s, $3.5$ s, and $4.5$ s.}
		\label{fig:PoroVelOutlet}
\end{figure}

Initially, the flow permeates the geometry along several main pathways, as can be seen via regions of higher fluid velocity magnitude in the topmost leftmost panel of Fig.~\ref{fig:PoroEvoInTime}. As time evolves, there is an initial gradual increase in porosity accompanied by an increased flow rate as bodies shrink and the pore space becomes larger, as illustrated in Fig.~\ref{fig:PoroAndQ}. Irregular geometries begin to emerge as a result of the local pore size and resulting regions of higher or lower flow velocity and thus heat transport. For example, bodies $1$ and $3$ both have very close neighbors ($22$ and $28$) in the upstream direction, diverting the flow and thus heat transport along their surfaces parallel to the flow. This results in a higher rate of melting along these surfaces parallel to the flow, whilst the upstream-facing side of both bodies remains closer to its initial size. However, at later times $t = 2.5$ s and $3.5$ s, the neighbors have disappeared and the melting process smooths out these regions of high curvature along the front of both bodies. Body $13$ sees a similar effect -- the flow transports more heat through the pores along its left and top sides resulting in a flatter surface in these regions. 
Additionally, we begin to observe channelization effects as pores grow larger due to the disappearance of several bodies and the flattening of surfaces near regions of higher fluid velocity. This is notable when considering the transition between $t = 1.5$ s and $2.5$ s in Fig.~\ref{fig:PoroEvoInTime}, when bodies $22$ and $30$ disappear and the pore between $4$ and $16$ widens, leading to one visible main channel which branches into two when the flow reaches body $15$ . By $t = 4.5$ s, bodies $15$ and $29$ have also vanished, and the space between $6$ and $16$ continues to widen, resulting in even higher flow velocities in the region. This diversion of the flow is also reflected in Fig.~\ref{fig:PoroVelOutlet}, which plots the fluid velocity profile $u/u_{\textrm{max}}$ across the outlet for each snapshot in time, where $u_{\textrm{max}}$ is the maximum fluid velocity for the given time. 
At the outlet initially, there are $6$ peaks in fluid velocity corresponding to channels formed between body $4$ and the wall, bodies $4$ and $16$, bodies $6$ and $29$, bodies $29$ and $8$, bodies $8$ and $17$, and body $17$ and the wall. As time increases, higher fractions of the flow at the outlet are diverted to y locations corresponding with channels between bodies $4$ and $16$ and bodies $6$ and $8$. This relatively rapid increase in prominence of main channels and resulting higher fluid velocities corresponds with the exponential increase of the flow rate as seen in Fig.\ref{fig:PoroAndQ}(b). 

This example highlights the promising nature of the method for simulation of reactive porous media flow. Future work could include application of the method to problems with much larger porous media, and extension of the approach to 3D. 

\section{Conclusion}\label{sec:conclusion}
In summary, in this paper we present a sharp numerical method for the solution of the Stefan problem coupled with an incompressible fluid flow, which utilizes finite-difference and finite-volume discretizations on adaptive quadtree grids, level-set methods for sharp interface capturing, and a pressure-free projection method for solution of the incompressible Navier Stokes equations. The numerical approach provides an advantage in accuracy and efficiency by combining three main elements -- (a) use of adaptive grids for efficient computation of a problem which is multiscale in nature, (b) accurate computation of gradients at the interface which govern the interfacial velocity, and (c) sharp interface representation which allows for accurate application of interfacial boundary conditions. The method is first verified with convergence tests using a synthetic numerical solution, and then validated for ice growth on a cylinder in cross flow for which we find good quantitative agreement between the simulation and experimental results. 
Then, we use the numerical tool to investigate the role of the Reynolds and the Stefan numbers on the evolution of interface morphologies, flow dynamics, and heat transfer near the interface. We recapture the qualitative phenomena of a self-reinforcing relationship between boundary layer separation and interface shape reported by previous experiments in both freezing and dissolution. The effects of $Re$ and $St$ are explored by discussing their uniform and nonuniform effects on the resulting shape and heat transfer, which have interesting implications about the possibility to control the shape of the interface. For example, in this case the effect of the $St$ was likely largely uniform due to the uniformity of the cooled cylinder temperature; however, it might be interesting to explore how a nonuniform temperature distribution applied to the cylinder could modify the way that the heat transport via solid diffusion and fluid advection interact. Additionally, we highlight the ways in which simulation data can leveraged to provide more insight into the interfacial temperature gradients, and offer a new scaling relation for average Nusselt number at the interface $\overline{Nu_d}$ as it relates with $St$ and $Re$. \revised{Lastly, we apply the presented method to the case of a melting porous media, illustrating the approach utilized on a problem with 30 evolving bodies. The method is able to capture relevant features of the problem with relative ease, such as vanishing of bodies and channelization effects. }

It is worth noting that the current method focuses on the effects of forced convection; however, it may be adapted in the future to include a Boussinesq approximation for variable density effects associated with natural convection \cite{huang2021a}. Future work may also aim to extend the current approach in a more general sense to study other types of interfacial growth phenomena coupled with flow, like erosion/deposition, as well as to study cases with more complex geometries (\emph{i.e.} porous media). Additionally, the method may be extended in the future to capture interfacial growth governed by multiple scalar fields (\emph{i.e.} temperature and concentration fields), which may have relevant applications to problems such as multialloy solidification.

\section{Acknowledgements}
This research was funded by ONR N00014-11-1-0027. Additionally, use was made of computational facilities purchased with funds from the National Science Foundation (CNS-1725797) and administered by the Center for Scientific Computing (CSC). The CSC is supported by the California NanoSystems Institute and the Materials Research Science and Engineering Center (MRSEC; NSF DMR 1720256) at UC Santa Barbara. 

\section{References}
\bibliography{./ref_stefan_fluids_paper}
\bibliographystyle{unsrtnat}
\nocite{*}
\end{document}